\documentclass[twocolumn]{aastex63}
\usepackage{graphicx}
\usepackage{natbib}
\usepackage{amsmath}
\usepackage{color}
\usepackage{ulem}
\usepackage{lineno}
\usepackage[caption=false]{subfig} 
\usepackage{subfloat}
\usepackage{CJK}

\makeatother
\begin{document}
\begin{CJK*}{UTF8}{gbsn}

\newcommand{\tocite}[1]{{\tt cite:} #1}.

\makeatletter

\newcommand{\logg}{{\rm log}\,g}
\newcommand{\Teff}{T_{\rm eff}}
\newcommand{\feh}{[{\rm Fe}/{\rm H}]}
\newcommand{\ohh}{[{\rm O}/{\rm H}]}
\newcommand{\mgh}{[{\rm Mg}/{\rm H}]}
\newcommand{\afe}{[\alpha/{\rm Fe}]}
\newcommand{\ofe}{[{\rm O}/{\rm Fe}]}
\newcommand{\mgfe}{[{\rm Mg}/{\rm Fe}]}
\newcommand{\femg}{[{\rm Fe}/{\rm Mg}]}
\newcommand{\mgfepl}{[{\rm Mg}/{\rm Fe}]_{\rm pl}}
\newcommand{\femgpl}{[{\rm Fe}/{\rm Mg}]_{\rm pl}}

\newcommand{\xh}{[{\rm X}/{\rm H}]}
\newcommand{\xxh}[1]{[{\rm #1}/{\rm H}]}
\newcommand{\xfe}{[{\rm X}/{\rm Fe}]}
\newcommand{\xxfe}[1]{[{\rm #1}/{\rm Fe}]}
\newcommand{\xmg}{[{\rm X}/{\rm Mg}]}
\newcommand{\xxmg}[1]{[{\rm #1}/{\rm Mg}]}
\newcommand{\xoverh}{\left({\rm X} \over {\rm H}\right)}
\newcommand{\mgoverh}{\left({\rm Mg} \over {\rm H}\right)}
\newcommand{\Xj}{{\rm X}_j}
\newcommand{\Xhj}{[{\rm X}_j/{\rm H}]}
\newcommand{\Xmgj}{[{\rm X}_j/{\rm Mg}]}

\newcommand{\Acc}{A_{\rm cc}}
\newcommand{\AIa}{A_{\rm Ia}}
\newcommand{\Accst}{A_{\rm cc,*}}
\newcommand{\AIast}{A_{\rm Ia,*}}
\newcommand{\pxcc}{p_{\rm cc}^X}
\newcommand{\pxIa}{p_{\rm Ia}^X}
\newcommand{\pxccs}{p_{\rm cc,\odot}^X}
\newcommand{\pxj}{p_j^X}
\newcommand{\pxIas}{p_{\rm Ia,\odot}^X}
\newcommand{\pmgcc}{p_{\rm cc}^{\rm Mg}}
\newcommand{\pmgIa}{p_{\rm Ia}^{\rm Mg}}
\newcommand{\pfecc}{p_{\rm cc}^{\rm Fe}}
\newcommand{\pfeIa}{p_{\rm Ia}^{\rm Fe}}
\newcommand{\qxcc}{q^X_{\rm cc}}
\newcommand{\qxIa}{q^X_{\rm Ia}}
\newcommand{\qxj}{q^X_j}
\newcommand{\qmgcc}{q_{\rm cc}^{\rm Mg}}
\newcommand{\qmgIa}{q_{\rm Ia}^{\rm Mg}}
\newcommand{\qfecc}{q_{\rm cc}^{\rm Fe}}
\newcommand{\qfeIa}{q_{\rm Ia}^{\rm Fe}}
\newcommand{\qmuj}{q_{\mu,j}}

\newcommand{\Aagb}{A_{\rm AGB}}
\newcommand{\AOhat}{\hat{A}_1}
\newcommand{\AThat}{\hat{A}_2}
\newcommand{\xjhat}{\hat{x}_j}

\newcommand{\fxcc}{f^X_{\rm cc}}
\newcommand{\fxccsun}{f^X_{\rm cc,\odot}}
\newcommand{\ffeccsun}{f^{\rm Fe}_{\rm cc,\odot}}
\newcommand{\ffecc}{f^{\rm Fe}_{\rm cc}}
\newcommand{\Rhigh}{R_{\rm high}}
\newcommand{\Rlow}{R_{\rm low}}

\newcommand{\SNR}{{\rm SNR}}

\newcommand{\Gyr}{\,{\rm Gyr}}
\newcommand{\Myr}{\,{\rm Myr}}
\newcommand{\yr}{\,{\rm yr}}
\newcommand{\kpc}{\,{\rm kpc}}
\newcommand{\kms}{\,{\rm km}\,{\rm s}^{-1}}
\newcommand{\K}{\,{\rm K}}

\title{Chemical Cartography with APOGEE: Mapping Disk Populations
with a Two-Process Model and Residual Abundances}
\author[0000-0001-7775-7261]{David H.~Weinberg}
\affiliation{Department of Astronomy and Center for Cosmology
and AstroParticle Physics, The Ohio State University, 
Columbus, OH 43210, USA}
\affiliation{Institute for Advanced Study, Princeton, NJ 08540, USA}
\author{Jon A.\ Holtzman}
\affiliation{Department of Astronomy, New Mexico State University,
Las Cruces, NM 88003, USA}
\author{Jennifer A.\ Johnson}
\affiliation{Department of Astronomy and Center for Cosmology
and AstroParticle Physics, The Ohio State University, 
Columbus, OH 43210, USA}
\author{Christian Hayes}
\affiliation{Department of Astronomy, University of Washington,
Seattle, WA 98195, USA}
\author{Sten Hasselquist}
\affiliation{Department of Physics \& Astronomy, University of Utah,
Salt Lake City, UT 84112,USA}
\author{Matthew Shetrone}
\affiliation{University of California, Santa Cruz, UCO/Lick Observatory,
1156 High St., Santa Cruz, CA 95064, USA}
\author{Yuan-Sen Ting (丁源森)}
\affiliation{Institute for Advanced Study, Princeton, NJ 08540, USA}
\affiliation{Department of Astrophysical Sciences, Princeton University, Princeton, NJ 08540, USA}
\affiliation{Observatories of the Carnegie Institution of Washington, 813 Santa Barbara Street, Pasadena, CA 91101, USA}
\affiliation{Research School of Astronomy \& Astrophysics, Australian National University, Cotter Rd., Weston, ACT 2611, Australia}
\author{Rachael L.\ Beaton}
\affiliation{The Observatories of the Carnegie Institution for Science, 813 Santa Barbara Street, Pasadena, CA 91101, USA}
\author{Timothy C.\ Beers}
\affiliation{Department of Physics and JINA Center for the Evolution of the Elements, University of Notre Dame, Notre Dame, IN 46556, USA}
\author{Jonathan C. Bird}
\affiliation{Department of Physics and Astronomy, Vanderbilt University, VU Station 1807, Nashville, TN 37235, USA}
\author{Dmitry Bizyaev}
\affiliation{Apache Point Observatory, P.O. Box 59, Sunspot, NM 88349}
\author{Michael R.\ Blanton}
\affiliation{Center for Cosmology and Particle Physics, Department of Physics, 726 Broadway, Room 1005, New York University, New York, NY 10003, USA}
\author{Katia Cunha}
\author{Jos\'e G. Fern\'andez-Trincado}
\affiliation{Instituto de Astronom\'ia y Ciencias Planetarias, 
Universidad de Atacama, Copayapu 485, Copiap\'o, Chile}
\affiliation{Instituto de Astronom\'ia, Universidad Cat\'olica del Norte, Av.
Angamos 0610, Antofagasta, Chile}
\author{Peter M.\ Frinchaboy}
\affiliation{Department of Physics \& Astronomy, Texas Christian University, Fort Worth, TX 76129, USA}
\author[0000-0002-1693-2721]{D.\ A.\ Garc\'ia-Hern\'andez}
\affiliation{Instituto de Astrof\'isica de Canarias, 38205 La Laguna, Tenerife, Spain}
\affiliation{Universidad de La Laguna (ULL), Departamento de Astrofísica, E-38206 La Laguna, Tenerife, Spain}
\author{Emily Griffith}
\affiliation{Department of Astronomy and Center for Cosmology
and AstroParticle Physics, The Ohio State University, 
Columbus, OH 43210, USA}
\author{James W.\ Johnson}
\affiliation{Department of Astronomy and Center for Cosmology
and AstroParticle Physics, The Ohio State University, 
Columbus, OH 43210, USA}
\author[0000-0002-4912-8609]{Henrik J\"onsson}
\affiliation{Materials Science and Applied Mathematics, Malm\"o University, SE-205 06 Malm\"o, Sweden}
\author{Richard R.\ Lane}
\affiliation{Centro de Investigaci\'on en Astronom\'ia, Universidad Bernardo O'Higgins, Avenida Viel 1497, Santiago, Chile}
\author{Henry W. Leung}
\affiliation{Dunlap Institute for Astronomy and Astrophysics, 
University of Toronto, 50 St. George Street, Toronto, ON M5S 3H4, Canada}
\author{J. Ted Mackereth}
\affiliation{Dunlap Institute for Astronomy and Astrophysics, 
University of Toronto, 50 St. George Street, Toronto, ON M5S 3H4, Canada}
\affiliation{David A. Dunlap Department for Astronomy and Astrophysics, 
University of Toronto, 50 St. George Street, Toronto, ON M5S 3H4, Canada}
\affiliation{Canadian Institute for Theoretical Astrophysics, 
University of Toronto, 60 St. George Street, Toronto, ON, M5S 3H8, Canada}
\author{Steven R.\ Majewski}
\affiliation{Department of Astronomy, University of Virginia, Charlottesville, VA 22904, USA}
\author{Szabolcz M{\'e}sz{\'a}ros}
\affiliation{ELTE E\"otv\"os Lor\'and University, Gothard Astrophysical Observatory, 9700 Szombathely, Szent Imre H. st. 112, Hungary} 
\affiliation{MTA-ELTE Lend{\"u}let Milky Way Research Group, Hungary} 
\affiliation{MTA-ELTE Exoplanet Research Group, Hungary}
\author{Christian Nitschelm}
\affiliation{Centro de Astronom{\'i}a (CITEVA), Universidad de Antofagasta, Avenida Angamos 601, Antofagasta 1270300, Chile}
\author{Kaike Pan}
\affiliation{Apache Point Observatory, P.O. Box 59, Sunspot, NM 88349}
\author{Ricardo P.\ Schiavon}
\affiliation{Astrophysics Research Institute, Liverpool John Moores University, Liverpool, L3 5RF, UK}
\author{Donald P. Schneider}
\affiliation{Department of Astronomy and Astrophysics, The Pennsylvania 
State University, University Park, PA 16802, USA}
\affiliation{Institute for Gravitation and the Cosmos, 
The Pennsylvania State University, University Park, PA 16802, USA}
\author{Mathias Schultheis}
\affiliation{Observatoire de la Côte d'Azur, Laboratoire Lagrange, 06304 Nice Cedex 4, France}
\author{Verne Smith}
\affiliation{NSF's National Optical-Infrared Astronomy Research Laboratory, 950 North Cherry Avenue, Tucson, AZ 85719, USA}
\author{Jennifer S.\ Sobeck}
\affiliation{Department of Astronomy, University of Washington,
Seattle, WA 98195, USA}
\author{Keivan G.\ Stassun}
\affiliation{Department of Physics and Astronomy, Vanderbilt University, VU Station 1807, Nashville, TN 37235, USA}
\author{Guy S. Stringfellow}
\affiliation{Center for Astrophysics and Space Astronomy, Department of Astrophysical and Planetary Sciences, University of Colorado, 389 UCB, Boulder, CO 80309-0389, USA}
\author{Fiorenzo Vincenzo}
\affiliation{Department of Astronomy and Center for Cosmology
and AstroParticle Physics, The Ohio State University, 
Columbus, OH 43210, USA}
\author{John C. Wilson}
\affiliation{Department of Astronomy, University of Virginia, Charlottesville, VA 22904, USA}
\author{Gail Zasowski}
\affiliation{Department of Physics and Astronomy, University of Utah, 115 S. 1400 E., Salt Lake City, UT 84112, USA}


\begin{abstract}
We apply a novel statistical analysis to measurements of 16 elemental 
abundances in 34,410 Milky Way disk stars from the final data release 
(DR17) of APOGEE-2.  Building on recent work, we fit median abundance 
ratio trends [X/Mg] vs.\ [Mg/H] with a 2-process model, which decomposes 
abundance patterns into a ``prompt'' component tracing core collapse 
supernovae and a ``delayed'' component tracing Type Ia supernovae.  
For each sample star, we fit the amplitudes of these two components, 
then compute the residuals $\Delta\xh$ from this two-parameter fit.  
The rms residuals range from $\sim 0.01-0.03$ dex for the 
most precisely measured APOGEE abundances to $\sim 0.1$ dex for Na, V, and Ce. 
The {\it correlations} of residuals reveal a complex underlying structure,
including a correlated element group comprised of Ca, Na, Al, K, Cr, and Ce
and a separate group comprised of Ni, V, Mn, and Co.  Selecting stars
poorly fit by the 2-process model reveals a rich variety of physical 
outliers and sometimes subtle measurement errors.  Residual abundances
allow comparison of populations controlled for differences in metallicity and
[$\alpha$/Fe].  Relative to the main disk ($R=3-13\kpc$), we find
nearly identical abundance patterns in the outer disk ($R=15-17\kpc$), 0.05-0.2
dex depressions of multiple elements in LMC and Gaia Sausage/Enceladus stars,
and wild deviations (0.4-1 dex) of multiple elements in $\omega\,$Cen.  Residual
abundance analysis opens new opportunities for discovering chemically 
distinctive stars and stellar populations, for empirically constraining
nucleosynthetic yields, and for testing chemical evolution models that include 
stochasticity in the production and redistribution of elements.
\end{abstract}



\section{Introduction}
\label{sec:intro}

Over the past decade, large and systematic spectroscopic surveys have
mapped elemental abundance patterns of hundreds of thousands of stars
across much of the Galactic disk, bulge, and halo, including
RAVE, SEGUE, LAMOST, Gaia-ESO, APOGEE, GALAH, and H3 
\citep{Steinmetz2006,Yanny2009,Luo2015,Gilmore2012,DeSilva2015,Majewski2017,
Conroy2019}.
The APOGEE survey of SDSS-III \citep{Eisenstein2011} and SDSS-IV 
\citep{Blanton2017} is especially well suited to mapping the inner disk and 
bulge because it observes at near-IR wavelengths where dust obscuration
is dramatically reduced, because it targets luminous evolved stars that
can be observed at large distances, and because its high spectral
resolution allows separate determinations of 15 or more elemental 
abundances per target star.\footnote{SDSS = Sloan Digital Sky Survey.  
APOGEE = Apache Point Observatory Galactic Evolution Experiment.
We use APOGEE to refer to both the SDSS-III program and its SDSS-IV
extension (a.k.a.\ APOGEE-2).
In SDSS-V \citep{Kollmeier2017} the Milky Way Mapper program is
using the APOGEE spectrographs to observe a sample ten times larger than
that of SDSS-III + IV.}
These surveys share two primary goals: to understand the astrophysical
processes that govern the synthesis of the elements, and to trace the
chemical evolution of the Milky Way, which is itself shaped
by many processes including gas accretion, star formation, outflows,
and radial migration of stars.  This paper introduces a novel approach
to characterizing and mapping abundance patterns in APOGEE, one that
opens new avenues to addressing both of these goals.

Our study builds on a series of investigations that have used APOGEE
data to characterize the multi-element abundance distributions of 
the Galactic disk and bulge \citep{Anders2014,Hayden2014,Hayden2015,
Nidever2014,Ness2016,Ting2016,Mackereth2017,Schiavon2017,Bovy2019,
Fernandez-Trincado2019,Fernandez-Trincado2020b,
Weinberg2019,Zasowski2019,Griffith2021a,Ting2021,Vincenzo2021}.
Its most direct predecessors are the papers of 
Hayden et al.\ (\citeyear{Hayden2015}, hereafter H15),
who mapped the distribution of stars in $\afe-\feh$ as a function of
Galactocentric radius $R$ and midplane distance $|Z|$, and
Weinberg et al.\ (\citeyear{Weinberg2019}, hereafter W19), who
examined the median trends of other abundance ratios as a function 
of $R$ and $|Z|$.
Because $\alpha$ elements such as O, Mg, and Si are produced mainly
by core collapse supernovae (CCSN), while Fe is produced by both
CCSN and Type Ia supernovae (SNIa), the $\afe$ ratio is a diagnostic of 
the relative contribution of these two sources to a star's chemical
enrichment.  Many studies have shown that stars in the solar 
neighborhood have a bimodal distribution of $\afe$, with ``thin disk''
stars having roughly solar abundance ratios and ``thick disk'' stars
(which have larger vertical velocities and consequently larger excursions
from the disk midplane) having elevated $\afe$
(e.g., \citealt{Fuhrmann1998,Bensby2003,Adibekyan2012,Vincenzo2021}).
H15 showed that the locus of the ``high-$\alpha$'' sequence in
the $\afe-\feh$ plane is nearly constant throughout the disk
(see also \citealt{Nidever2014}) but the relative number of
high-$\alpha$ and low-$\alpha$ stars and the distribution of those
stars in $\feh$ changes systematically with $R$ and $|Z|$.
W19 advocated the use of Mg rather than Fe as a reference element
because it traces a single enrichment source (CCSN), and they showed
that the median trends of $\xmg$ for nearly all of the elements
measured by APOGEE are universal throughout the disk, provided that
one separates the high-$\alpha$ and low-$\alpha$ populations.
\cite{Griffith2021a} showed that this universality of abundance ratio
trends extends to the bulge.

Since the star formation and enrichment histories do change across the
Galaxy, W19 argued that the universal median sequences must be determined
mainly by IMF-averaged nucleosynthetic yields.\footnote{IMF = Initial Mass 
Function} 
They interpreted these sequences in
terms of a ``2-process model,'' which describes APOGEE abundances as
the sum of a core collapse process representing the IMF-averaged yields
of CCSN and a Type Ia process reflecting the IMF-averaged yields of SNIa.
The elemental abundances of a given star can be summarized by
the two parameters $\Acc$ and $\AIa$ that scale the amplitudes of
these processes.
The success of the 2-process model means that all of a disk or bulge
star's APOGEE abundances can be predicted to surprisingly high accuracy
from its Mg and Fe abundances alone.  They can be predicted to similar
accuracy from the combination of $\feh$ and age \citep{Ness2019}.
Nonetheless, the residual abundances of other elements
at fixed $\feh$ and $\mgfe$
contain rich information, as demonstrated empirically by
Ting \& Weinberg (\citeyear{Ting2021}, hereafter TW21), who show that
one must condition on at least {\it seven} APOGEE elements (e.g.,
Fe, Mg, O, Si, Ni, Ca, Al) before the correlations among the
remaining abundances are reduced to a level consistent with observational
uncertainties.  In this paper, therefore, we turn our attention from
median trends to the star-by-star abundance patterns described by
2-process model parameters {\it and the residuals} from this description.
As argued by TW21, the correlations of residual abundances encode
crucial information about nucleosynthetic processes and stochastic
effects in chemical evolution.

Although ``high-$\alpha$'' stars have elevated $\afe$ compared to the Sun,
this difference really arises because they have a lower contribution of
SNIa to Fe rather than enhanced production of $\alpha$ elements by CCSN
\citep{Tinsley1980,Matteucci1986,McWilliam1997}.
Adopting this physical interpretation, we will refer to high-$\alpha$
and low-$\alpha$ stars in this paper as the ``low-Ia'' and ``high-Ia''
populations, respectively, following terminology introduced by 
\cite{Griffith2019}.

In \S\ref{sec:2process}
we describe the 2-process model, which is similar to that of W19 and 
\cite{Griffith2019}
but with adjustments that make the model more flexible and easier to 
generalize.  
In \S\ref{sec:data}
we describe our selection of APOGEE stars from SDSS Data Release 17
(DR17): red giants in a restricted range of $\logg$,
$\Teff$, and $\mgh$ intended to minimize statistical and differential
systematic errors while sampling the disk in the range
$3\leq R \leq 13\kpc$ and $|Z| \leq 2\kpc$.  
Section~\ref{sec:median}
presents median abundance trends from this sample and uses them to
infer the CCSN and SNIa 2-process vectors, i.e., the abundance of each
of the APOGEE elements associated with these two processes at a given
metallicity.  
In \S\ref{sec:residuals}, 
the heart of the paper, we fit each sample star's abundances with
the 2-process model and examine the distributions and correlations
of the abundance residuals.  As in TW21, we find a rich correlation
structure among these residuals, and we further examine the correlation
of these residuals with stellar age and kinematics.
In \S\ref{sec:outliers} we investigate stars whose abundance patterns
deviate unusually far from the 2-process model fits, a group that
includes both genuine physical outliers and stars with measurement errors that
exceed the reported uncertainties.
In \S\ref{sec:populations}
we examine the residual abundances of a few special populations,
such as likely halo stars that reside within the geometrical boundaries
of the disk and members of the rich cluster $\omega\,$Cen, which is thought to
be the stripped core of an accreted dwarf spheroidal galaxy.
Section~\ref{sec:beyond}
discusses ways to go beyond the 2-process model, first with a conceptual
$N$-process formulation, then with an empirical approach that fits
two additional components to the APOGEE abundance residuals.
We review our conclusions and outline prospects for future studies in
\S\ref{sec:conclusions}.

This is a long paper covering many interconnected topics.  Readers who want
to start with a bird's-eye view can read the introduction to the 2-process
model at the start of \S\ref{sec:2process}, look through figures with 
particular attention to the median trends 
(Figure~\ref{fig:dataq_alpha}-\ref{fig:dataq_peakodd}), distributions
and covariance of residual abundances (Figures~\ref{fig:delta_dist}
and~\ref{fig:covar}), examples of high-$\chi^2$ stars 
(Figure~\ref{fig:elem_highchi2}), and residual patterns of selected
populations (Figure~\ref{fig:populations}), then read the conclusions
and loop back to earlier sections as needed.  For readers interested
in overall abundance trends and their interpretation, \S\ref{sec:median}
is the most relevant.  For readers interested in unusual stars and stellar
populations, \S\ref{sec:outliers} and~\S\ref{sec:populations} are the
most relevant.  Readers interested in the dimensionality of the stellar
distribution in abundance space and its connection to the physics of
nucleosynthesis and chemical evolution should pay particular attention
to \S\ref{sec:residuals} and~\S\ref{sec:beyond}.  The challenge of 
determining accurate, precise, robust abundances for many elements in
large stellar samples is a running theme throughout the paper.

\section{The 2-process model}
\label{sec:2process}

\begin{figure*}
\centerline{\includegraphics[width=5.5truein]{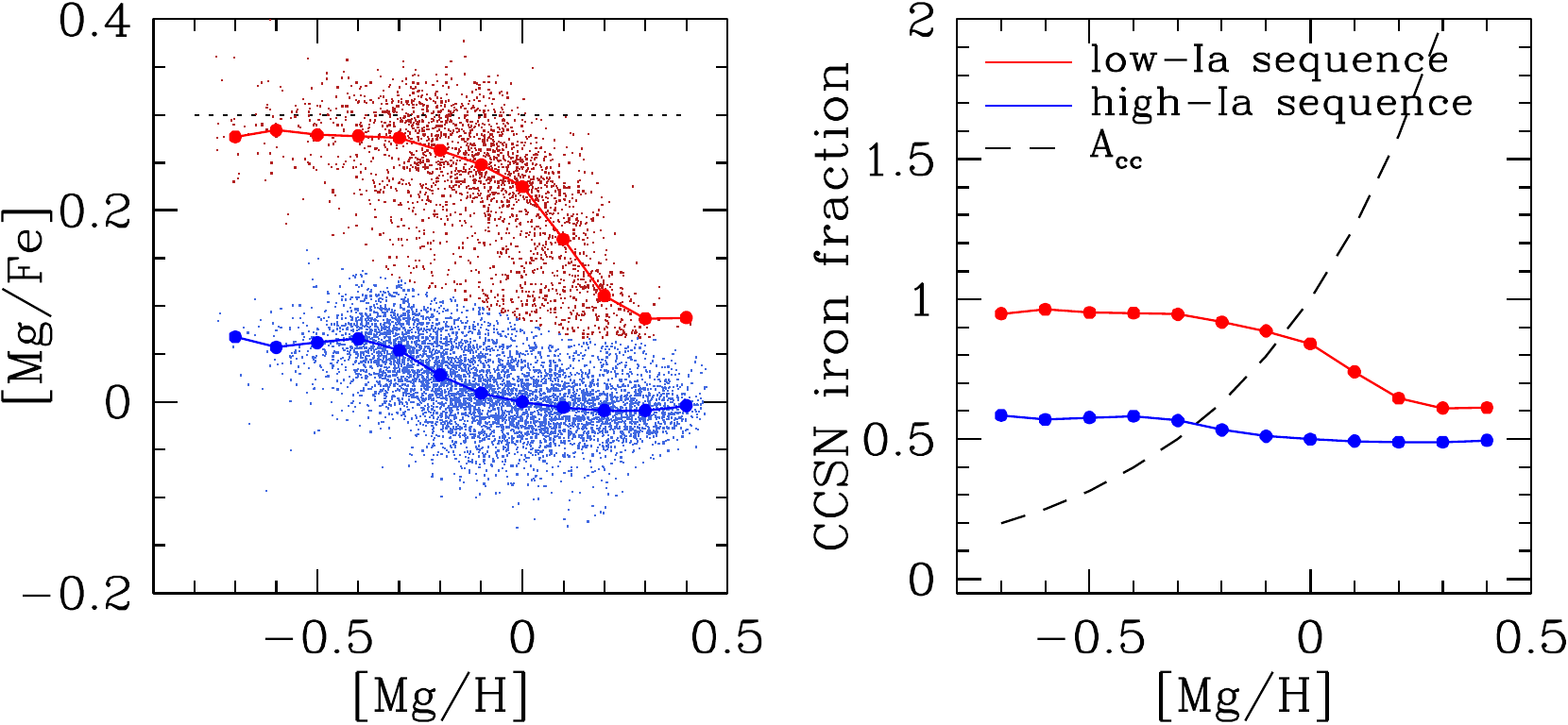}}
\caption{(Left) $\mgfe$ vs. $\mgh$ for sample stars in the low-Ia
(red) and high-Ia (blue) populations.  Points are randomly downsampled
by a factor of four to reduce crowding.  Connected large points show
the median $\mgfe$ in bins of $\mgh$.  An offset of 0.053 dex has
been applied to the APOGEE Fe abundances (decreasing [Mg/Fe] by 0.053)
so that the median high-Ia sequence passes through $\mgfe=0$ at $\mgh=0$.
The dotted horizontal line shows the ratio $\mgfepl=0.30$ that is assumed
to correspond to pure CCSN enrichment in our 2-process modeling.
(Right) The fraction of iron inferred (via the 2-process model) 
to arise from CCSN at points along the median low-Ia (red) and high-Ia (blue)
sequences (eq.~\ref{eqn:fxcc}).  The dashed curve shows $\Acc = 10^{\mgh}$.
}
\label{fig:mgfe}
\end{figure*}

We begin with a conceptual introduction to the 2-process model.
The left panel of Figure~\ref{fig:mgfe} shows the distribution of our
APOGEE sample (described in \S\ref{sec:data}) in the familiar plane
of $\mgfe$ vs.\ $\mgh$, with the low-Ia and high-Ia populations 
as red and blue points, respectively.  
Like W19,
we adopt $\mgh$ as our reference abundance on the $x$-axis because Mg is well
measured in APOGEE and, unlike Fe, it is thought to come from 
a single nucleosynthetic source (CCSN).\footnote{We choose Mg in preference
to O because the observed trends for O are significantly different between
optical and near-IR surveys.  We choose Mg in preference to Si or Ca
because these have non-negligible SNIa contributions.}
In the conventional interpretation of this diagram, which we adopt
in this paper, the ``plateau'' in the abundances of metal-poor low-Ia
stars at $\mgfe \approx 0.3$ represents the Mg/Fe ratio of CCSN yields,
and stars that lie below this plateau do so primarily because they
have additional Fe from SNIa.  While the relative number of low-Ia
and high-Ia stars depends strongly on Galactic location, the median
$\mgfe-\mgh$ tracks of these populations, shown by red and blue
lines in Figure~\ref{fig:mgfe}, are nearly universal throughout
the disk (\citealt{Nidever2014}; H15; W19).  The median $\xmg-\mgh$ 
tracks for other APOGEE elements are also universal 
throughout the disk (W19) and bulge \citep{Griffith2021a}, provided
that one separates the low-Ia and high-Ia populations.   This universality
motivates the hypothesis that these tracks are governed by 
stellar yields and that differences between the two [X/Mg] tracks reflect
the contribution of SNIa enrichment to element X.

In the 2-process model, the position of a star in the $M$-dimensional space
of its measured abundances is approximated as the weighted sum of two
``vectors'' that represent the contributions from CCSN and SNIa:
\begin{equation}
{({\rm X}/{\rm H})_* \over ({\rm X}/{\rm H})_\odot} = 
\Accst\qxcc + \AIast\qxIa~.
\label{eqn:2processq}
\end{equation}
The 2-process vectors $\qxcc$ and $\qxIa$ are taken to be universal
for the stellar sample under study, though they may depend on metallicity.
The amplitudes $\Acc$ and $\AIa$ vary from star to star, and they are
normalized such that $\Acc=\AIa=1$ for solar abundances.
For notational compactness we define metallicity by
\begin{equation}
z \equiv 10^{\mgh} ~,
\label{eqn:zdef}
\end{equation}
i.e., the Mg abundance in solar units.
As discussed in \S\S\ref{sec:2process_basic}-\ref{sec:2process_stars} below,
we infer $\qxcc(z)$ and $\qxIa(z)$ from median abundance ratios of
low-Ia and high-Ia stars, then determine $\Acc$ and $\AIa$ for all
stars in the observational sample by a $\chi^2$ fit to a subset of
their measured abundances.

\begin{figure}
\centerline{\includegraphics[width=2.8truein]{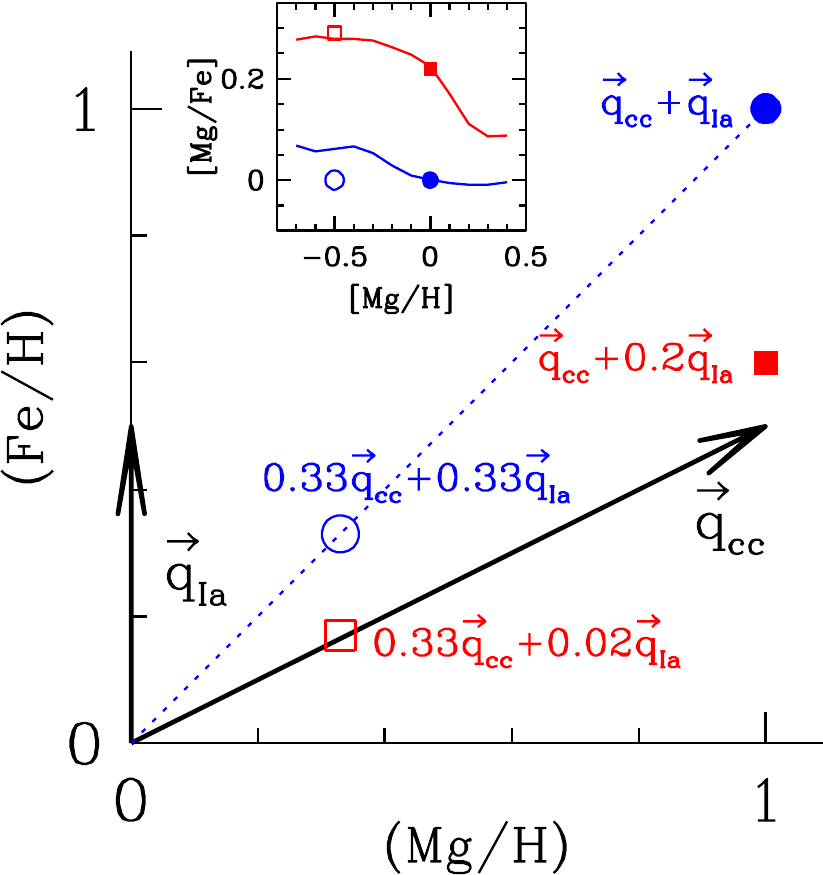}}
\caption{Illustration of the 2-process model for the simple case of
two abundances, (Mg/H) and (Fe/H), both linear ratios scaled to solar values.
The (Mg,Fe) components of the two-process vectors are 
$\vec{q}_{\rm cc}=(1,0.5)$ and
$\vec{q}_{\rm Ia}=(0,0.5)$, respectively. 
Large points show the abundances of stars with
$(\Acc,\AIa)=(1.0,1.0)$ (filled blue circle),
(0.33,0.33) (open blue circle),
(1.0,0.2) (filled red square),
(0.33,0.02) (open red square).
The inset marks the position of these four stars in the [Mg/Fe]-[Fe/H] plane,
with red and blue curves showing the observed median sequences of low-Ia
and high-Ia stars from Fig.~\ref{fig:mgfe}.
The dotted blue line marks the locus of $\mgfe=0$.
}
\label{fig:2pro_vector}
\end{figure}

Figure~\ref{fig:2pro_vector} illustrates the simple case of $M=2$ dimensions, 
with
points marking the location of four representative stars in (Fe/H) vs.\ (Mg/H),
expressed in solar units.  The 2-process (Mg,Fe) vectors are
$\vec{q}_{\rm cc}=(1,0.5)$ and $\vec{q}_{\rm Ia}=(0,0.5)$, reflecting our
model assumptions that Mg is produced entirely by CCSN and that solar Fe
comes equally from CCSN and SNIa.  The filled blue circle has solar abundances,
with $\Acc=\AIa=1$ and 
$({\rm Mg/H},{\rm Fe/H})=\Acc\vec{q}_{\rm cc} + \AIa\vec{q}_{\rm Ia}=(1,1)$.
The open blue circle has $\Acc=\AIa=1/3$, so its (Mg/H) and (Fe/H) 
abundances are 1/3 solar but its (Mg/Fe) ratio is solar.
Filled and open red squares represent low-Ia stars with $\Acc=1$ and $1/3$,
respectively.  These stars have $\AIa<\Acc$, so they have (Mg/Fe) ratios
above solar (``$\alpha$-enhanced'' or ``iron poor'').  With $M=2$, the
two parameters $\Acc$ and $\AIa$ suffice to fit each star's abundances
perfectly, but they recast the information from the space of individual
elements to the space of the processes that produce those elements.

\begin{figure}
\centerline{\includegraphics[width=3.2truein]{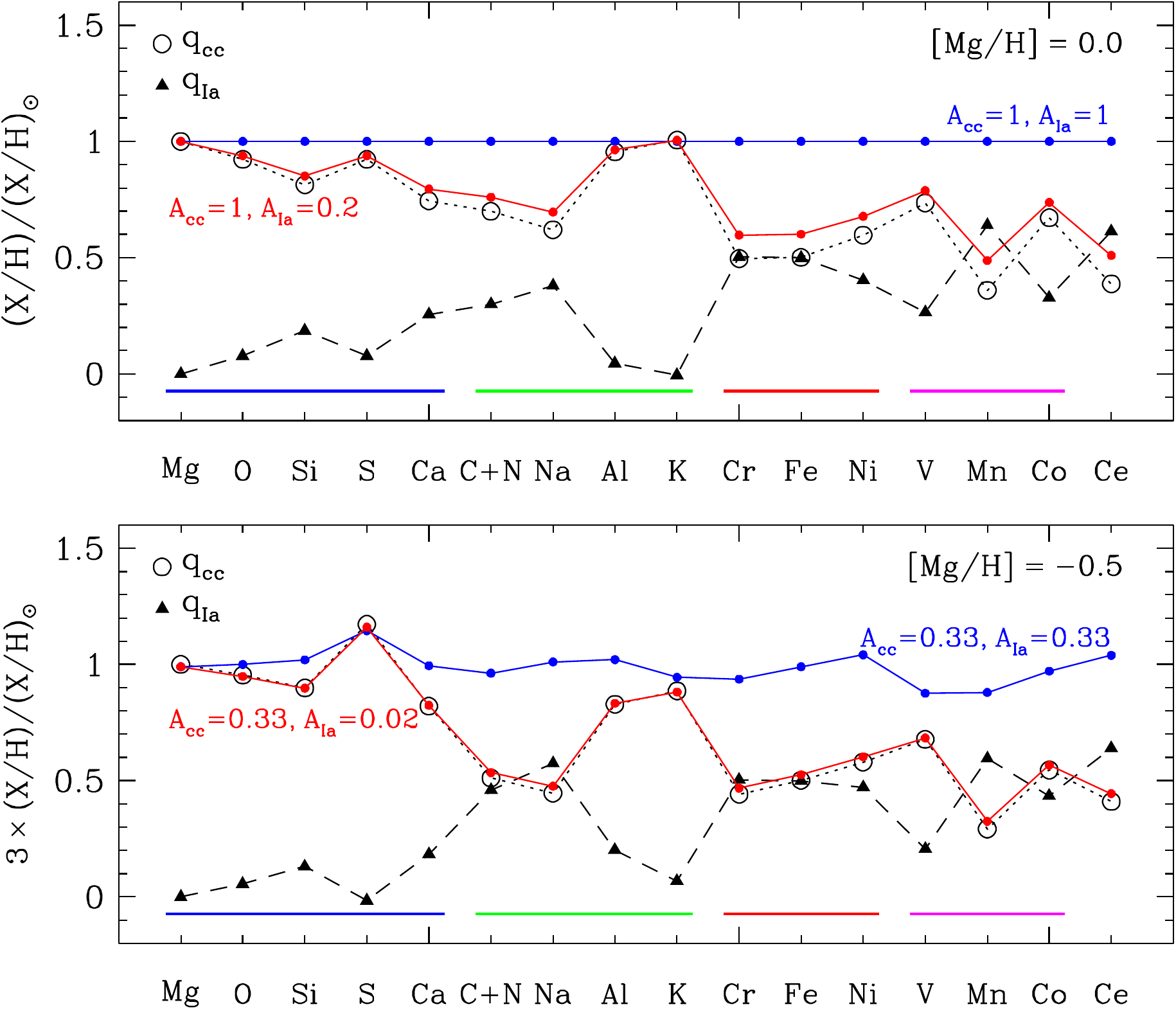}}
\caption{Illustration of the 2-process model for the 16 elements
considered in our analysis (one of which is the element combination C+N).
In the top panel, open circles and filled triangles show the 2-process
vectors $\qxcc$ and $\qxIa$ at $\mgh=0$ inferred from the APOGEE data
in \S\ref{sec:median}.  For a star with $\Acc=\AIa=1$, the predicted
abundances are the sum of these two vectors, which in this case yield
exactly solar values (blue points) by construction.  Red points show
the predicted abundances for a low-Ia star with $\Acc=1$, $\AIa=0.2$.
The lower panel shows analogous results at $\mgh=-0.5$, with predicted
abundances multiplied by a factor of three in this panel 
to aid visual comparison with the top panel.  
In each panel, colored horizontal lines group elements with similar
physical properties.
The four combinations of $(\Acc,\AIa)$ in this
Figure are the same cases illustrated in Figure~\ref{fig:2pro_vector}.
}
\label{fig:2pro_explain}
\end{figure}

Using the same four combinations of $(\Acc,\AIa)$ as 
Figure~\ref{fig:2pro_vector},
Figure~\ref{fig:2pro_explain} illustrates the 2-process model for the full
set of 16 abundances that we consider in this paper, one of which is the
element combination C+N (eq.~\ref{eqn:cndef} in \S\ref{sec:data}).
In the upper panel, open circles and filled triangles show the components
of $\qxcc$ and $\qxIa$ that we derive from the APOGEE median abundance
trends in \S\ref{sec:median} below, at metallicity $z=1$.
For $\alpha$-elements (on the left), $\qxcc$ values are much larger than
$\qxIa$ values, while for iron-peak elements (on the right) they are 
roughly equal.  For $\Acc=\AIa=1$, the predicted abundances are exactly
solar by construction (blue points).  For $\Acc=1$ and $\AIa=0.2$,
the predicted abundances (red points) are only slightly above $\qxcc$ 
(black open circles),
with sub-solar (X/H) values for all elements that have a substantial
SNIa contribution in the Sun.

The lower panel shows our inferred 2-process vectors for $\mgh=-0.5$
($\Acc=z=1/3$).  These vectors are similar to those found at $z=1$, but
they are not identical because some elements have metallicity dependent
yields.  As a result, the predicted abundances for $\AIa=\Acc=1/3$ have
element ratios that are approximately but not exactly solar (blue points).
For a star on the low-Ia (high-$\alpha$) plateau, the predicted abundances
(red points) are just slightly above $\Acc\qxcc$.
The predicted abundances (blue and red points) are multiplied by a
factor of three so that the patterns can be visually compared to
those of the $\mgh=0$ stars shown in the upper panel.

With 16 abundances, any given star will not be perfectly reproduced by a
2-parameter $(\Acc,\AIa)$ fit, in part because of measurement errors, but
also because the 2-process model is not a complete physical description
of stellar abundances.  For example, the model does not allow for stochastic
variations around IMF-averaged yields or for varying
contributions from other sources
such as AGB enrichment.  The 2-process vectors themselves are useful tests
of supernova nucleosynthesis predictions 
(e.g., \citealt{Griffith2019,Griffith2021b}), and the distributions of
$(\Acc,\AIa)$ and their correlations with stellar age and kinematics are
useful diagnostics of Galactic chemical evolution.
However, our primary focus in this paper will be the star-by-star departures
from the 2-process predictions, and what these departures can tell us about
the astrophysical sources of the APOGEE elements, about distinct stellar
populations within the geometric boundaries of the Milky Way disk, and about
rare stars with distinctive abundance patterns.  Disk stars span a range
of $>1$ dex in [Fe/H] and typically 0.3 dex or more in [X/Fe].  
With two free parameters, the 2-process
model fits the measured APOGEE abundances of most disk stars to within
$\sim 0.1$ dex, and for the best measured elements to within $\sim 0.01-0.04$
dex, so focusing on these residual abundances allows us to discern subtle
patterns that might be lost within the much larger dynamic range of a
conventional [X/Fe]-[Fe/H] analysis.

We now proceed to a more formal definition of the 2-process model and its
assumptions, and our methods for inferring the 2-process vectors and fitting
$\Acc$ and $\AIa$ values.  While our approach is similar to that of W19,
here we define the model in a way that is more general and allows natural
extension to include other processes as discussed in \S\ref{sec:beyond}.

\subsection{Model assumptions and basic equations}
\label{sec:2process_basic}

We can express equation~(\ref{eqn:2processq}) in the alternative form
\begin{equation}
\xoverh_* = \Accst\pxcc(z) + \AIast\pxIa(z)~,
\label{eqn:2process}
\end{equation}
with 
\begin{equation}
\qxj(z) \equiv {\pxj(z) \over ({\rm X}/{\rm H})_\odot}~.
\label{eqn:qdef}
\end{equation}
While it may seem gratuitous to introduce both $p$ and $q$, many of our
equations can be written more compactly in terms of $p$, and these two 
forms of the process vectors respond differently to changes in the adopted
solar abundance values.  For example, if stellar abundances are inferred by 
purely {\it ab initio} model-fitting then the $p$ vectors are directly
determined while the $q$ vectors depend on adopted solar abundances.
Conversely, if zero-point offsets are used to calibrate the abundance scale
to reproduce solar values, then the $q$ vectors are directly determined
and the conversion to $p$ vectors depends on the adopted solar abundances.
At a given $z$, $\pxcc$ is a set of discrete values, one for each 
element X being modeled, and likewise for $\pxIa$.  For brevity,
we will frequently drop the explicit $z$-dependence of $\pxcc$ and $\pxIa$
in our equations if it is not needed for clarity, but it is only for Mg and Fe
that we assume that these processes are actually independent of metallicity.

We {\rm define} $\Acc = \AIa$ in the Sun.  Therefore, if we ignore
the possible contribution of other processes,
\begin{equation}
\pxcc(z=1)+\pxIa(z=1) = \xoverh_\odot~
\label{eqn:2process_sum}
\end{equation}
for all X.
In a star with metallicity $z$ and amplitudes $\Accst$ and $\AIast$,
the fraction of element $X$ that arises from CCSN is
\begin{eqnarray}
\fxcc &=& {\Accst \pxcc(z) \over \Accst\pxcc(z) + \AIast\pxIa(z)} 
          \label{eqn:fxcc}\\
      &=& \left[1+(\AIast/\Accst)(\qxIa/\qxcc)\right]^{-1}~, 
          \label{eqn:fxcc2}
\end{eqnarray}
where we have used the fact that $\qxIa/\qxcc = \pxIa/\pxcc$.
More generally, the denominator of equation~(\ref{eqn:fxcc}) should be
the sum of all processes that contribute to element X 
(see \S\ref{sec:beyond_model}).
For the Sun we have $\AIa=\Acc=1$ and $\qxIa=1-\qxcc$, which simplifies
equation~(\ref{eqn:fxcc2}) to
\begin{equation}
\fxccsun = \qxcc(z=1).
\label{eqn:fxccsun}
\end{equation}

For our implementation of the 2-process model, we assume that the Mg and
Fe processes are independent of metallicity and that Mg is a pure
core collapse element:
\begin{eqnarray}
\pmgcc(z) &=& \pmgcc, \qquad \pmgIa=0, \label{eqn:pmg} \\
\pfecc(z) &=& \pfecc, \qquad \pfeIa(z) = \pfeIa \label{eqn:pfe}~.
\end{eqnarray}
Standard supernova models predict that Mg and Fe yields are approximately
independent of metallicity (see, e.g., Fig.~20 of \citealt{Andrews2017})
and that the SNIa contribution to Mg is negligible.  At low metallicity,
the high-$\alpha$ population in APOGEE and other surveys exhibits
a nearly flat plateau in $\mgfe$ at
\begin{equation}
\mgfepl \approx 0.3 
\label{eqn:mgfepl}
\end{equation}
(see, e.g., \citealt{Adibekyan2012,Bensby2014,Buder2018,Griffith2019}; and
Figure~\ref{fig:mgfe} above).
This flatness provides
empirical support for metallicity independence of
the Mg and Fe CCSN processes.  A flat plateau could also
arise if CCSN yields of these elements have the same metallicity
dependence while keeping the Mg/Fe ratio constant.
Our formalism could be adapted to
metallicity-dependent Mg and Fe processes if there were motivation
to do so, but this would introduce some mathematical complication
so we do not consider this generalization here.

Combining equations~(\ref{eqn:pmg}) and~(\ref{eqn:2process}) implies
\begin{equation}
\mgoverh_* = \Accst\pmgcc = \Accst\mgoverh_\odot
\end{equation}
and thus
\begin{equation}
\Accst = 10^{\mgh_*}~.
\label{eqn:Accst}
\end{equation}
Equation~(\ref{eqn:Accst}) provides a simple way to estimate $\Accst$,
from a star's Mg abundance alone, though in practice we will use a
multi-element fit as described below (\S\ref{sec:2process_stars}).

For iron, equations~(\ref{eqn:pfe}) and~(\ref{eqn:2process}) imply
\begin{equation}
{({\rm Fe}/{\rm Mg})_* \over ({\rm Fe}/{\rm Mg})_\odot} = 
  {\Accst\pfecc + \AIast\pfeIa \over \Accst\pmgcc} \cdot 
  {\pmgcc \over \pfecc + \pfeIa}~,
\label{eqn:femgstar}
\end{equation}
which can be rearranged to yield
\begin{equation}
10^{\femg_*} = {\pfecc + (\AIast/\Accst)\pfeIa \over \pfecc+\pfeIa}~.
\label{eqn:femgstar2}
\end{equation}
Our third key assumption is that iron in stars on the $\mgfe$ plateau 
comes from CCSN alone, implying $\AIast=0$ and thus
\begin{equation}
10^{\femgpl} = {\pfecc \over \pfecc+\pfeIa} = \ffeccsun~.
\end{equation}
By definition, if $\AIast=\Accst=1$ we are at solar abundances
for Mg and Fe (because they are assumed to have no contributions from
other processes), and therefore $\femg=0$ as implied by 
equation~(\ref{eqn:femgstar2}).

With a bit of manipulation one can write
\begin{eqnarray}
  {\AIast \over \Accst} &=&
  {({\rm Fe}/{\rm Mg})_*     - ({\rm Fe}/{\rm Mg})_{\rm pl}  \over
   ({\rm Fe}/{\rm Mg})_\odot - ({\rm Fe}/{\rm Mg})_{\rm pl} } \\
   &=&
   { 10^{\femg_*} - 10^{\femgpl} \over 1 - 10^{\femgpl} } ~.
\label{eqn:Aratio}
\end{eqnarray}
In the first equation, the numerator is the amount SNIa Fe in the
star relative to Mg, and the denominator is the amount of SNIa Fe
at solar $\femg$, for which $\AIast/\Accst=1$.
The second equation relates $\AIast/\Accst$ to the displacement
of $\femg_*$ below the CCSN plateau.
We adopt $\femgpl = -\mgfepl = -0.3$ as the observed level of the
plateau, and thus $10^{\femgpl} \approx 0.5$.

Equation~(\ref{eqn:Aratio}) provides a simple way to estimate
$\AIast$ after estimating $\Accst$ from equation~(\ref{eqn:Accst}).
In the right panel of Figure~\ref{fig:mgfe}, 
red and blue curves show the inferred values of $\ffecc$
(equation~\ref{eqn:fxcc2})
for points along the low-Ia and high-Ia median sequences shown in 
the left panel.  The values of $\Acc$ and $\AIa$ are derived from the
$\mgh$ and $\mgfe$ values along these sequences as described above.
On the high-Ia
sequence the inferred core collapse iron fraction is about 0.5 at all $\mgh$.
On the low-Ia sequence the fraction declines from nearly 100\% at 
low $\mgh$ to about 0.6 at the highest $\mgh$.
The dashed curve shows the value of $\Acc$ corresponding to $\mgh$
via equation~(\ref{eqn:Accst}).

In terms of the solar-scaled process vectors, our model assumptions
and $\femgpl$ value correspond to $\qmgcc=1$ and $\qfecc=\qfeIa=0.5$.
For an element X that is produced entirely by CCSN and SNIa, the
solar-scaled abundances are
\begin{equation}
\xh_* = \log_{10} \left[\Accst\qxcc(z) + \AIast\qxIa(z)\right]~.
\label{eqn:xhratios}
\end{equation}
Subtracting $[{\rm Mg}/{\rm H}]_* = \log_{10}\Accst$ gives
\begin{equation}
\xmg_* = \log_{10} \left[\qxcc(z)+\qxIa(z)\AIast/\Accst\right]~.
\label{eqn:xmgratios}
\end{equation}
More generally, an element may have contributions from CCSN or other
``prompt'' enrichment sources (e.g., massive star winds) 
that rapidly follow star formation,
and additional contributions from enrichment sources with a
distribution of delay times (e.g., SNIa, AGB stars).
When modeled with the 2-process formalism, $\qxcc$ represents
the prompt contributions and $\qxIa$ represents the contributions
that follow SNIa iron enrichment, with the implicit assumption
that the ISM is sufficiently well mixed to average out the diverse
properties of individual supernovae or other sources.

To apply the 2-process model to APOGEE data, we must first 
determine the values of $\qxcc(z)$ and $\qxIa(z)$ from the ensemble
of measurements, then determine the amplitudes $\Accst$ and $\AIast$
for each star.  We can then predict each star's abundances and
measure the residuals, i.e., the difference between the observed
abundances and the 2-process predictions.  

\subsection{Inferring the 2-process vectors from median sequences}
\label{sec:2process_vectors}

Similar to W19,
we infer the process vectors $\qxcc(z)$ and $\qxIa(z)$ from the observed
{\it median sequences} of $\xmg$ vs.\ $\mgh$ for the low-Ia and high-Ia
stellar populations.  We do this separately in each bin of $\mgh$, and
we will henceforth drop the $z$-dependence from our notation with the
understanding that $\qxcc$ and $\qxIa$ can change from bin to bin.
In principle we could perform a global $\chi^2$ fit to the abundances
of stars in each $\mgh$ bin,
but inferring the process vectors from the median sequences is much
easier, and it is also more robust because outliers (whether physical
or observational) have minimal impact on median values in a large data set.
The median values of $\femg$ are significantly different between the 
two populations even at high $\mgh$, so there is sufficient leverage
to separate the CCSN and SNIa contributions.  
Statistical errors on
the median abundance ratios are very small because there are many
stars in each bin, though we are still sensitive to systematic errors
in the abundance measurements.

W19 assumed a power-law $z$-dependence of the process vectors, but
here we allow a general metallicity dependence.  For each element and
each $\mgh$ bin, there are two measurements, $\xmg_{\rm med}$ of the
low-Ia and high-Ia populations, to fit with two parameters,
$\qxcc$ and $\qxIa$, so the 2-process model can exactly reproduce
the observed median sequences by construction.
We adopt a general (bin-by-bin) $z$-dependence in part to capture
possibly complex trends, but for purposes of this paper our primary
motivation is to ensure that the mean star-by-star residuals from the 
2-process predictions are close to zero at all $\mgh$.
Although the more restrictive power-law formulation usually allows a
good fit to the observed median sequences, there are departures for
some elements in some $\mgh$ ranges, and residuals could easily be
dominated by these global differences rather than star-to-star variations.

Using the observed values of $\femgpl$ and of the median values
of $\femg$ on the high-Ia and low-Ia sequences in the $\mgh$ bin
under consideration, we define
\begin{equation}
\Rhigh \equiv \left({\AIa\over\Acc}\right)_{\rm high} =
  {10^{\femg_{\rm high}} - 10^{\femgpl} \over 1 - 10^{\femgpl}}
\label{eqn:rhigh}
\end{equation}
and
\begin{equation}
\Rlow \equiv \left({\AIa\over\Acc}\right)_{\rm low} =
  {10^{\femg_{\rm low}} - 10^{\femgpl} \over 1 - 10^{\femgpl}}~.
\label{eqn:rlow}
\end{equation}

From equation~(\ref{eqn:xmgratios}) we have
\begin{eqnarray}
10^{\xmg_{\rm high}} &=& \qxcc + \Rhigh \qxIa \\
10^{\xmg_{\rm low}} &=& \qxcc + \Rlow \qxIa~.
\end{eqnarray}
Solving these equations yields
\begin{equation}
\qxIa = {10^{\xmg_{\rm high}}-10^{\xmg_{\rm low}} \over \Rhigh-\Rlow}
\label{eqn:qxIa}
\end{equation}
and
\begin{equation}
\qxcc = 10^{\xmg_{\rm low}} - 
{10^{\xmg_{\rm high}}-10^{\xmg_{\rm low}} \over \Rhigh/\Rlow - 1}~.
\label{eqn:qxcc}
\end{equation}
If there is no difference between $\xmg$ on the two sequences we get
$\qxcc=10^{\xmg}$ and $\qxIa=0$.
For a point with $\femg_{\rm low} = \femgpl$, we get $\Rlow=0$ and
$\qxcc = 10^{\xmg_{\rm low}}$, which is as expected because such
a point has no SNIa contribution.  We use equations~(\ref{eqn:qxIa}) 
and~(\ref{eqn:qxcc}) to infer $\qxIa$ and $\qxcc$ for each element X
in each bin of $\mgh$ 
(see Figures~\ref{fig:dataq_alpha}-\ref{fig:dataq_peakodd} below).

\subsection{Fitting stellar values of $\Acc$ and $\AIa$}
\label{sec:2process_stars}

Equations~(\ref{eqn:Accst}) and~(\ref{eqn:Aratio}) provide a simple
way to estimate a star's 2-process amplitudes $\Acc$ and $\AIa$
from its Mg and Fe abundances.  This is the method used by W19,
and because Mg and Fe are well measured by APOGEE it is accurate
enough for many purposes.  However, for our goal of studying
the correlations of residual abundances it has an important 
disadvantage: random measurement errors in $\mgh$ and $\femg$
will induce spurious apparent correlations in the residuals of
other elements.  For example, if a star's measured $\mgh$ fluctuates
low, its $\Acc$ will be underestimated, and all of the star's other
$\alpha$-elements will tend to lie above the 2-process prediction.
TW21 examined the closely connected question of residual abundances
after conditioning on $\feh$ and $\mgfe$.  They described the spurious
correlations that arise from random Mg and Fe abundance errors as
``measurement aberration,'' caused by defining the residual abundances
relative to a (randomly) incorrect reference point.

We can mitigate the effects of measurement aberration by estimating
a star's $\Acc$ and $\AIa$ from multiple abundances, since the random
errors in these abundances tend to average out.  As discussed in 
\S\ref{sec:residuals} below, we choose to infer $\Acc$ and $\AIa$ from
the abundances of six elements (Mg, O, Si, Ca, Fe, Ni) that have
small statistical errors in APOGEE and that collectively provide
good leverage on the 2-process amplitudes because they have a range
of relative contributions from SNIa vs.\ CCSN.
These elements are not expected to have significant contributions from
sources other than CCSN and SNIa.
We fit each star's $\Acc$ and $\AIa$ by $\chi^2$ minimization using the
observational measurement uncertainties reported by APOGEE.

In practice, we take the parameter estimates from Mg and Fe as
an initial guess, then iterate between optimizing $\Acc$ and $\AIa$,
an approach that is computationally cheap and quickly
converges to a 2-d $\chi^2$ minimum.  To avoid fit parameters being
affected by outlier abundances (which could well be observational errors),
we eliminate O, Si, Ca, or Ni from the fit if their abundance differs
by more than $5\sigma$ from the value predicted based on the
initial guess.  This criterion leads to the elimination of 206 O measurements,
118 Si measurements, 279 Ca measurements, and 625 Ni measurements from
our sample of 34,410 stars.  Fitting six abundances with
two parameters does not add any more freedom to the model, but
instead of fitting Mg and Fe exactly it chooses compromise values
that give the best overall fit to the selected elements.
We demonstrate the reduced measurement aberration from six-element
fitting in Fig.~\ref{fig:covar} below.

\section{APOGEE data sample}
\label{sec:data}

We use data from the 17th data release (DR17) of the 
SDSS/APOGEE survey \citep{Majewski2017}.  
The APOGEE disk sample consists primarily of evolved stars with
2MASS \citep{Skrutskie2006} magnitudes $7<H<13.8$, sampled largely on a
grid of sightlines at Galactic latitudes
$b=0^\circ$, $\pm 4^\circ$, and $\pm 8^\circ$
and many Galactic longitudes.
Targeting for APOGEE is described in detail by
\cite{Zasowski2013,Zasowski2017}, 
Beaton et al.\ (2021), and Santana et al.\ (2021).
APOGEE obtains high-resolution ($R\sim 22,500$) H-band spectra
(1.51-1.70\,$\mu$m) using 300-fiber spectrographs
\citep{Wilson2019} on the 2.5m Sloan Foundation telescope
\citep{Gunn2006} at Apache Point Observatory in New Mexico and the 2.5m
du Pont Telescope \citep{Bowen1973} at Las Campanas Observatory in Chile.
The great majority of
spectra in the main APOGEE sample have signal-to-noise ratio 
per pixel $\SNR>80$ (with a typical pixel width $\approx 0.22$\AA).
Spectral reductions and calibrations are performed by the
APOGEE data processing pipeline \citep{Nidever2015}, which provides
input to the APOGEE Stellar Parameters and Chemical Abundances Pipeline
(ASPCAP; \citealt{Holtzman2015,Garcia2016}).  ASPCAP uses a grid
of synthetic spectral models \citep{Meszaros2012,Zamora2015} and
H-band linelists \citep{Shetrone2015,Hasselquist2016,Cunha2017,Smith2021} 
compiled from a variety of laboratory, theoretical, and astrophysical sources,
fitting effective temperatures, surface gravities, and elemental abundances.

A detailed description of the APOGEE DR17 data will be presented by
J.\ Holtzman et al.\ (in preparation), updating the comparable description
of the APOGEE DR16 data by \cite{Jonsson2020}.
These papers explain the spectral fitting and calibration procedures,
the estimation of observational uncertainties, and comparisons to literature
values.  Notably, the DR17 abundances used here employ a synthetic spectral
grid generated by Synspec \citep{Hubeny2017} with NLTE treatments
of Na, Mg, K, and Ca \citep{Osorio2020}.  
These spectra are based on MARCS atmospheric models
\citep{Gustafsson2008}, with spherical geometry in the $\logg$
range used for our analysis.  
The Synspec synthesis uses these structures but assumes plane parallel 
geometry.
DR17 uses improved H-band wavelength windows for the s-process element Ce
\citep{Cunha2017}, providing higher precision measurements than previous
APOGEE data releases.
We do not distinguish isotopes for any elements, as APOGEE does
not have the resolution to clearly separate different isotopic lines.

Stellar abundance measurements are subject to statistical errors arising
from photon noise and data reduction 
and to systematic errors that arise because one is
fitting the data with imperfect models.  These imperfections include
incomplete or inaccurate linelists, astrophysical effects such as 
departures from local thermodynamic equilibrium (LTE), and observational
effects such as inexact spectral linespread functions.
The systematic effects will change with stellar parameters such
as $\Teff$, $\logg$, and metallicity, but if the range of parameters in
the sample is small then the {\it differential} systematics within the
sample will be limited, so the systematics will produce zero-point
offsets but will not add much in the way of scatter or correlated
abundance deviations for stars of the same $\mgh$ and $\mgfe$.

For the analyses of this paper, we have several goals that affect the
choice of sample selection criteria:
\begin{enumerate}
\item Minimize statistical errors to improve measurements of residuals
      from 2-process predictions.
\item Minimize differential systematic errors across the sample so that
      scatter and correlated residuals are minimally affected by
      systematics.
\item Cover a substantial fraction of the disk to probe populations with
      a range of enrichment histories.
\item Retain a large enough sample to enable accurate measurements of
      median trends, scatter, and correlations.
\end{enumerate}
Plots of $\xmg$ vs. $\mgh$ show that DR17 ASPCAP abundances still have
systematic trends with $\logg$ (see \citealt{Griffith2021a}).
However, to get a large sample one cannot afford to take too narrow a 
range of $\logg$.  Luminous giants provide the best coverage of a wide 
range of the disk.  

From the DR17 data set, we remove stars with the ASPCAP
{\tt STAR\_BAD} or {\tt NO\_ASPCAP\_RESULT} flags set, and we remove
stars with flagged $\feh$ or $\mgfe$ measurements.  We use
only stars targeted as part of the main APOGEE survey
(flag {\tt EXTRATARG=0}) to avoid any selection biases associated
with special target classes.  
We use the DR17 ``named'' abundance tags {\tt X\_FE}, which apply
additional reliability cuts for each element
(see \S 5.3.1 of \citealt{Jonsson2020}).
As a compromise among the considerations above, we have adopted
the following sample selection cuts:
\begin{enumerate}
\item $R = 3-13\kpc$, $|Z|\leq 2\kpc$ (399,573 stars)
\item $-0.75 \leq \mgh \leq 0.45$ (387,218 stars)
\item $\SNR \geq 200$ for $\mgh > -0.5$; $\SNR \geq 100$ for $\mgh < -0.5$
      (160,133 stars)
\item $\logg = 1-2.5$ (65,611 stars)
\item $\Teff = 4000-4600$ (34,410 stars)~.
\end{enumerate}
Numbers in parentheses indicate the number of sample stars remaining
after each cut.   
Spectroscopic distances for computing $R$ and $Z$ are taken from the 
DR17 version of the AstroNN catalog (see \citealt{Leung2019b}); at distances
of many kpc, these spectroscopic estimates are more precise than those
from {\it Gaia} parallaxes.
We use a lower $\SNR$ threshold below $\mgh=-0.5$
to retain a sufficient number of low metallicity stars.
The combination of cuts 4 and 5 eliminates red clump (core helium burning)
stars (see \citealt{Vincenzo2021}), which
might have different measurement systematics from red giant branch (RGB)
stars and could thus artificially add scatter or correlated deviations.  
The APOGEE red clump stars are themselves a well controlled
and powerful sample \citep{Bovy2014}, and it would be useful to repeat
some of our analyses below for the red clump sample and to understand
the origin of any differences.

We compute $\xh$ values as the sum of the ASPCAP quantities
{\tt X\_FE} and {\tt FE\_H}.  We take the quantity {\tt X\_FE\_ERR} as the
statistical measurement uncertainty in $\xh$.  
Although {\tt FE\_H} has its own statistical uncertainty, we are
primarily interested in differential scatter among elements, and all
abundances in a given star use the same value of {\tt FE\_H}.
ASPCAP abundance uncertainties are estimated empirically as a
function of SNR, $\Teff$, and metallicity using repeat observations
of a subset of stars (see \S 5.4 of \citealt{Jonsson2020}).
These empirical errors are usually larger (by a factor of several)
than the $\chi^2$ model-fitting uncertainty.
This procedure means that the adopted observational uncertainty for a given
element is representative of that for stars with the same global properties
and SNR but does not reflect the specifics of the individual star's spectrum 
near the element's spectral features.  
In the rare cases where 
the $\chi^2$ model-fitting 
uncertainty exceeds the empirical uncertainty, the fitting uncertainty is
reported instead.
Some stars have flagged values of individual elements, in which case
we keep the star in the sample but omit the star from any calculations involving
those elements.  These cuts eliminate 562 Ce values but no more than two
values for other elements.

The C and N surface abundances of RGB stars differ from
their birth abundances because the CNO cycle preferentially converts
$^{12}$C to $^{14}$N and some processed material is dredged up to
the convective envelope (e.g., \citealt{Iben1965,Shetrone2019}).
However, because the extra N nuclei come almost entirely from C nuclei,
leaving the O abundance much less perturbed, the number-weighted C+N
abundance is nearly equal to the birth abundance, with theoretically
predicted differences $\sim 0.01$ dex over most of the $\logg$ range
considered here \citep{Vincenzo2021b}.  
We therefore take C+N as an ``element'' in our analysis, computing
\begin{eqnarray}
\xxh{(C+N)} = &&\log_{10}\left(10^{\xxh{C}+8.39}+10^{\xxh{N}+7.78}\right) - 
                \nonumber\\
              && \log_{10}\left(10^{8.39}+10^{7.78}\right)~,
\label{eqn:cndef}
\end{eqnarray}
where 8.39 and 7.78 are our adopted logarithmic values of the solar
C and N abundances \citep{Grevesse2007}
on the usual scale where the hydrogen number density is 12.0.
We somewhat arbitrarily set the uncertainty in $\xxh{(C+N)}$ equal
to the ASPCAP uncertainty in $\xxfe{C}$, i.e., to {\tt C\_FE\_ERR}.
While the fractional error in N may exceed the fractional error in C,
N contributes only 20\% to C+N for a solar C/N ratio.

\begin{deluxetable}{lrrclrr}[]
\tablecaption{Zero-point offsets and $T_{\rm eff}$ trend slopes
\label{tbl:offsets}}
\tablehead{
\colhead{Elem} & \colhead{Offset} & \colhead{$10^3\alpha_T$} & 
\colhead{\phantom{blank}} &
\colhead{Elem} & \colhead{Offset} & \colhead{$10^3\alpha_T$}
}
\startdata
Mg	 & $ 0.000 $ & $0.94$  & &
K	 & $ 0.002 $ & $1.68$  \\
O 	 & $-0.016 $ & $2.28$  & &
Cr	 & $ 0.048 $ & $4.35$  \\
Si	 & $ 0.038 $ & $-3.22$ & &
Fe 	 & $ 0.053 $ & $0.76$  \\
S	 & $ 0.008 $ & $5.29$  & &
Ni	 & $ 0.030 $ & $1.33$  \\
Ca	 & $ 0.071 $ & $-6.01$ & &
V	 & $ 0.222 $ & $14.9$  \\
C+N      & $ 0.022 $ & $4.12$  & &
Mn	 & $ 0.002 $ & $16.3$  \\
Na	 & $ 0.043 $ & $8.89$  & &
Co	 & $-0.032 $ & $8.86$  \\
Al	 & $ 0.050 $ & $-12.3$ & &
Ce	 & $ 0.125 $ & $-2.64$ \\
\enddata
\end{deluxetable}

As discussed by \cite{Jonsson2020} and Holtzman et al.\ (in prep.), 
the APOGEE abundances include
zero-point shifts of up to 0.2 dex (though below 0.05 dex for most
elements) chosen to make the mean abundance ratios of solar metallicity
stars in the solar neighborhood satisfy $\xfe=0$.
These zero-point shifts are computed separately for giant and dwarf stars.
Here we use a particular set of $\logg$ and $\Teff$ cuts and a
sample that spans the Galactic disk.  We have therefore chosen to apply
additional zero-point offsets that force the median abundance ratio
trends of the high-Ia population in our sample to run through
$\xmg=0$ at $\mgh=0$.  These offsets are reported in Table~\ref{tbl:offsets};
the Mg offset is zero by definition.  The order of elements in the Table
follows that used in plots below, based on dividing elements into
related physical groups.
The V and Ce offsets are 0.222 dex and 0.125 dex, while others are
below 0.1 dex and mostly below 0.05 dex.
Since $\logg$ trends are also present in APOGEE at this level,
we regard it as reasonable to treat these as calibration offsets
rather than assume that the Sun is atypical of stars with similar
$\mgh$ and $\mgfe$.  
However, this is a debatable choice.
The most important offset is the one applied to Fe because the $\femg$ 
abundances determine the values of $\Acc$ and $\AIa$, though we note
that our choice of $\femgpl$ has a similar impact and is uncertain
at a similar level.  Furthermore, we identify $\femgpl$ from data that
have the Fe offset applied (Figure~\ref{fig:mgfe}), and much of the impact
of a different offset would be absorbed by the associated change in $\femgpl$.
The zero-point offsets for other elements
have a small but not negligible impact on our derived values of $\qxcc$ 
and $\qxIa$.  They should have minimal impact on residual abundances,
since the 2-process model is calibrated to reproduce the observed
median sequences.
Table~\ref{tbl:offsets} also lists slopes of trends with $\Teff$ that
are discussed in \S\ref{sec:residuals_teff} below 
(see equation~\ref{eqn:tefftrend}).

We adopt a high, $\SNR\geq 200$ threshold for most of our analyses
because we want to minimize the impact of observational errors on
our results, especially the statistics and correlations of residual
abundances.  However, for some purposes we want to improve our coverage
of the inner Galaxy, where distance and extinction leave fewer stars bright
enough to pass this high threshold.  For these analyses we lower
the SNR threshold to 100 at all $\mgh$ values, which increases the
sample to 55,438 (a factor of 1.6) and increases the number of stars
at $R=3-5\kpc$ by a factor of 4.3.  We refer to this as the SN100 sample,
but our calculations and plots use the higher threshold sample unless
explicitly noted otherwise.

\section{Median sequences and 2-process vectors}
\label{sec:median}

We separate our sample into low-Ia and high-Ia populations 
(conventionally referred to as ``high-$\alpha$'' and ``low-$\alpha$,''
respectively), using the same dividing line as W19:
\begin{equation}
\begin{cases}
\mgfe > 0.12 - 0.13\feh, & \feh<0 \cr
\mgfe > 0.12,                & \feh>0. \cr
\end{cases}
\label{eqn:boundary}
\end{equation}
For consistency with W19, we apply this separation to the APOGEE abundances
{\it before} adding the zero-point offsets in Table~\ref{tbl:offsets}, but
we include the offsets in all of our subsequent calculations and plots.
The distribution of our sample stars in $\mgfe$ vs.\ $\mgh$, together
with the median sequences and inferred CCSN iron fractions, have been
shown previously in Figure~\ref{fig:mgfe}.

\begin{figure*}
\centerline{\includegraphics[width=5.5truein]{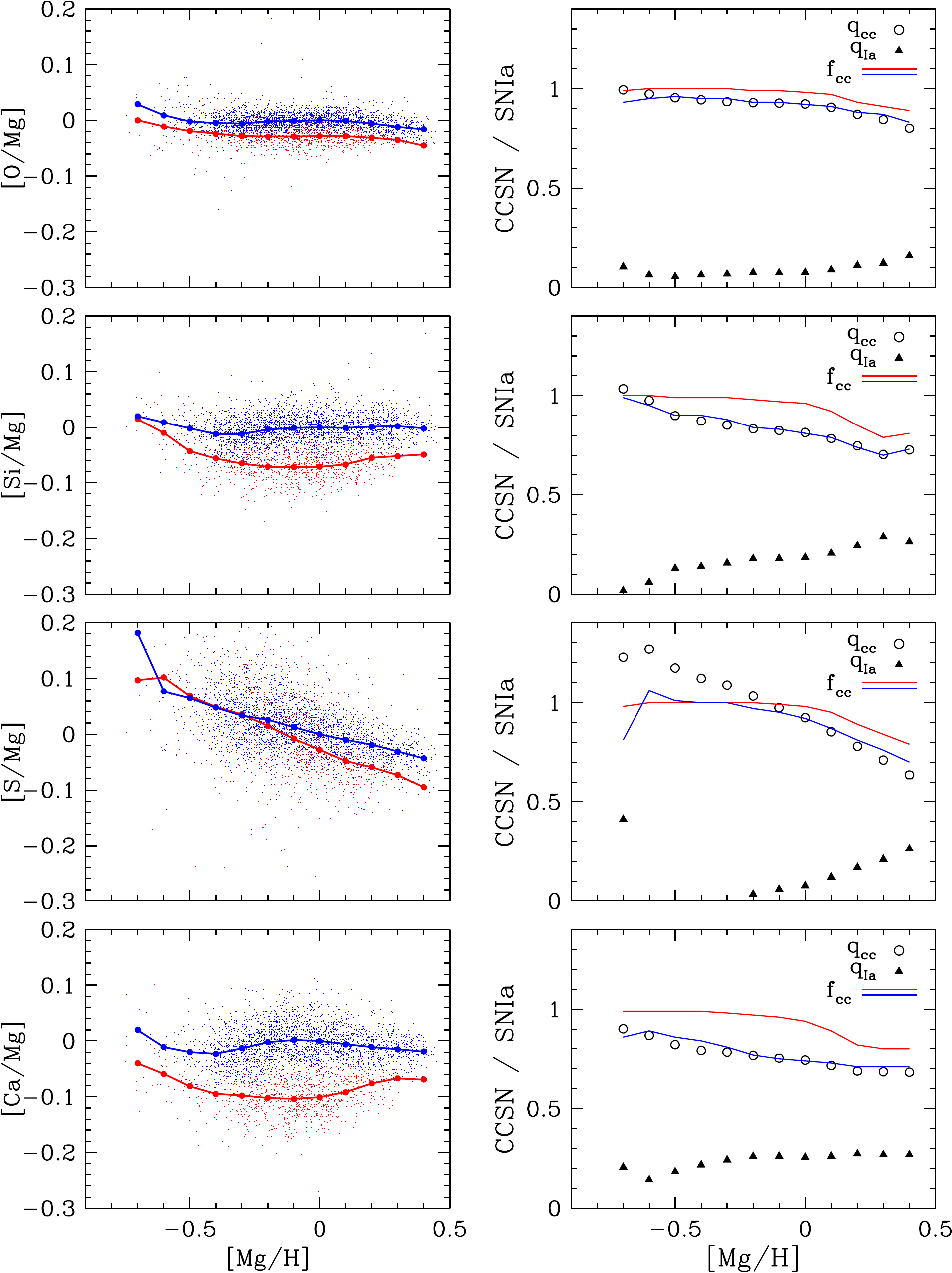}}
\caption{(Left) $\xmg$ vs. $\mgh$ for the $\alpha$ elements of
sample stars, color-coded as low-Ia (red) or high-Ia (blue).
Stars are randomly downsampled by a factor of four to reduce crowding.
Connected large points show the median values in bins of $\mgh$,
which the 2-process model fits exactly by construction.
(Right) Solar-scaled values of the CCSN and SNIa process vector components
for each element, $\qxcc$ (circles) and $\qxIa$ (triangles), 
inferred by fitting the observed median sequences.  
Red and blue curves show the CCSN fractions $\fxcc$,
which depend on the values of $\qxcc$ and $\qxIa$ and on the amplitude
ratio $\AIa/\Acc$ at the corresponding point on the median sequence
(eq.~\ref{eqn:fxcc2}).  Our abundances include zero-point calibrations
(Table~\ref{tbl:offsets}) that force the high-Ia sequence to pass through
$\xmg=0$ at $\mgh=0$.  For solar abundances, $\fxcc=\qxcc$, so blue curves
in the right panels always pass through the open circle at $\mgh=0$.
}
\label{fig:dataq_alpha}
\end{figure*}

\begin{figure*}
\centerline{\includegraphics[width=5.5truein]{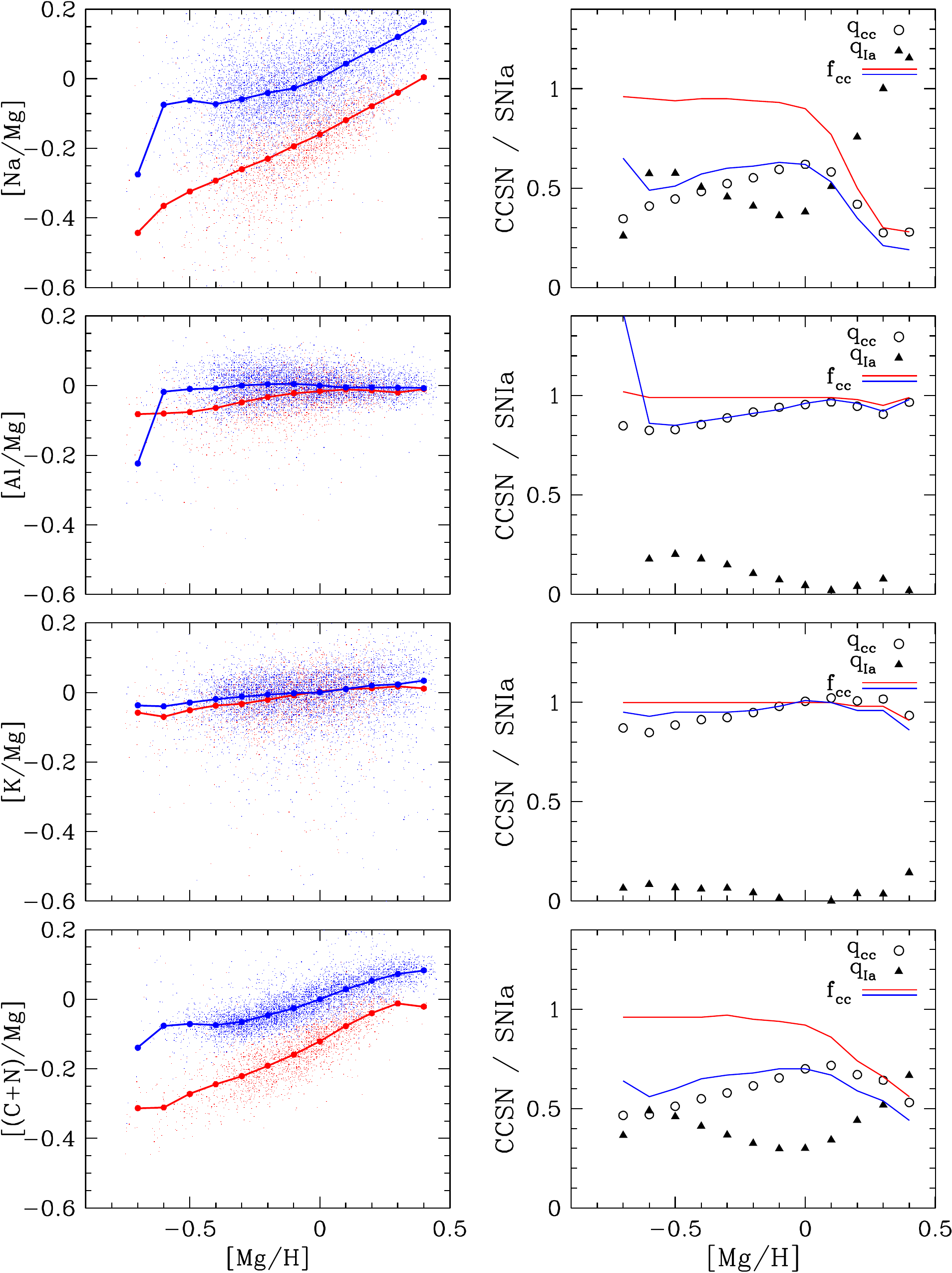}}
\caption{Same as Fig.~\ref{fig:dataq_alpha} but showing the
light odd-$Z$ elements Na, Al, and K and the element combination C+N.
}
\label{fig:dataq_oddz}
\end{figure*}

\begin{figure*}
\centerline{\includegraphics[width=5.5truein]{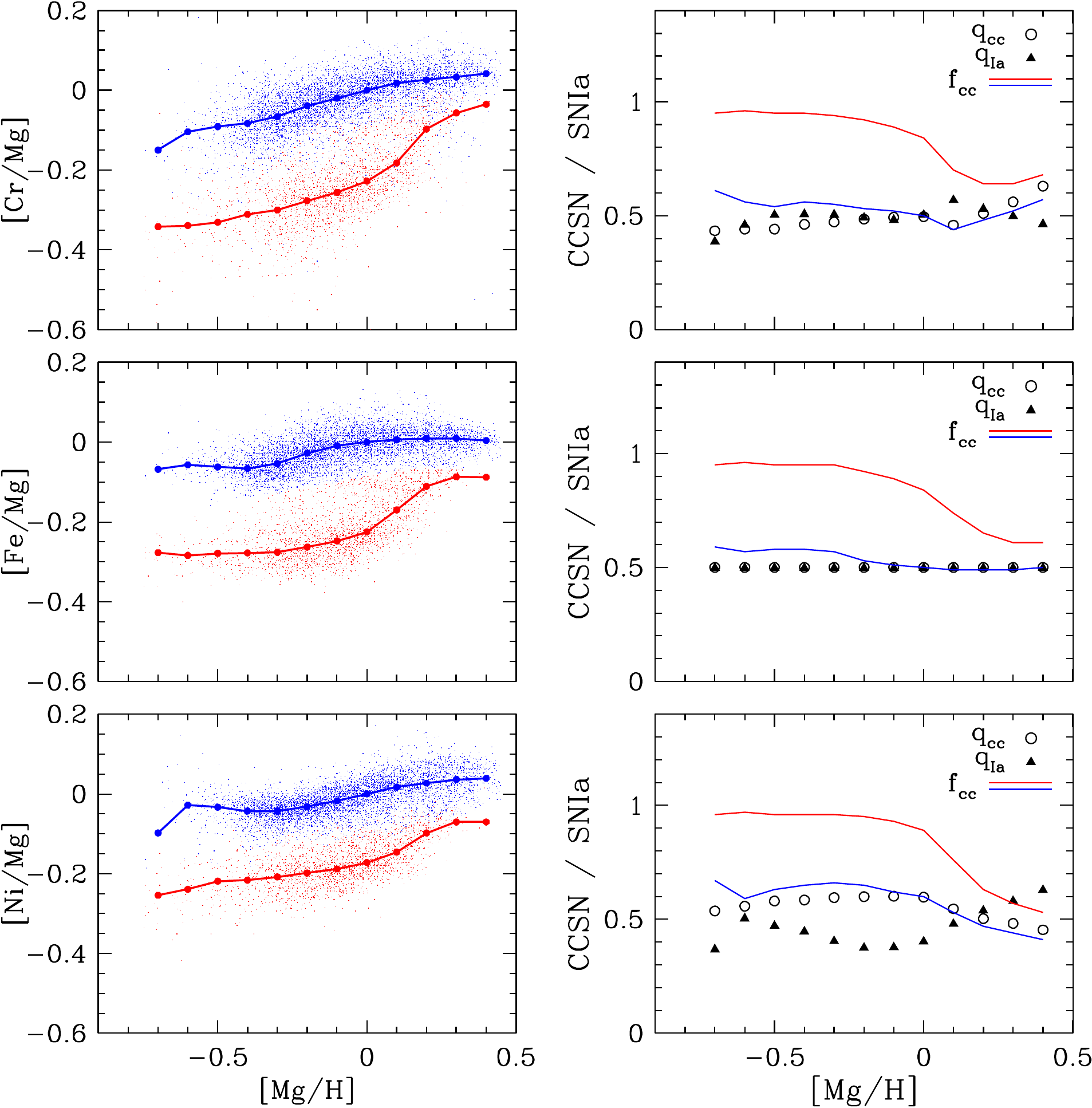}}
\caption{Same as Fig.~\ref{fig:dataq_alpha} but for even-$Z$ 
iron-peak elements.
}
\label{fig:dataq_peakeven}
\end{figure*}

\begin{figure*}
\centerline{\includegraphics[width=5.5truein]{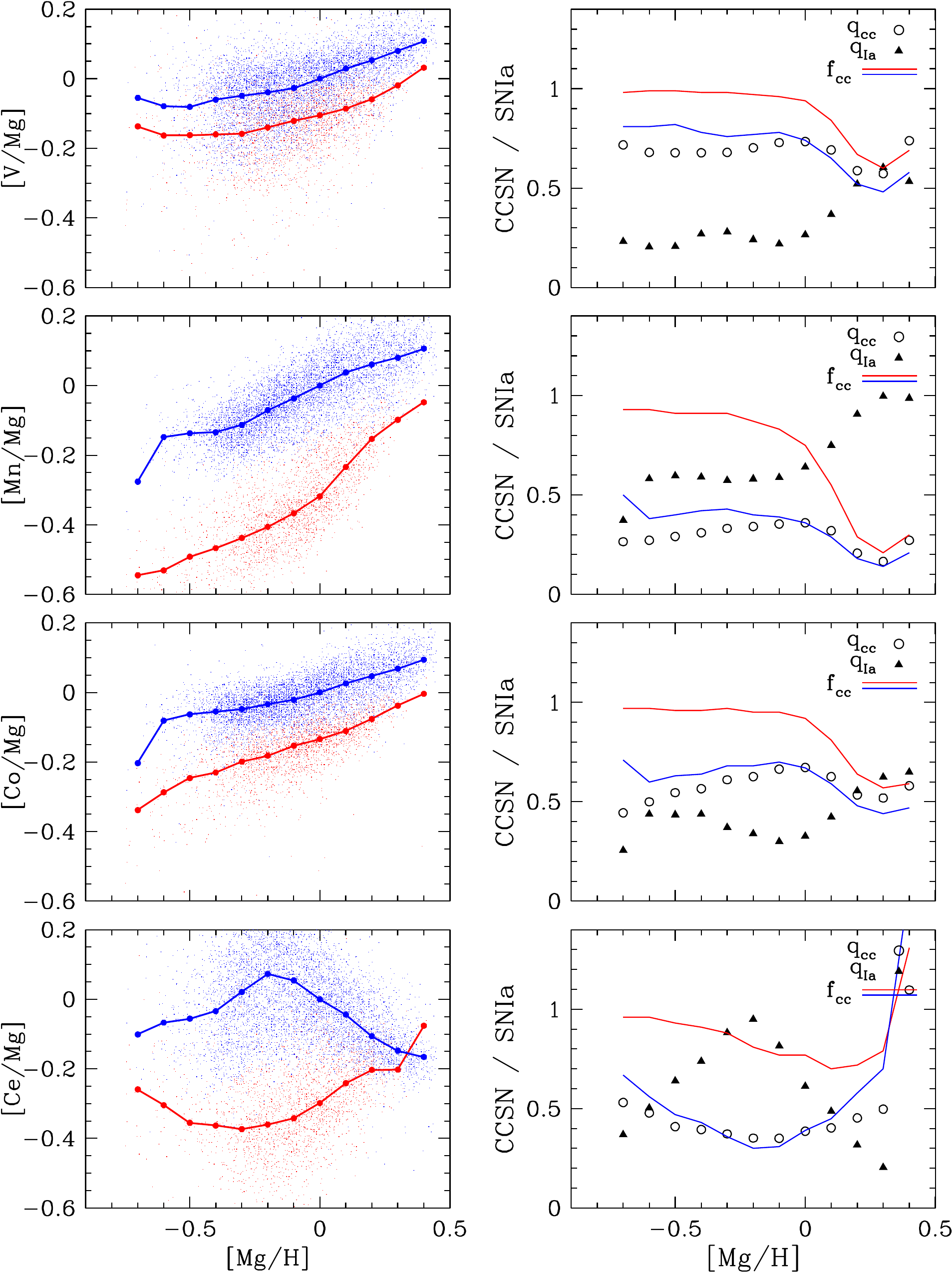}}
\caption{Same as Fig.~\ref{fig:dataq_alpha} but for odd-$Z$ 
iron-peak elements and the s-process element Ce.
}
\label{fig:dataq_peakodd}
\end{figure*}

The left panels of Figure~\ref{fig:dataq_alpha} show $\xmg$ vs.\ $\mgh$
distributions for other $\alpha$-elements (O, Si, S, Ca) in
the low-Ia and high-Ia populations, with median 
sequences shown by connected large points.
These can be compared to corresponding distributions in Figure 8 of W19,
based on DR14 data; not surprisingly, the results are similar.
The median [O/Mg] trends are nearly flat, with only a small separation
between the low-Ia and high-Ia median sequences, as expected if the
production of both O and Mg is dominated by CCSN with metallicity-independent
yields.  If real, the $\approx 0.05$-dex separation between the sequences
suggests a small contribution to O abundances from a delayed source,
perhaps AGB stars rather than SNIa.  For Si and Ca the sequence separations
are progressively larger, implying larger SNIa contributions to these
elements as predicted by nucleosynthesis models 
(e.g., \citealt{Andrews2017,Rybizki2017}).
For S the two median sequences are very close, implying minimal SNIa
contribution to S production, and they are sloped, implying S yields
from CCSN that decrease with increasing metallicity.  They diverge
slightly at high $\mgh$, implying a growing $\qxIa$.

The right panels of Figure~\ref{fig:dataq_alpha} show the values of
$\qxcc$ and $\qxIa$ derived from these median sequences via 
equations~(\ref{eqn:qxcc}) and~(\ref{eqn:qxIa}), using the ratios
$\AIa/\Acc$ along the two sequences shown in Figure~\ref{fig:mgfe}.
As discussed in \S\ref{sec:2process_vectors}, with a general metallicity
dependence the 2-process model fits the observed median sequences
exactly; empirical evidence for the qualitative validity of the model
will come from the reduction of scatter and correlations in residual
abundances shown below in \S\ref{sec:residuals}.
For O the inferred $\qxcc \approx 0.9$ at all $\mgh$, declining
slightly at the highest metallicity.
For Si and Ca the inferred values of $\qxIa$ at solar $\mgh$
are 0.19 and 0.26, respectively.  For S the low-Ia median sequence
crosses above the high-Ia median sequence at low metallicity, 
a violation of the 2-process model assumptions that leads to
slightly negative values of $\qxIa$.  Given the large observational
scatter in S abundances, this sequence-crossing appears compatible
with observational fluctuations, so we do not regard it as a serious
problem.  The inferred $\qxcc$ for S declines continuously with
increasing $\mgh$, tracking the sloped median sequences.
Because our zero-point offsets are chosen to give
$\xmg=0$ on the high-Ia sequence at $\mgh=0$, and because stars
with $\mgfe=\mgh=0$ have $\AIa=\Acc=1$ by definition, our fits
always yield $\qxcc+\qxIa=1$ at $\mgh=0$ 
(see equation~\ref{eqn:xmgratios}).  However, this constraint does
not apply at other metallicities.  Red and blue curves in these panels
show the fraction of each element that is inferred to come from CCSN
in stars on the low-Ia or high-Ia sequence (equation~\ref{eqn:fxcc}).
Even if $\qxIa > 0$, the low-Ia population has $\fxcc \approx 1$ 
at low $\mgh$ because these stars have $\mgfe \approx \mgfepl$,
implying (at least according to the 2-process model assumptions)
that nearly all enrichment is from CCSN.

Figure~\ref{fig:dataq_oddz} shows $\xmg$ vs. $\mgh$
distributions, median sequences, and inferred 2-process vectors for
Na, Al, K, and the element combination C+N.  
Because they have odd atomic numbers, the nucleosynthesis of Na, Al, and K
is fundamentally different from that of $\alpha$-elements even within
massive stars.  C and N are both
expected to have significant contributions from AGB stars in addition
to prompt contributions from CCSN and massive star winds, and the
AGB yields of N are predicted to have substantial metallicity dependence
(e.g., \citealt{Karakas2010,Ventura2013,Cristallo2015}).
Similar to W19, the median sequences for K and Al show little separation
between the low-Ia and high-Ia populations, indicating a dominant 
contribution from CCSN; in detail, the DR17 data show a slightly larger
sequence separation for Al and slightly smaller for K.
More significantly, the DR17 data show [Al/Mg] trends that are essentially
flat over this $\mgh$ range while the DR14 data showed an increasing
trend that implied CCSN yields increasing with metallicity.

As in W19, [Na/Mg] trends show a large separation between low-Ia and
high-Ia populations implying a substantial delayed contribution to Na
enrichment.  Standard nucleosynthesis models predict that SNIa and AGB
contributions are small compared to CCSN \citep{Andrews2017,Rybizki2017},
so this evidence for delayed enrichment comes as a surprise.
The Na features in APOGEE spectra are weak, making the abundance 
measurements noisy and susceptible to systematic errors, but there is
no obvious effect that would cause artificially boosted Na measurements
at this level
for high-Ia stars relative to low-Ia stars of the same $\mgh$.
A comparable sequence separation is also found in GALAH DR2
\citep{Griffith2019}.
The $\qxcc$ and $\qxIa$ values inferred from the 2-process fit are
comparable in magnitude over most of the $\mgh$ range.  The inferred
metallicity-dependence is complex, but given the scatter and uncertainties
of the Na measurements it should be treated with caution.

While W19 considered P as an additional odd-$Z$ elements, the
analyses in \cite{Jonsson2020} suggest that APOGEE's P measurements
in DR16 are not robust.  The P abundances are improved in DR17,
but they remain subject to significant systematics, 
and we have elected to omit P from this paper.

The two [(C+N)/Mg] sequences show a separation nearly as large as
the two [Na/Mg] sequences, again implying a substantial delayed contribution.
Nucleosynthesis models predict a moderate AGB contribution to C and
a dominant AGB contribution to N \citep{Andrews2017,Rybizki2017}, 
so this result is qualitatively expected.  The inferred metallicity
dependence is complex, with $\qxcc$ rising with $\mgh$ before leveling
out and dropping at high metallicity, and the opposite trend for $\qxIa$.
We caution, however, that the separation into prompt and delayed
contributions is not quantitatively accurate for elements that have
large AGB contributions because it is based on tracking Fe from SNIa,
and the delay time distributions for AGB enrichment and SNIa enrichment
are different (see, e.g., Figure~5 of \citealt{Johnson2020}).
Our present analysis does not allow us to separate the
roles of C and N in these trends, though for stars with asteroseismic
mass measurements one can apply corrections from stellar evolution models
to infer birth abundances of the two elements (see \citealt{Vincenzo2021b}).
The observed [(C+N)/Mg] sequences can themselves provide a quantitative 
test of chemical evolution models that track both elements.
The high-Ia medians of [(C+N)/Mg], [Na/Mg], and [Al/Mg] all show drops
in the lowest metallicity bin $-0.75 \leq \mgh < -0.65$.  This bin contains
just 53 stars, so this drop could simply be a statistical fluctuation,
though it might also be affected by accreted halo stars becoming a
significant fraction of the high-Ia population at this low metallicity
(see \S\ref{sec:populations}).

Figures~\ref{fig:dataq_peakeven} and~\ref{fig:dataq_peakodd} show
sequences and 2-process parameters for iron peak elements with
even atomic number (Cr, Fe, Ni) and odd atomic number (V, Mn, Co), respectively.
Trends are similar to those shown for DR14 data by W19, though
in W19 the $\xxmg{Cr}$ trends are flatter with $\mgh$ and the
$\xxmg{V}$ trends are steeper and with somewhat larger separation
between the low-Ia and high-Ia sequences.
For Fe, $\qxIa=\qxcc=0.5$ at all $\mgh$ as a consequence of adopting
$\mgfepl=0.3\approx \log_{10}2$ and assuming metallicity independence.
For Cr we also infer $\qxIa\approx\qxcc\approx 0.5$ at $\mgh=0$.
APOGEE Cr abundances exhibit apparent systematics for a significant
fraction of stars above solar metallicity \citep{Griffith2021a},
so we regard the median sequences and inferred 2-process parameters
as unreliable in the super-solar regime.  
While $\xxmg{Ni}$ trends are similar to $\femg$ trends, the
separation of low-Ia and high-Ia sequences is smaller, implying
that CCSN contribute 60\% of the Ni at solar abundances vs. 50\%
for iron.  We find somewhat higher CCSN fractions at solar abundances
for Co and V, 67\% and 74\%, respectively.  (To phrase things still
more precisely, in a star with [Fe/Mg]=[Ni/Mg]=[Co/Mg]=[V/Mg]=[Mg/H]=0,
we infer that 50/60/67/74\% of the star's Fe/Ni/Co/V atoms 
were produced in CCSN.)  

As in W19 we find that Mn has the largest SNIa contribution
of any APOGEE element.  Note that \cite{Bergemann2019} find a 0.15-dex
difference between 1D LTE and 3D NLTE abundances from H-band Mn I lines,
with little dependence on $\feh$, $\Teff$, or $\logg$; here we take
the ASPCAP Mn determinations at face value.
Although the low-Ia median sequence in 
$\xxmg{Mn}$ is steeply rising, this slope can be largely explained
by the increasing SNIa enrichment fraction along the sequence,
so that the inferred metallicity dependence of $\qxcc$ is weak.
We infer a sharp rise in $\qxIa$ at super-solar metallicity,
needed to explain the rising $\xxmg{Mn}$ trend on the high-Ia sequence.
Given the spectral modeling and calibration uncertainties in the super-solar 
regime, the rising trend of $\qxIa$
should be viewed with some caution, but it could suggest
a change in the physical properties of SNIa progenitors or explosion
mechanisms in super-solar stellar populations.
A similar pattern of rising $\qxIa$ at $\mgh > 0$ is seen for
C+N, Na, V, Co, and (more weakly) Ni.  These common trends could
indicate a delayed source (SNIa or AGB) that becomes important for
all of these elements at high metallicity, though they could also
be a sign that our assumptions for separating prompt and delayed
components are breaking down in this regime.
Like Na, Al, and C+N, the median trend of the high-Ia population
drops sharply in the lowest $\mgh$ bin for [Mn/Mg] and [Co/Mg],
and to a lesser extent for [Ni/Mg].

Figure~\ref{fig:dataq_peakodd} also presents results for the s-process 
element Ce.  Like Na, V, and K, APOGEE Ce abundances have relatively
large statistical uncertainties (mean of 0.043 dex in our sample),
and the large scatter about the median trends is likely dominated
by observational errors.  DR14 did not include Ce abundances, so we 
cannot compare to W19.   However, the large separation between the
low-Ia and high-Ia sequences and the non-monotonic metallicity
dependence of the high-Ia sequence are qualitatively similar to
results from GALAH DR2 for the neutron capture elements Y and Ba
\citep{Griffith2019}.  A rising-then-falling metallicity dependence
is expected for AGB nucleosynthesis of heavy s-process elements:
at low [Fe/H] the number of seeds available for neutron capture
increases with increasing metallicity, but at high [Fe/H] the number
of neutrons per seed becomes too low to produce the heavier s-process
elements \citep{Gallino1998}.  As with C+N, the decomposition into
prompt and delayed components implied by the $\qxcc$ and $\qxIa$
values should be regarded as qualitative because the delay time 
distribution for AGB Ce production should differ from that of 
SNIa Fe production.

We report the values of $\qxcc$ and $\qxIa$ for all elements
in Tables~\ref{tbl:qcc} and~\ref{tbl:qIa} in the Appendix , along with
the values of $\AIa/\Acc$ along the two median sequences 
(Table~\ref{tbl:aratio}).
The value of $\Acc$ follows from equation~(\ref{eqn:Accst}).
These quantities can be used in equation~(\ref{eqn:xmgratios}) to exactly
reproduce the median sequences shown in 
Figures~\ref{fig:dataq_alpha}-\ref{fig:dataq_peakodd}.
The values of $\fxccsun=\qxcc(z=1)$ (equation~\ref{eqn:fxccsun})
can be used to correct observed solar abundances
to the abundances produced by CCSN, which can then be used to test
the predictions of supernova models, as done by W19 (their fig.~20),
by \cite{Griffith2019} (their fig.~17), and most comprehensively
by \cite{Griffith2021b}, who carefully investigate the interplay between
IMF-averaged supernova yields and black hole formation scenarios.
The solar values of $\qxIa=1-\qxcc$ 
can be similarly used as a test of SNIa yield
models, a task we defer to future work.

\section{Residual abundances and their correlations}
\label{sec:residuals}

With the 2-process vectors determined by fitting the median $\xmg$
sequences of the low-Ia and high-Ia populations, we proceed to fit
the values of $\Acc$ and $\AIa$ for all individual sample stars as described
in \S\ref{sec:2process_stars}.  We perform a $\chi^2$-minimization
fit to the abundances $\mgh$, [O/H], [Si/H], [Ca/H], [Fe/H], and [Ni/H],
using the reported ASPCAP observational uncertainties for each measurement.
We choose these six elements because they have small mean observational
uncertainties, ranging from 0.0084 dex (Fe) to 0.0136 dex (Ca), 
because they have no major known systematic uncertainties in APOGEE data,
because the production of these elements is theoretically expected to
be dominated by CCSN and SNIa, and because their $\qxIa/\qxcc$ ratios 
span a wide range, giving strong collective leverage on $\AIa$ and $\Acc$.  
The only other abundance with a mean observational error in this 
range is [(C+N)/Mg], but we do not use this quantity in our
fits because it does not represent a single element and because the
production of C and (especially) N is expected to have significant
AGB contributions.  The [Mn/H] abundance also has a small mean
observational uncertainty (0.0144 dex), but the strong and
unusual metallicity dependence of $\qxIa$ above $\mgh=0$ could
distort inferred $\AIa/\Acc$ ratios in this regime if it is incorrect.
The 2-process amplitudes inferred from this six-element fit are close
to those inferred from $\mgh$ and $\femg$ alone via 
equations~(\ref{eqn:Accst}) and~(\ref{eqn:Aratio}), but they have
smaller statistical uncertainties, are robust to observational errors
in these two abundances alone, and mitigate artificial correlations
among residual abundances from measurement aberration
(see Fig.~\ref{fig:covar} below).

\begin{figure*}
\centerline{\includegraphics[width=6.5truein]{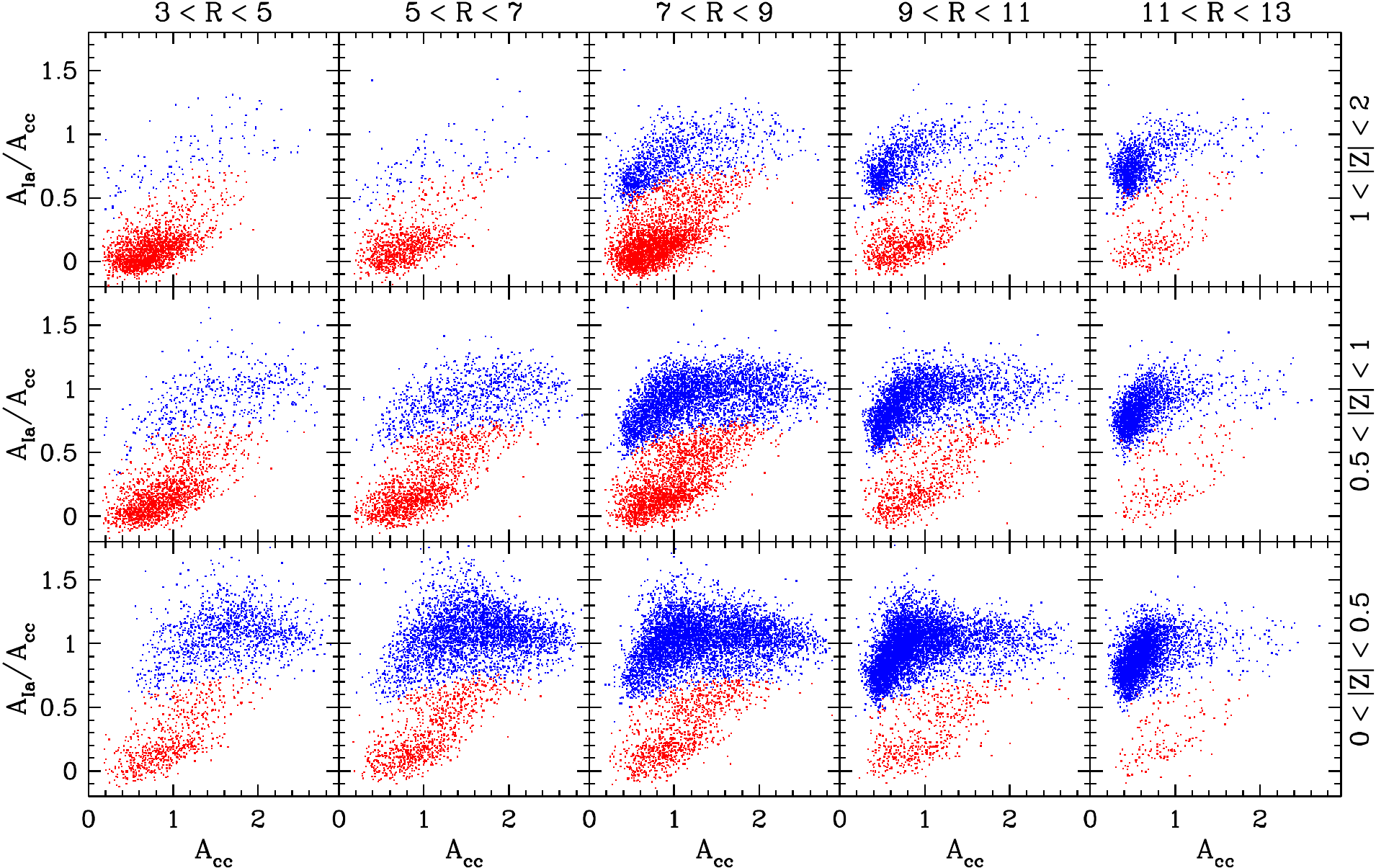}}
\caption{Distribution of stars in the 2-process parameters 
$\AIa/\Acc$ vs.\ $\Acc$, in zones of Galactocentric radius 
(columns) and midplane distance (rows) as labeled.
$\Acc$ measures the abundance of CCSN elements (e.g., Mg) relative
to solar, and $\AIa/\Acc$ measures the ratio of SNIa to CCSN enrichment;
$\AIa=\Acc=1$ for solar abundances.  Red and blue points show stars in
the low-Ia and high-Ia population, respectively.  To improve coverage
of the inner Galaxy, this plot uses the SN100 sample, which has a
SNR threshold of 100 at all $\mgh$ values.
Although we use six elements to fit stellar values of $\Acc$ and $\AIa$,
they are generally close to the values implied by Mg and Fe, so this plot
resembles a plot of $\femg$ vs.\ $\mgh$ but with transformed variables
that are linearly proportional to the inferred CCSN and SNIa content.
}
\label{fig:aratio_acc_map}
\end{figure*}

Figure~\ref{fig:aratio_acc_map} plots the distribution of stars in the
plane of $\AIa/\Acc$ vs. $\Acc$ in zones of Galactic $R$ and $|Z|$,
with red and blue points denoting stars in the low-Ia and high-Ia
populations, respectively.  For this plot we have used our SN100 sample
to improve coverage of the inner Galaxy.  Although we use our six-element
fits for $\AIa$ and $\Acc$, this map does not look noticeably different
if we use the values inferred from $\mgh$ and $\femg$ alone.
The $x$-axis quantity $\Acc$ is simply a linear measure of metallicity
as traced by CCSN elements, in solar units.
This plot is analogous to
Figure~4 of H15, showing $\afe$ vs. $\feh$, and still more
closely analogous to Figure~3 of W19, showing $\femg$ vs. $\mgh$.
Similar to those element-ratio maps, the low-Ia population is more 
prominent at small $R$ and large $|Z|$, and the metallicity ($\Acc$)
distribution of the high-Ia population shifts towards lower values at
larger $R$.  However, the non-linear relations between $\mgh$ and $\Acc$
(equation~\ref{eqn:Accst})
and between $\femg$ and $\AIa/\Acc$ (equation~\ref{eqn:Aratio}) 
highlight three features that are
less obvious in these earlier maps.  First, the median trend of $\AIa/\Acc$
in the low-Ia population rises continuously and approximately linearly
with $\Acc$ up to $\Acc \approx 1.5$, reaching $\AIa/\Acc \approx 0.5$.
Second, for $\Acc < 1$ the high-Ia stars also show a clear trend of
increasing $\AIa/\Acc$ with $\Acc$, especially evident in the 
$R \geq 7\kpc$ annuli.  Both of these trends follow
from the fact that the low-Ia and high-Ia sequences in $\mgfe$ are
sloped below $\mgh=0$ (see Fig.~\ref{fig:mgfe}), and their persistence
in a plot based on six-element fits implies that these slopes are not
caused by vagaries of $\mgfe$ abundance ratio measurements.
Third, the $\approx 0.04$-dex scatter of $\afe$ along the low-Ia
and high-Ia sequences, which is dominated by intrinsic scatter rather
than measurement noise \citep{Bertran2016,Vincenzo2021},  
translates to substantial scatter in $\AIa/\Acc$ at a given
metallicity within each population.  In the solar annulus
($R=7-9\kpc$) the distribution of $\AIa/\Acc$ is clearly bimodal
at sub-solar metallicity, with typical values of $0-0.3$ for the
low-Ia population and $0.6-1.2$ for the high-Ia population.
This bimodality reflects the bimodality of $\femg$ values,
which \cite{Vincenzo2021} demonstrate is an intrinsic feature
of the underlying stellar populations that is robust to $|Z|$-dependent
and age-dependent selection effects in the APOGEE sample.

\begin{figure*}
\centerline{\includegraphics[width=5.5truein]{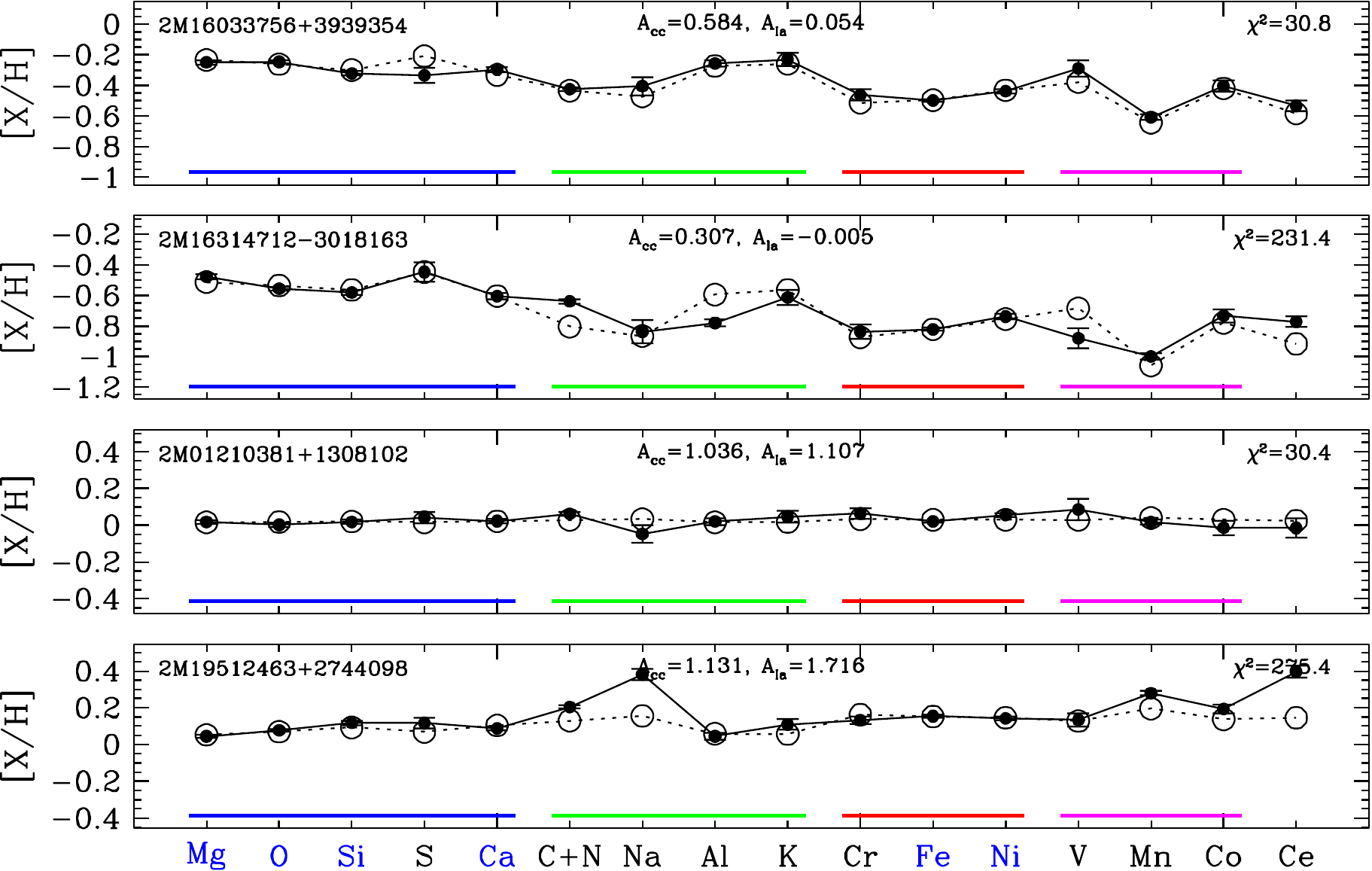}}
\caption{Examples of 2-process fits to element abundance ratios, for
two low-Ia stars (rows one and two)
with sub-solar $\mgh$ and two high-Ia stars (rows three and four) with
near-solar $\mgh$ and $\mgfe$.  The first and third stars have $\chi^2$
values near the median of the distribution for sample stars, while the
second and fourth stars have $\chi^2$ values near the 98th-percentile
of this distribution.  The model has two free parameters for each star
and is fit to the six elements listed in blue on the horizontal axis 
(Mg, O, Si, Ca, Fe, Ni).  In each panel, filled points with error bars 
show the measured value of [X/H] and the ASPCAP error.  
Open circles show the abundances predicted by the
2-process fit.  Solid and dotted lines are present to guide the eye.
Colored bars along the bottom of each panel group $\alpha$ elements
(blue), light odd-$Z$ elements (green), even-$Z$ iron-peak elements
(red), and odd-$Z$ iron-peak elements (magenta).
}
\label{fig:elem_stars1}
\end{figure*}

Turning to residual abundances,
Figure~\ref{fig:elem_stars1} shows examples of measured 
vs.\ predicted abundances for four stars.  
Recall that for each sample star we fit the two free parameters
$\Acc$ and $\AIa$ using the measured abundances of six elements,
then predict {\it all} of the abundances using these two process amplitudes
and the global values of $\qxcc$ and $\qxIa$ that have been inferred
from the median trends of the full sample.
The first two stars in Figure~\ref{fig:elem_stars1} are low
metallicity members ($\Acc=0.584$ and 0.307) of the low-Ia population
($\AIa=0.054$ and -0.005), one with a $\chi^2$ value near the median
for all sample stars and one with a high $\chi^2$ value near the 
98th-percentile of the $\chi^2$ distribution.
(Negative $\AIa$ values can arise for stars with $\afe$ values above
the low metallicity plateau.)
The third and fourth stars are solar metallicity ($\Acc=1.036$, 1.131)
stars from the high-Ia population, again one that is near the median
of the $\chi^2$ distribution and one near the 98th-percentile.  

For the first star, the 2-process model reproduces the observed abundance
pattern quite well, though the $\chi^2$ value of 30.8 for 14 degrees of freedom
(16 elements fit with two free parameters) is inconsistent with a purely
statistical fluctuation for Gaussian measurement errors with the reported
observational uncertainties.  The largest residuals are for 
S (0.13 dex), V (0.09 dex), Na (0.07 dex), and Cr (0.05 dex), 
elements with relatively large observational uncertainties in APOGEE.
The second star shows $\sim 0.2$-dex residuals for several
elements, including C+N, Al, V, and Ce, some overpredicted and
some underpredicted.  The predicted abundances of the third star are
all nearly solar, since $\AIa \approx \Acc \approx 1$, and the observed 
abundances are also near solar, with the largest residual being 0.08 dex
for Na.
The fourth star shows large residuals ($\sim 0.3$ dex) for Na and Ce
and smaller ($\sim 0.1$ dex) but statistically significant residuals
for C+N and Mn.  We will discuss other examples of high-$\chi^2$ outliers
in \S\ref{sec:outliers}.

\subsection{Removing trends with $\Teff$}
\label{sec:residuals_teff}

\begin{figure*}
\centerline{\includegraphics[width=5.5truein]{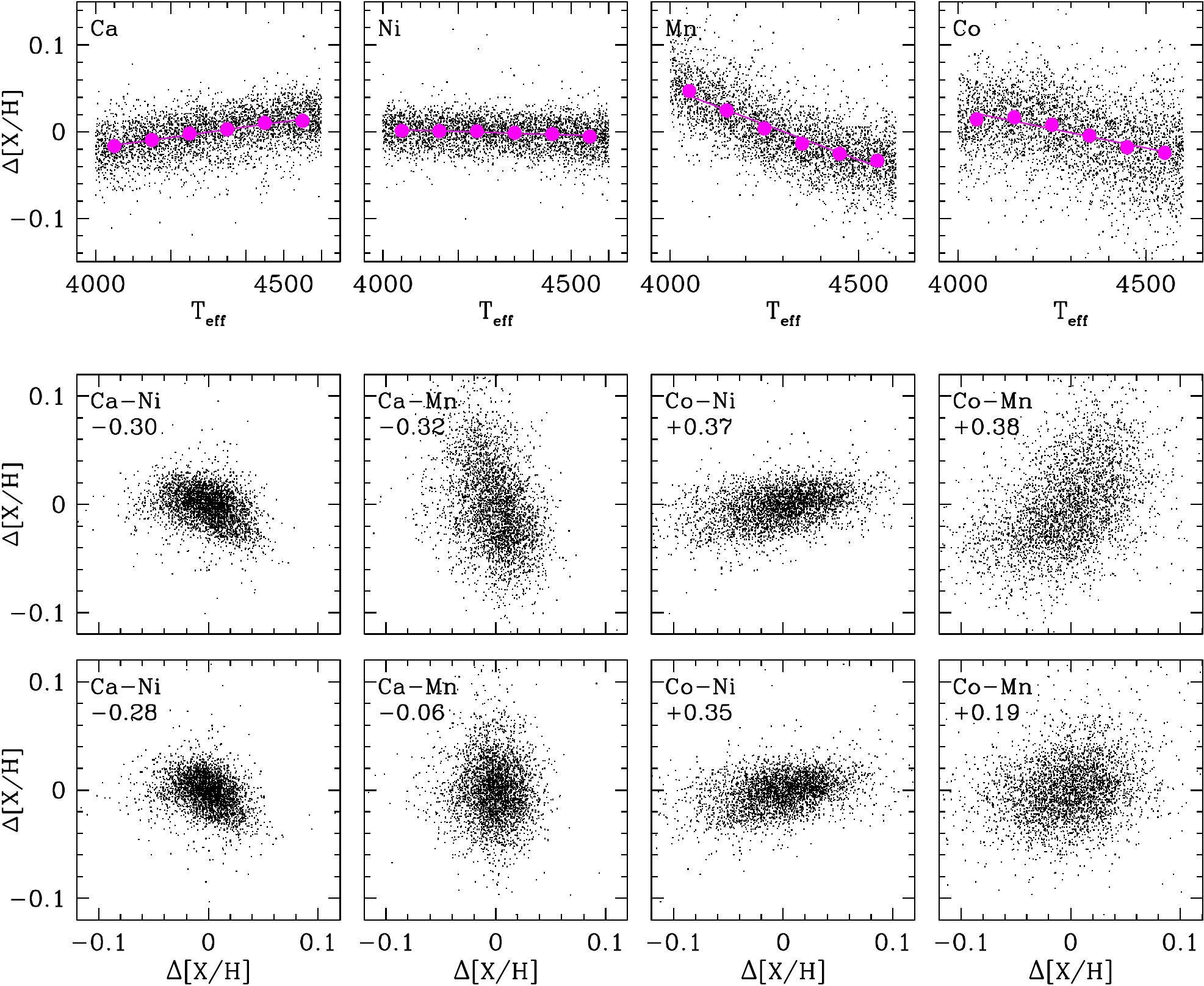}}
\caption{(Top) Correlation of abundance residuals 
(measured abundance - 2-process prediction) with $\Teff$ for 
Ca, Ni, Mn, and Co.  Magenta points and lines show the median residuals
in $100\K$ bins of $\Teff$ and the linear trends fit to these medians.
(Middle) Correlation of abundance residuals for the element pairs
Ca-Ni, Ca-Mn, Co-Ni, and Co-Mn (left to right).  Correlation coefficients
are listed in each panel.  
(Bottom) Same as middle using abundances that have been corrected for
the temperature trends as described in the text.  In all panels, points
have been downsampled by a factor of ten to reduce crowding, and
abundance residuals are shown in dex.
}
\label{fig:tefftrend}
\end{figure*}

We have limited the range of $\logg$ and $\Teff$ in our sample in order to
minimize the differential impact of abundance measurement systematics on
our results.  Nonetheless, there are subtle trends of residual abundances
with $\Teff$ over this range, as illustrated for four elements in the
top row of Figure~\ref{fig:tefftrend}.  In this and all subsequent plots
we adopt the sign convention
\begin{equation}
\Delta\xh \equiv \xh_{\rm data}-\xh_{\rm model}~.
\label{eqn:deltaxh}
\end{equation}
Mn residuals have the strongest
trend, with the median abundance residual changing from $0.047$ to $-0.033$
as $\Teff$ increases from $4050\K$ to $4550\K$.  Co residuals have a
weaker trend of the same sign, Ca residuals have a trend of similar 
magnitude but opposite sign, and Ni residuals have almost no trend
with $\Teff$.  

To avoid artificial correlations induced by these trends,
we fit them with linear relations and apply corrections to the DR17 APOGEE
abundance values:
\begin{equation}
\xh_{\rm corr} = \xh_{\rm APO} + {\rm Offset} + \alpha_T(\Teff-4300)/100~.
\label{eqn:tefftrend}
\end{equation}
The values of the zero-point offsets (discussed in \S\ref{sec:data}) and
the slopes $\alpha_T$ are listed in Table~\ref{tbl:offsets}, with
values of the latter ranging from $|\alpha_T| \sim 0.001$ 
(Mg, O, K, Fe, Ni) to $|\alpha_T| \sim 0.01$ (Ca, Na, Al, V, Mn, Co).
For Mn, for example, all abundances are increased by 0.002 dex, and the
abundances of stars with $\Teff=4600\K$ are further increased by
$0.0163\times 3 = 0.0489$ dex, thus increasing the median residual abundance
to near zero.  We have confirmed that residual trends with $\Teff$ and
$\logg$ are negligible for all elements after applying these corrections.
Note that the abundances plotted in 
Figures~\ref{fig:dataq_alpha}-\ref{fig:dataq_peakodd} include the zero-point
offsets but do {\it not} have the $\Teff$ corrections applied.
Median sequences and derived $\qxcc$ and $\qxIa$ values are negligibly
affected by these $\Teff$ trends.  They matter for our subsequent
analysis (and are used in all subsequent calculations and plots)
because they can affect correlations of residual abundances.

The lower half of Figure~\ref{fig:tefftrend} shows scatterplots of residual 
abundances for four pairs of these four abundances before (middle row) and
after (bottom row) removing the $\Teff$ trends.  The Ca-Ni correlation
is minimally affected because Ni residuals have almost no trend with $\Teff$.
The correlation coefficient changes from $-0.30$ before correction to $-0.28$
after correction.  However, Ca and Mn residuals have significant and 
opposite trends with $\Teff$ before correction, causing an artificial
anti-correlation with coefficient $-0.32$ that is almost entirely removed
by the $\Teff$ correction.  The Co-Ni correlation, like Ca-Ni, is minimally
affected by $\Teff$ trends.  However, the Mn-Co correlation is artificially
boosted because residuals of both elements are anti-correlated with $\Teff$,
and correcting the $\Teff$ trend lowers the correlation coefficient from 
0.38 to 0.19.  

In sum, we apply small ($\la 0.05$-dex) detrending corrections to the ASPCAP
abundances that remove weak correlations between residual abundances and 
$\Teff$.

\subsection{Simulating the impact of observational errors}
\label{sec:residuals_simulation}

If all stars were perfectly described by the 2-process model, the measured
abundances would still depart from the predicted abundances because of
statistical errors induced by observational noise.  It is tricky to predict
the distribution and correlation of these noise-induced residuals because
the observational uncertainties span a significant range from star-to-star
and element-to-element and because some of the measured values are
used to fit the 2-process amplitudes $\Acc$ and $\AIa$.  We have 
therefore created a simulated data set in which we take each star
from the sample, set its true abundances exactly equal to the 2-process
predictions given its measured $\Acc$ and $\AIa$, then add an ``observational''
error to each abundance, drawn from a Gaussian distribution with the
star's ASPCAP uncertainty for that element.  If an abundance measurement
is flagged in the APOGEE data, then we flag it in the simulation as well.
We can then apply the same analysis to the simulated data that we apply
to our observed sample to understand the results that would be expected
if all stars followed the 2-process model {\it and} all measurement
errors were described by Gaussian noise with the reported observational
uncertainties.

\subsection{Distribution of residuals}
\label{sec:residuals_distribution}

\begin{figure}
\centerline{\includegraphics{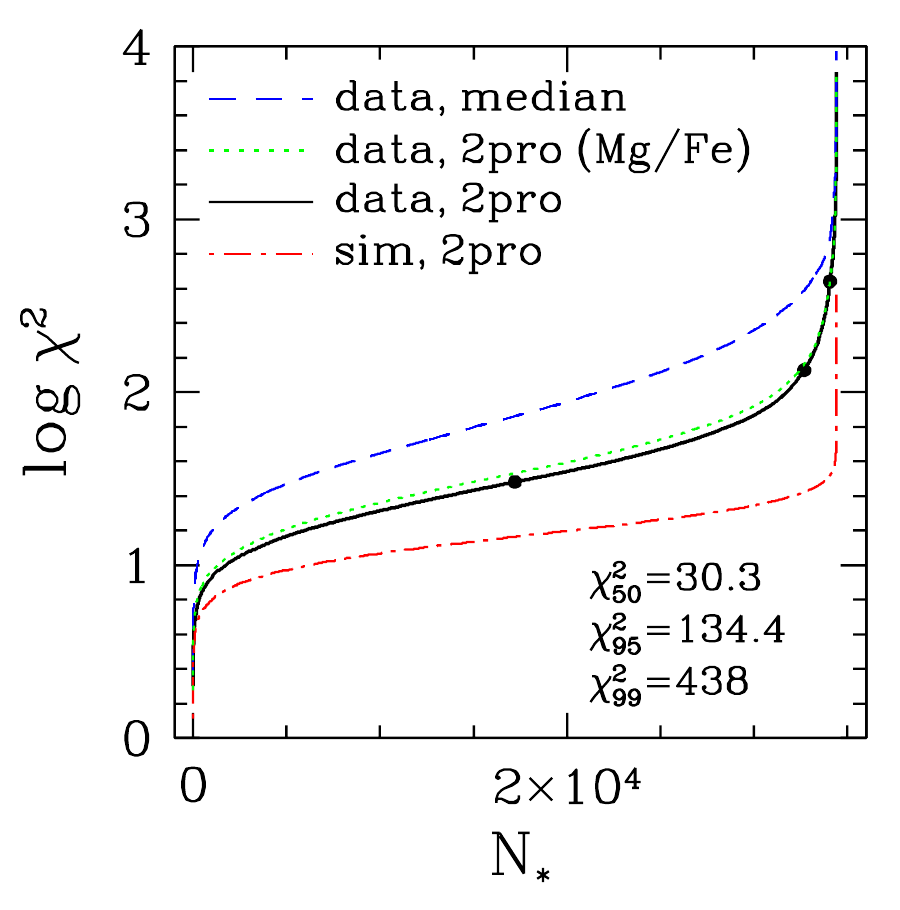}}
\caption{Cumulative distribution of $\chi^2$ values for model fits
to the sample of 34,110 stars.  Values of $\chi^2$ are computed using
all 16 elements shown in Figure~\ref{fig:elem_stars1}, omitting
flagged values.  The black solid curve shows results from applying
the full 2-process fit, with points marking the median, 95th-percentile,
and 99th-percentile values of the distribution (listed in the lower right).
The green dotted curve shows results when 
$\AIa$ and $\Acc$ values are inferred from [Mg/H] and [Fe/Mg] alone.
The blue dashed curve shows results when abundances are predicted
from the observed median trends (blue and red curves in the left panels
of Figures~\ref{fig:dataq_alpha}-\ref{fig:dataq_peakodd}), 
independent of the 2-process model.
The red-dashed curve shows results for a simulated data sample in
which all stars lie on the 2-process model and arise only
from Gaussian noise with the reported abundance errors.
}
\label{fig:chisq_dist}
\end{figure}

The black curve in
Figure~\ref{fig:chisq_dist} shows the cumulative distribution of $\chi^2$
values for the 34,410 sample stars, computed using all of the elements 
shown in Figure~\ref{fig:elem_stars1} but omitting flagged element values.  
The median, 95th, and 99th-percentiles of this
distribution are 30, 134, and 438, respectively.
Only 13\% of stars have $\chi^2$ values below 14, the number of
degrees of freedom for 16 elements and two free parameters, so either
the true distribution has intrinsic element scatter relative to the
2-process model or the observational abundance errors exceed those
predicted for Gaussian noise with the ASPCAP uncertainties, or both.
The red curve shows the $\chi^2$ distribution for the simulated data
set described in \S\ref{sec:residuals_simulation}, and in this case
45\% of stars have $\chi^2 < 14$.

The green curve shows the $\chi^2$ distribution if $\AIa$ and $\Acc$
are inferred from [Mg/H] and [Fe/Mg] alone instead of the six-element
2-process fit.  This leads to only a small increase in $\chi^2$ values.
The blue curve shows the distribution of $\chi^2$ if we compute 
element residuals from the observed median sequences instead of
the 2-process fits, interpolating the sequences in $\mgh$ to avoid 
any effects of metallicity variation across our 0.1-dex $\mgh$ 
bins.
These $\chi^2$ values are substantially higher (e.g., a median value
of 78, vs. 30 for the 2-process residuals), a first demonstration
that the 2-process model is removing physical scatter present in the
stellar abundance distribution.  In other words, a star's APOGEE abundances
are typically closer to those predicted by the 2-process model than
they are to the median abundances of stars of the same $\mgh$ and
the same population (low-Ia or high-Ia), by an amount that is highly
statistically significant.

\begin{figure*}
\centerline{\includegraphics[width=5.5truein]{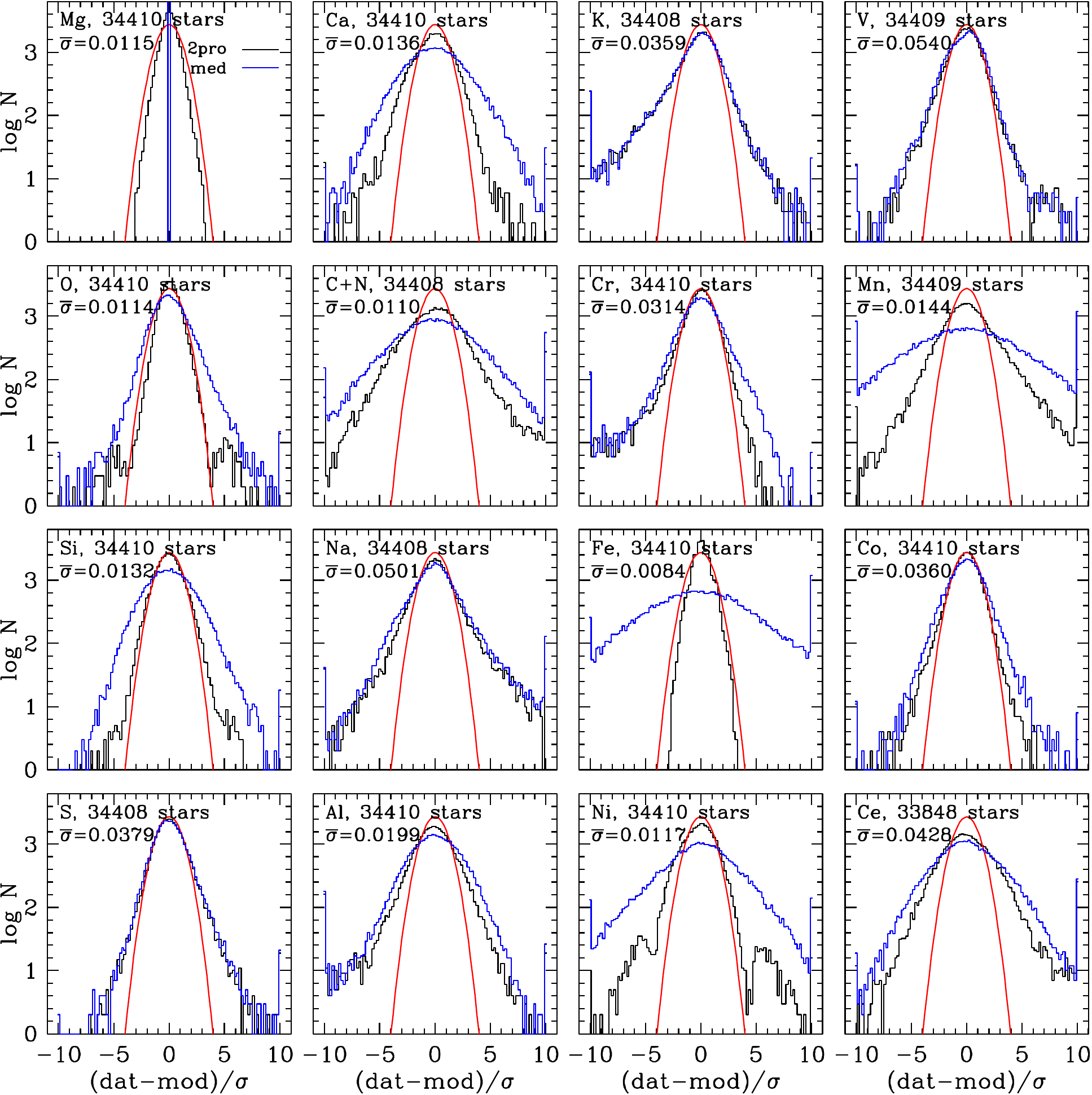}}
\caption{Distribution of [X/H] deviations from model predictions,
normalized by the reported abundance error.  Black histograms
show results using the full 2-process fit to six abundances.
Blue histograms show deviations from
the median sequences.  Red curves show a unit Gaussian for reference.
The mean ASPCAP error 
is listed in each panel.  For the median trend deviations, the $\mgh$ deviation
is zero by definition, giving a zero-width blue histogram.
}
\label{fig:delta_dist}
\end{figure*}

Figure~\ref{fig:delta_dist} examines the deviation distributions 
element-by-element.  For those elements that have a mean ASPCAP
abundance error smaller than 0.015 dex, the deviations from the 2-process
model predictions (black histograms) are significantly narrower 
than the deviations from the median sequence predictions (blue).
Some of this improvement arises because many of these elements are
used in the 2-process fit, but even if we use only Mg and Fe to determine
the 2-process parameters the deviation distributions for the other fit
elements are narrower than the distribution of deviations from the median
sequences.  Mn and C+N, which are not used in the 2-process fit,
both show narrower deviations from the 2-process predictions than
from the median sequences.  For other elements, with larger mean errors,
there are only small differences between the 2-process deviation
distribution and the median-sequence deviation distribution, presumably
because the deviations in both cases are dominated by observational errors. 
However, the extended tails of the distributions are still noticeably reduced
for Al, Cr, Co, and Ce.

Red curves show a unit Gaussian, which is narrower than the
2-process deviation distributions for all elements except Mg and Fe.
The simulated star sample drawn from the 2-process model yields
deviation distributions (not shown) very close to these unit Gaussians,
as expected.  For many elements the extended tails of the deviation
distribution appear more nearly exponential than Gaussian, and for
some elements there is a significant asymmetry between positive and
negative deviations in these extended tails.  Extended tails and asymmetries
could arise from either genuine physical deviations or measurement errors.
We investigate this question to a limited degree in \S\ref{sec:outliers},
though it is difficult to quantify the relative contribution of physical
and observational outliers without detailed investigation of the abundance
determinations for a large number of stars.

\begin{figure}
\centerline{\includegraphics[width=3.2truein]{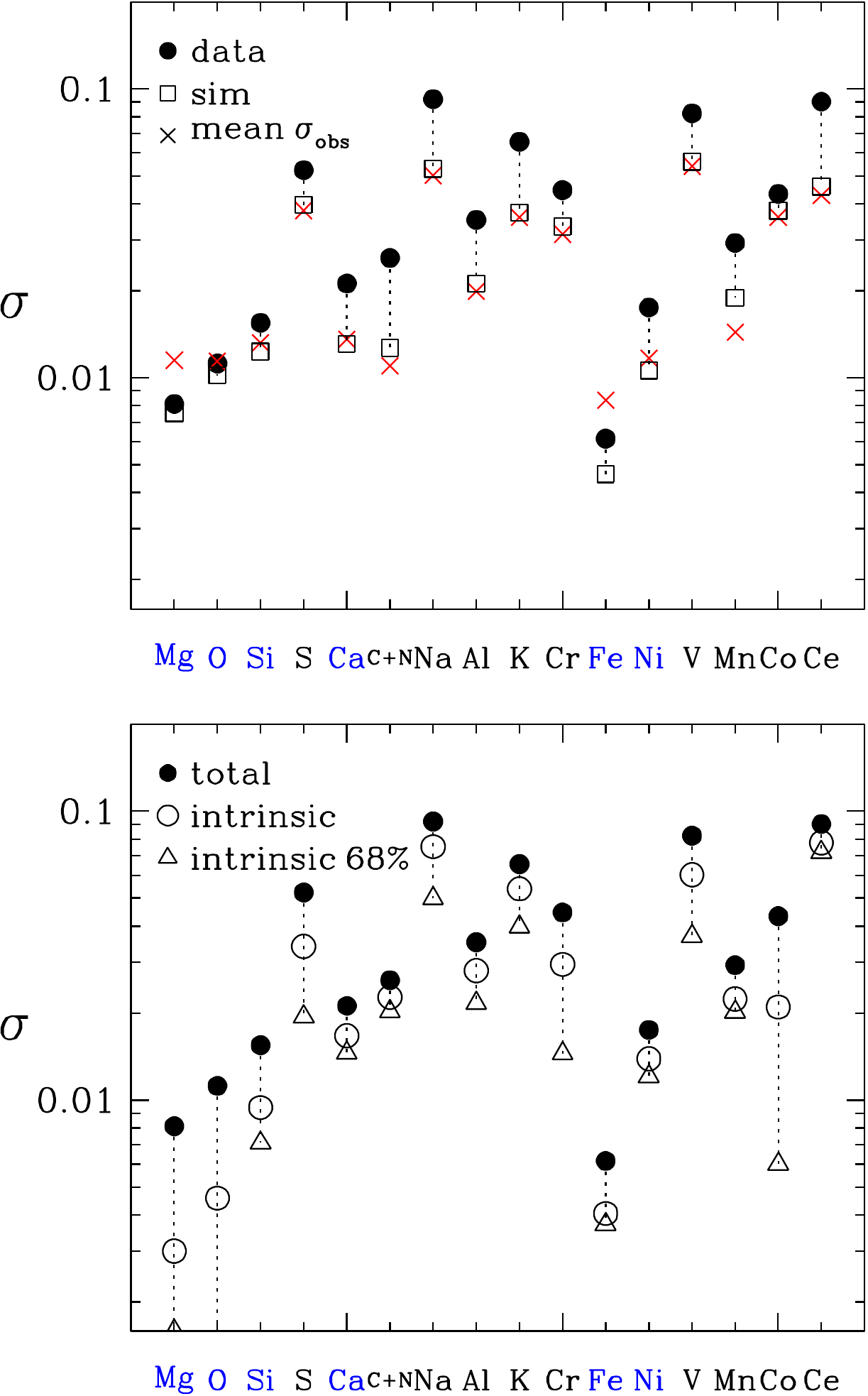}}
\caption{{\it Top}: Dispersion of [X/H] values relative to 2-process 
predictions, in dex.
Filled circles show results for the data.  Open squares show results
from the simulation in which stars lie on the 2-process model except
for Gaussian errors.  Red crosses show the mean observational error.  
Elements used in the 2-process fit are denoted in blue on the horizontal axis.
{\it Bottom:} Filled circles, repeated from the top panel, show the total
dispersion.  Open circles show the intrinsic dispersion estimated by
subtracting in quadrature the dispersion of the simulated data set
(i.e., the squares in the top panel).  Open triangles show an alternative
estimate of the intrinsic dispersion obtained by subtracting 
half of the 16-84\% percentile range 
($\pm 1\sigma$ for a Gaussian distribution), in quadrature, for the
simulated data from the same quantity for the observed data.
}
\label{fig:sigma}
\end{figure}

In both panels of Figure~\ref{fig:sigma}, filled circles show the rms 
difference between ASPCAP abundances and 2-process model predictions.  
In the upper panel, red crosses show the mean abundance error reported 
by ASPCAP for that element.  Open squares show the rms
deviations for the simulated data set.  These simulated dispersions
are generally very similar to the mean abundance errors, though they
are significantly lower for Mg and Fe, which carry high weight in the 
2-process fit.  In the lower panel, we estimate the intrinsic dispersion
by subtracting the dispersion of the simulated data from the dispersion
of the observational data.  If intrinsic and observational scatter
contributed equally to the variance, then the intrinsic dispersion 
would be $1/\sqrt{2} = 70.7\%$ of the total dispersion.
Half of the elements have an intrinsic/total dispersion
ratio near or below this value (Mg, O, Si, S, Cr, Fe, V, Co), and the other
half have higher ratios that imply intrinsic dispersion dominating
over the observational scatter.
However, the observational contribution could be underestimated, and the
intrinsic dispersion overestimated, if the reported measurement uncertainties
are systematically low or if non-Gaussian tails of the measurement errors 
inflate the dispersion.  As already noted, the residual distributions
for many elements exhibit exponential tails, which could represent real
deviations or non-Gaussian measurement errors.  As an alternative
estimate of dispersion, we have taken half of the 16-84\% percentile
range ($\pm 1\sigma$ for a Gaussian distribution) for the observed
residuals, then computed the same quantity for the simulated data
and subtracted in quadrature, obtaining the open triangles in 
Figure~\ref{fig:sigma}.  These alternative estimates characterize 
scatter in the core of the residual abundance distribution, with less
sensitivity to large deviations (whether physical or observational).

For Mg, O, and Fe we estimate rms intrinsic scatter
of only 0.003-0.005 dex, but given the weight of these elements
in the 2-process fit a low scatter is expected.
For other elements the rms intrinsic scatter ranges from $\approx 0.01-0.02$
dex (Si, Ca, C+N, Ni, Mn, Co) to $\approx 0.05-0.08$ dex (Na, K, V, Ce).  
The percentile-based
intinsic scatter estimate is lower for all elements, with Mg, O, Si, S, Ca,
C+N, Al, Cr, Fe, Ni, Mn, and Co having values $\la 0.02$ dex and the 
Na, K, V, and Ce scatter reduced to 0.04-0.07 dex.
Our estimates of intrinsic scatter, including the relative values of
different elements, are similar to those inferred by TW21 for scatter
in APOGEE abundances conditioned on [Mg/H] and [Mg/Fe].
We originally performed our analysis for the APOGEE DR16 data set,
and while the relative ranking of elements was nearly the same, the
total scatter and estimated intrinsic scatter were consistently higher.
The lower estimates of intrinsic scatter in DR17 likely reflect improvements
in the abundance analysis that reduce the number of large measurement errors.

In sum, the 2-process model predicts a star's APOGEE abundances better than
the median abundances of similar stars, demonstrating that much of the 
abundance scatter {\it within}
the low-Ia and high-Ia populations arises from scatter
in SNIa/CCSN ratios at fixed $\mgh$.  RMS residuals about 2-process predictions
range from $\sim 0.01$ dex for the most precisely measured elements to
$\sim 0.1$ dex for Na, V, and Ce.  These dispersions exceed those expected
from observational errors alone, implying intrinsic scatter at the
$\sim 0.01-0.05$ dex level, depending on element.

\subsection{Covariance of residuals}
\label{sec:residuals_covariance}

As emphasized by TW21, correlations are a more robust measure of residual
structure in elemental abundance patterns than dispersion, because estimating
the intrinsic dispersion requires accurate knowledge of the observational
error distribution.  The correlations also provide richer information 
about the sources of residual structure, which could include additional
enrichment processes, stochastic sampling of the IMF, binary mass transfer,
or even effects like variable depletion of refractory elements in 
proto-planetary disks or abundances boosted by giant planet engulfment.
We compute elements of the covariance matrix of element pairs X$_i$, X$_j$ as
\begin{equation}
C_{ij} = 
\langle (\Delta[{\rm X}_i/{\rm H}])(\Delta[{\rm X}_j/{\rm H}]) \rangle
\label{eqn:cij}
\end{equation}
with $\Delta[{\rm X}/{\rm H}]$ defined as the difference between the
observed abundance and the 2-process model prediction 
(equation~\ref{eqn:deltaxh}).

\begin{figure}
\centerline{\includegraphics[width=3.2truein]{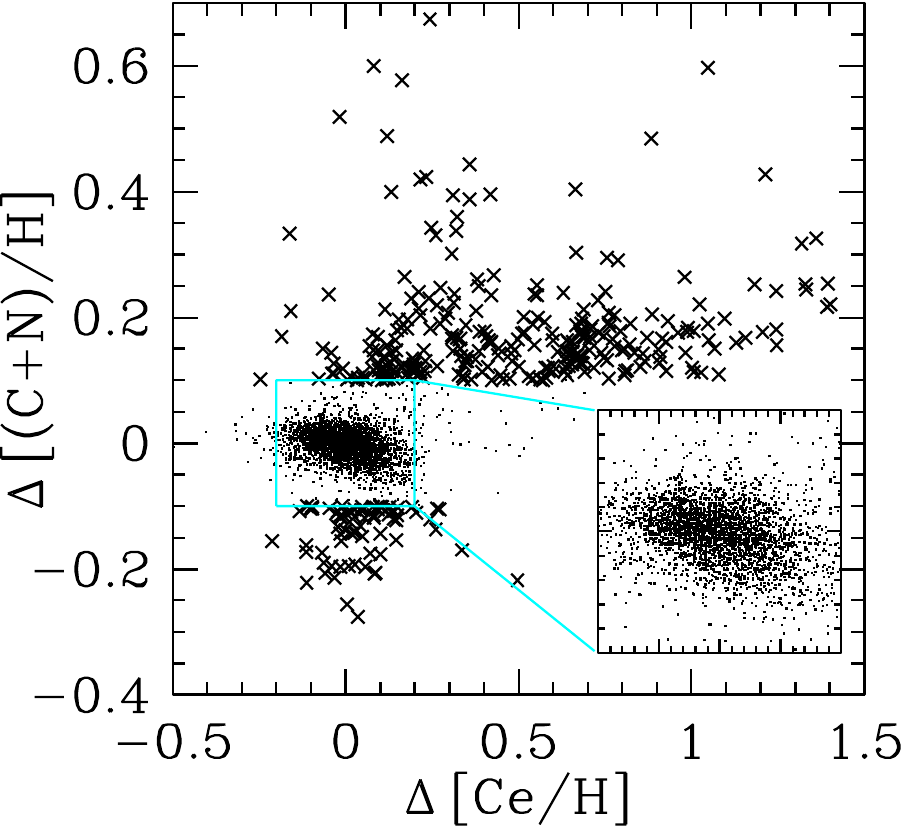}}
\caption{Bivariate distribution of 2-process residuals in 
[(C+N)/H] vs.\ [Ce/H].  Small dots show a random 10\% subset of
the full sample.  Crosses show all sample stars that have
$|\Delta[{\rm (C+N)}/{\rm H}]| > 0.1$.  The inset provides an expanded
view of the core of the distribution, from -0.2 to 0.2 in
$\Delta[{\rm Ce}/{\rm H}]$ and -0.1 to 0.1 in $\Delta[{\rm (C+N)}/{\rm H}]$.
Although residuals in the core of the distribution are anti-correlated,
there is a population of rare stars with large positive deviations
in (C+N) and Ce.  Two examples of such stars are shown in 
Figure~\ref{fig:elem_highchi2} below and discussed in \S\ref{sec:outliers}.
}
\label{fig:CN_Ce}
\end{figure}

Pairwise scatterplots of element residuals generally resemble the
examples shown in Figure~\ref{fig:tefftrend}.  In particular, for
element pairs with significant correlation or anti-correlation, 
the scatterplot shows a consistent slope between the core of the
distribution and the stars with large residuals.
The one exception is (C+N) vs.\ Ce: the core of the distribution 
shows a clear anti-correlation of the residual abundances, but there 
is a population of rare outlier stars with large positive residuals
in both (C+N) and Ce, as illustrated in Figure~\ref{fig:CN_Ce}.
We discuss this population further in 
\S\ref{sec:outliers}.  To avoid covariance estimates being driven
by rare outliers, we eliminate stars with element deviations 
$>10\sigma_{\rm obs}$ before computing covariances involving that element.
This censoring reverses the sign of the (C+N)-Ce covariance, which
would be positive instead of negative if we retained the extreme outliers.
It has a moderate impact on some other matrix elements involving Ce or (C+N)
and minimal effect on other element pairs.

\begin{figure*}
\centerline{
\includegraphics[width=3.2truein]{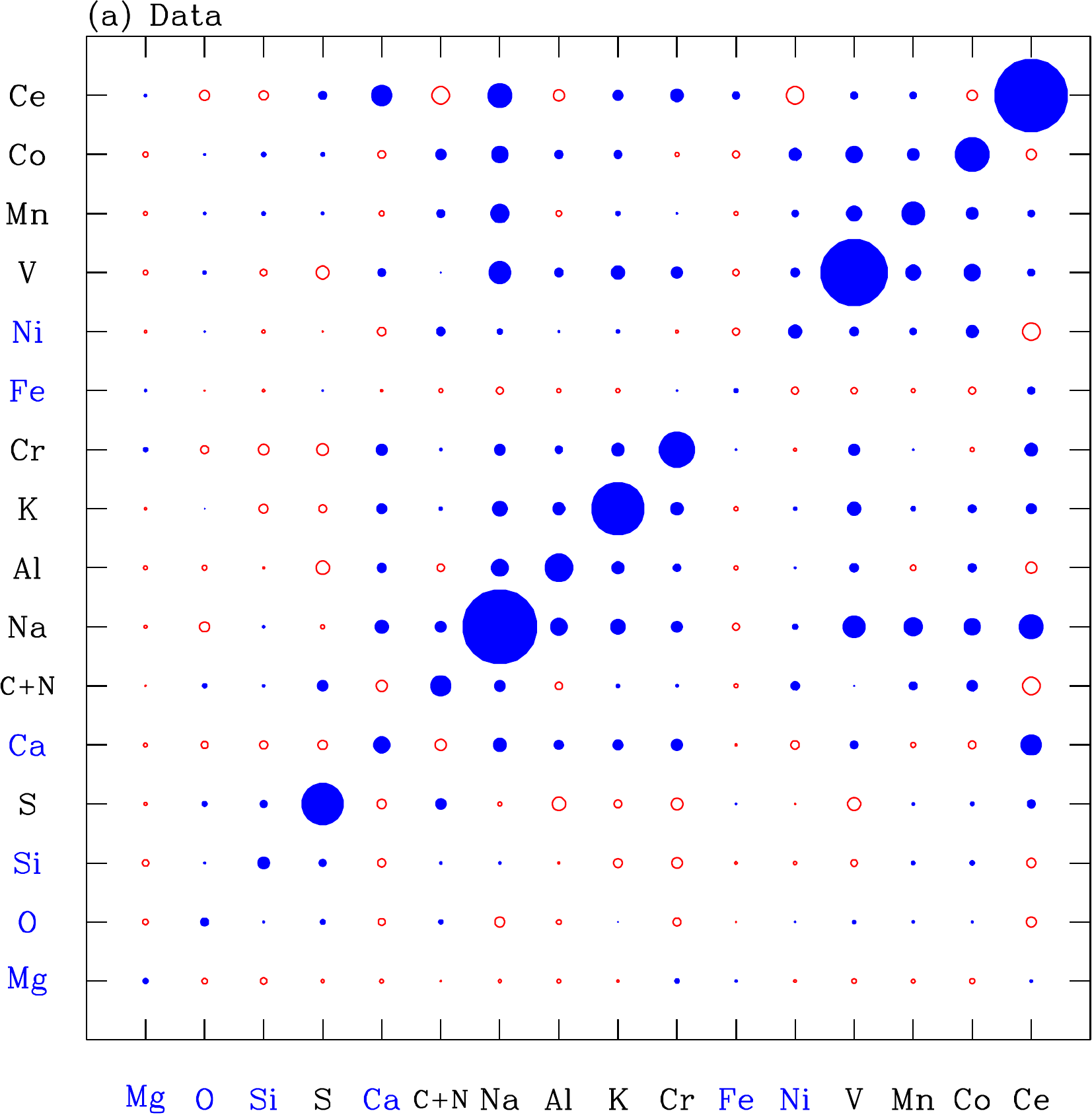}
\hskip 0.2truein
\includegraphics[width=3.2truein]{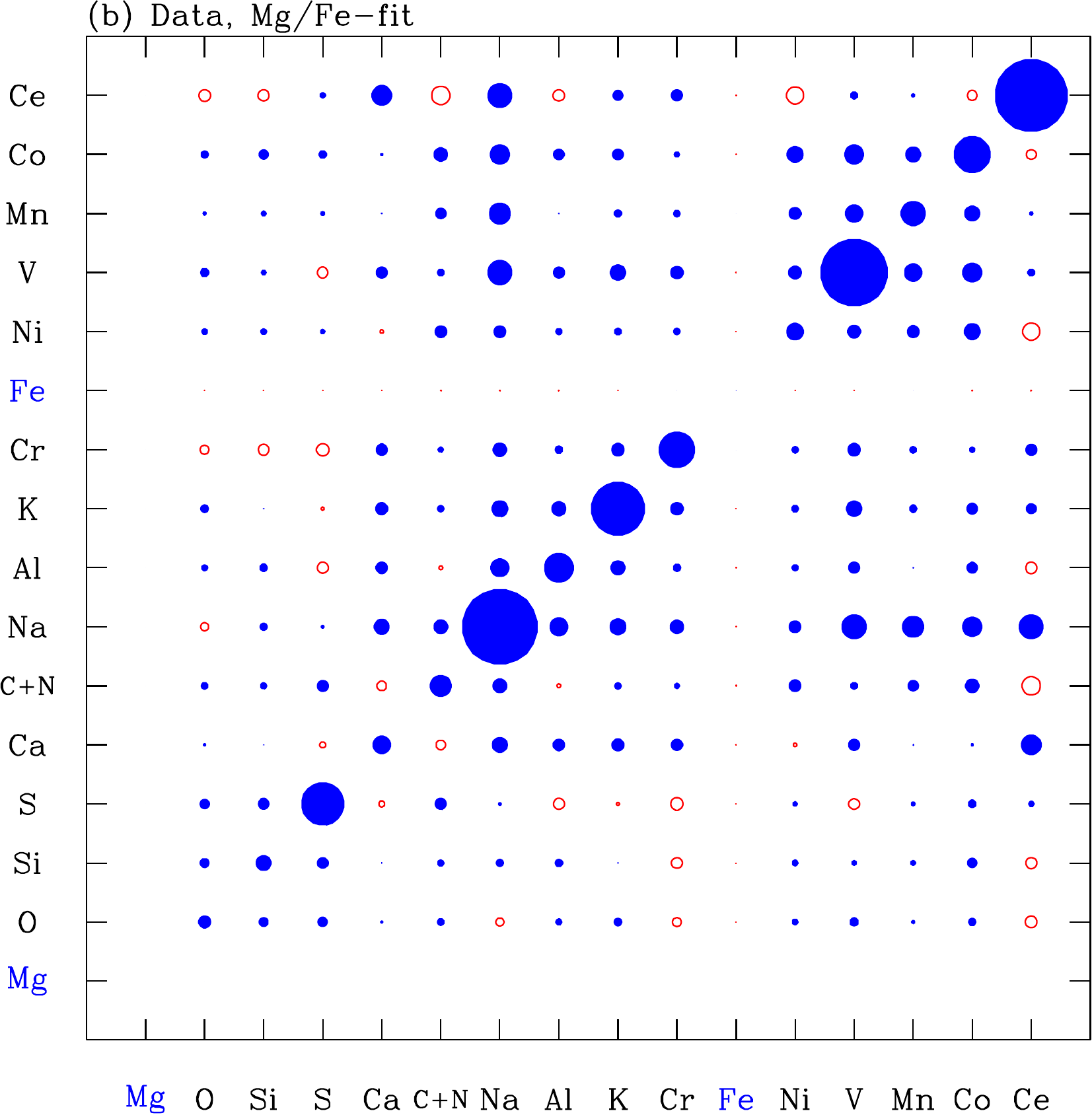}
}
\centerline{
\includegraphics[width=3.2truein]{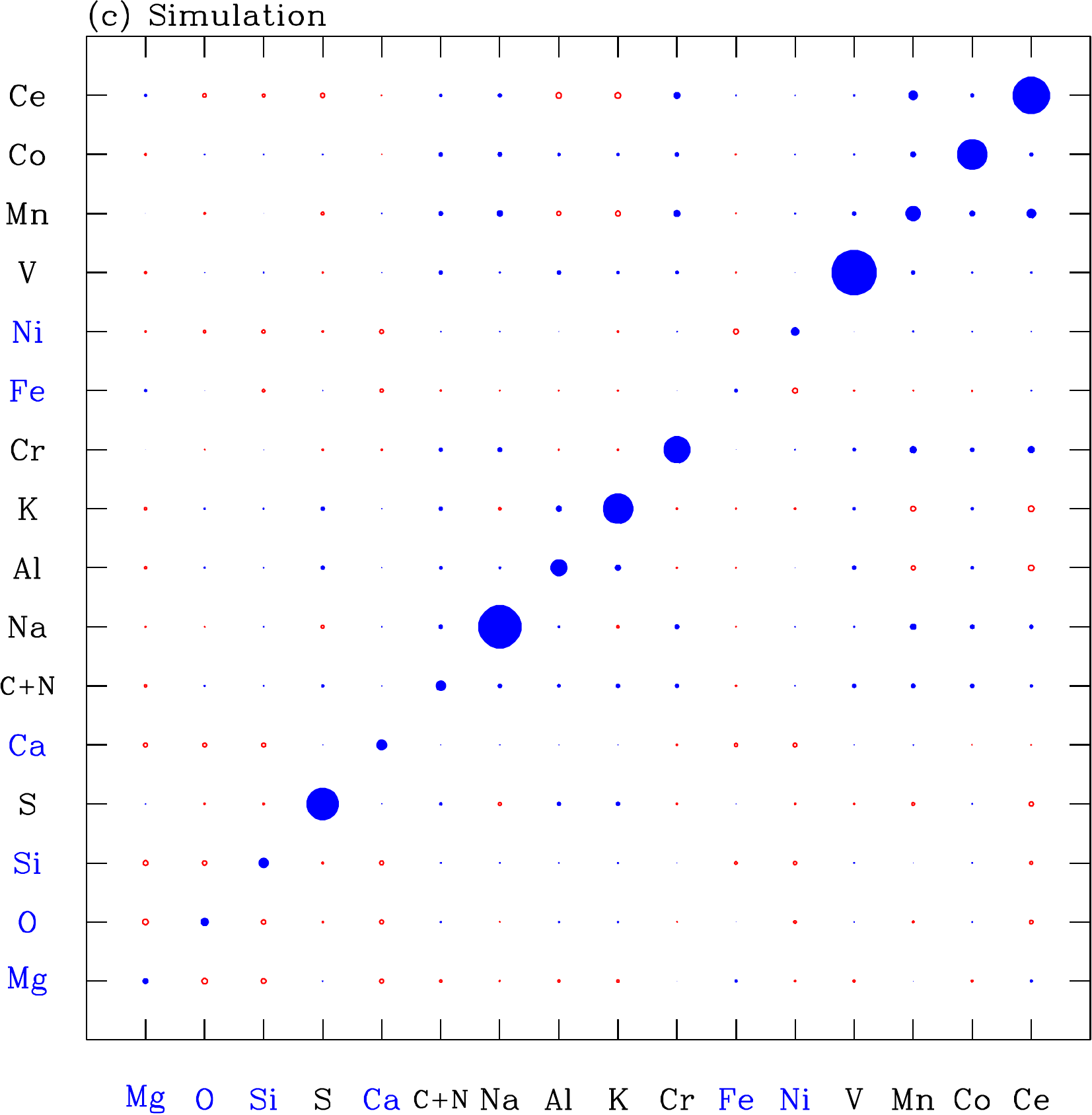}
\hskip 0.2truein
\includegraphics[width=3.2truein]{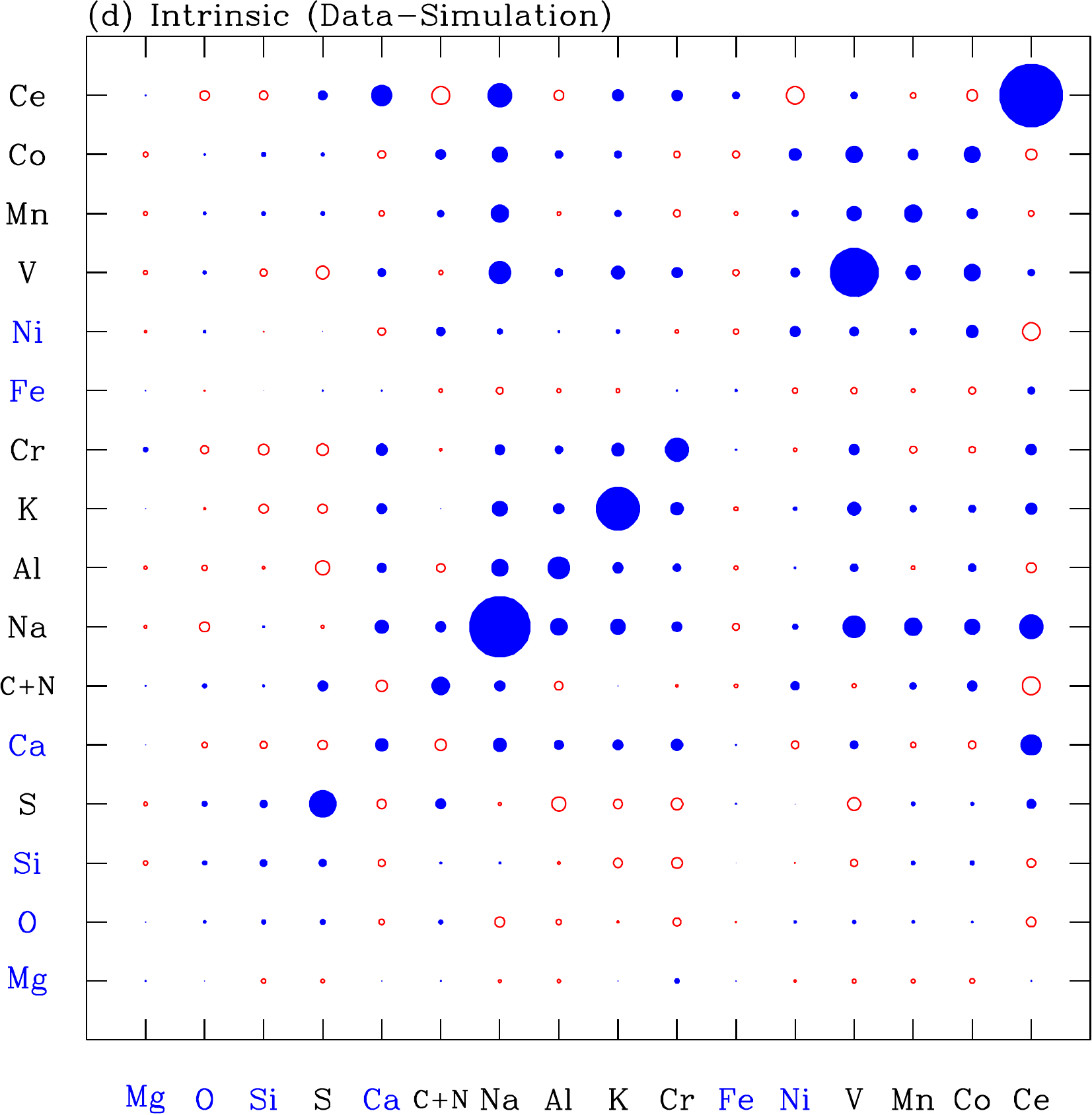}
}
\caption{Covariance matrix of deviations between measured [X/H]
values and 2-process model predictions.  In each panel, filled and open circles
denote positive and negative values, respectively, with area 
proportional to the magnitude of the matrix element and consistent
scaling used across all panels.
(a) Residual covariance for the APOGEE sample, with $\Acc$ and $\AIa$
fit using the six elements denoted in blue on the axes.
The diagonal elements are the squares of the rms deviations shown by filled
circles in Figure~\ref{fig:sigma}.  For visual scaling, note that the
diagonal elements (Na,Na) and (Ce,Ce) have values of about $(0.09)^2$,
(S,S) has a value of about $(0.05)^2$, and (O,O) has a value of about 
$(0.01)^2$.
(b) Residual covariance when $\Acc$ and $\AIa$ are fit using Mg and Fe only,
which increases the artificial correlations induced by measurement aberration.
(c) Residual covariance for the simulated data set in which all stars
lie on the 2-process model prediction and have Gaussian observational errors
at the level of the reported ASPCAP uncertainties.  Off-diagonal elements
arise from measurement aberration, but they are small compared to the 
observed covariances.
(d) Intrinsic covariance estimated by subtracting (c) from (a).
}
\label{fig:covar}
\end{figure*}

Figure~\ref{fig:covar}a shows the residual abundance
covariance matrix of our APOGEE sample.
Diagonal elements are the squares of the rms deviations shown by
the filled circles in Figure~\ref{fig:sigma}.
This covariance matrix would look qualitatively similar if we 
did not remove the 
temperature trends discussed in \S\ref{sec:residuals_teff}, but 
the covariances involving pairs of elements with the strongest
trends (largest values of $|\alpha_T|$ in Table~\ref{tbl:offsets})
would be noticeably affected.
Figure~\ref{fig:covar}b shows the residual covariance if we determine
$\Acc$ and $\AIa$ values from [Mg/H] and [Mg/Fe] alone, instead of 
fitting six elements.  Here the correlations are stronger and almost
all positive, except those involving Mg and Fe, which have vanishing
residuals by definition.  These artificial correlations span many
elements because random Fe and Mg measurement errors lead to errors 
in $\AIa$ and $\Acc$ and thus to correlated deviations from 
the 2-process predictions, the effect that TW21 describe as
``measurement aberration.''  Fitting six elements mitigates 
this effect, though it does not entirely remove it.  
Figure~\ref{fig:covar}c shows the covariance of the simulated data,
which has no {\it intrinsic} residual correlations by construction.
However, because $\Acc$ and $\AIa$ must be fit to abundances with
random statistical errors, measurement aberration still produces
off-diagonal covariances.  


The most important conclusion from comparing the simulation covariance
matrix to the data covariance matrix is that measurement aberration is
much too small to explain the observed covariances.  Our conclusion
agrees with that of TW21, who investigated the correlation of residual
abundances in the {\it conditional} probability distribution
$p(\xh)$ at fixed $\feh$ and $\mgfe$, using an APOGEE
sample nearly identical to ours.  TW21 also find that the measured
residual correlations are much larger than those arising from measurement
aberration.  Artificial correlations could also arise from the 
abundance determination itself, e.g., from random errors in $\Teff$
leading to correlated deviations in the abundances of multiple elements.
TW21 examine this issue by approximately modeling the ASPCAP measurement 
procedure and conclude that artificial correlations from the
measurement method are also much smaller than the observed correlations
(see their figure~9).

To estimate the covariance matrix of intrinsic deviations from the
2-process model, we subtract the simulated covariance matrix (c) from
the data covariance matrix (a).  This subtraction produces little change
because the $C_{ij}$ elements for the simulation are generally
much smaller than those of the data.  The result is shown in 
Figure~\ref{fig:covar}(d).  The diagonal elements of this covariance
matrix correspond to the open circles in Figure~\ref{fig:sigma}.
As discussed in \S\ref{sec:residuals_distribution}, the estimates
of the intrinsic variance are sensitive to knowledge of the observational
error distribution, so their magnitudes are uncertain.
However, the prominent off-diagonal structures in 
\S\ref{sec:residuals_distribution} likely represent genuine physical
correlations among abundance residuals.  The most obvious of these
structures are the block of correlations among the iron-peak elements
Ni, V, Mn, and Co, and another block of correlations among 
the elements Ca, Na, Al, K, and Cr.  Ce is also positively correlated
with all members of this latter group except Al.
There is also a noticeable positive correlation of Na with
V, Mn, and Co.

\begin{figure*}
\centerline{
\includegraphics[width=3.2truein]{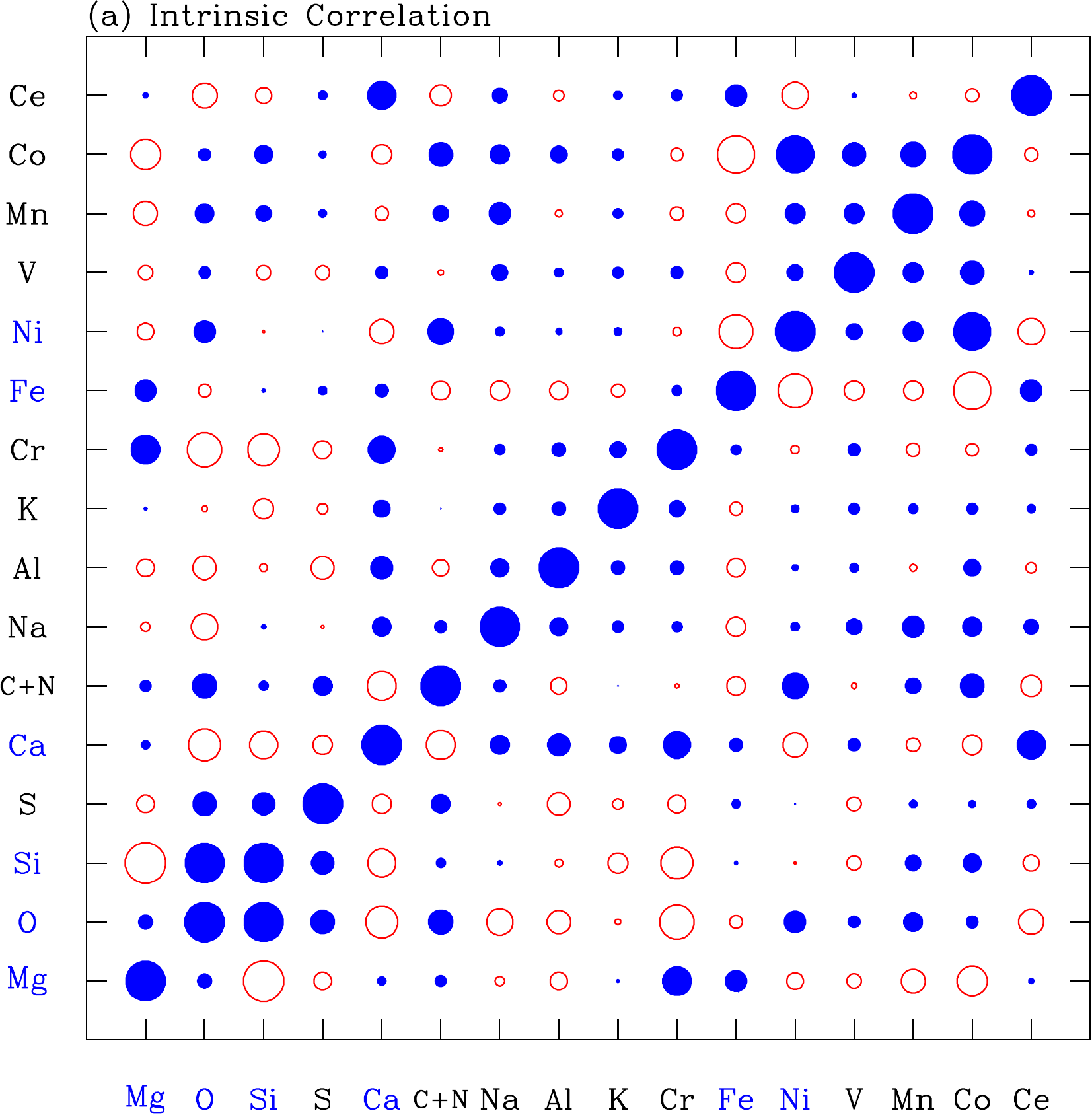}
\hskip 0.2truein
\includegraphics[width=3.2truein]{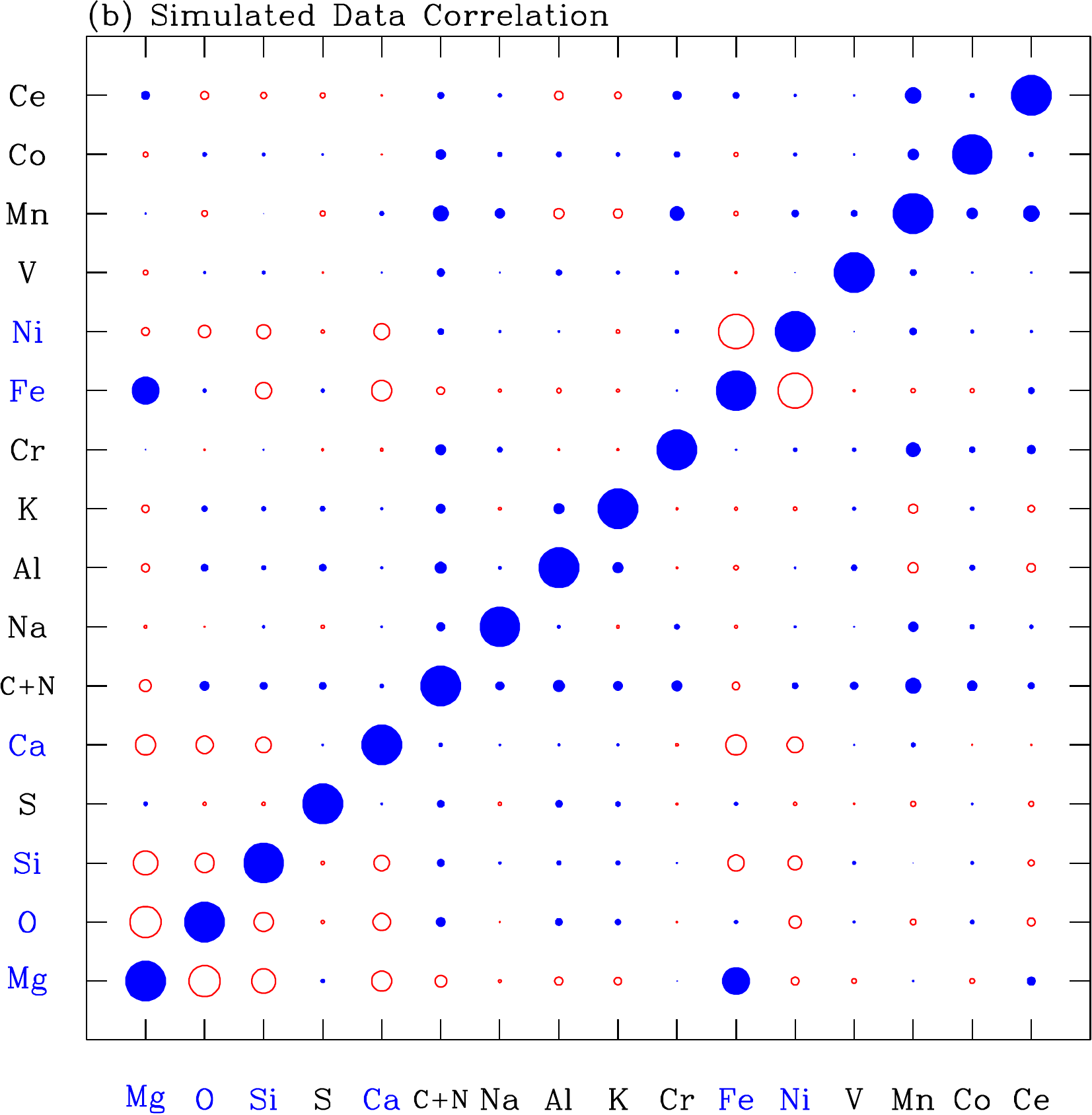}
}
\caption{(a) Correlation matrix (eq.~\ref{eqn:correlmatrix})
corresponding to the intrinsic covariance matrix in Fig.~\ref{fig:covar}d. 
For visual scaling, note that diagonal 
elements have a magnitude of 1.0 by definition, the O-S
coefficient is 0.37, the Ni-Mn coefficient is 0.26, and the
O-V coefficient is 0.10.
(b) Correlation matrix corresponding to the simulated data covariance
matrix (Fig.~\ref{fig:covar}c).
Off-diagonal coefficients in this matrix are caused by measurement
aberration.
}
\label{fig:correl}
\end{figure*}

Figure~\ref{fig:correl}a converts this intrinsic covariance matrix
to the corresponding correlation matrix,
\begin{equation}
c_{ij} = C_{ij}/(C_{ii} C_{jj})^{1/2},
\label{eqn:correlmatrix}
\end{equation}
thus normalizing all diagonal elements to unity.
This transformation depends on our estimate of the intrinsic variance,
and if we used the percentile-based estimates (triangles in 
Figure~\ref{fig:sigma}) then the off-diagonal correlations would all
be larger, though the structure would be similar.  Relative to the
covariance matrix, this conversion highlights the substantial correlations
among elements that have small observational and total scatter.
The intrinsic correlation matrix is similar in its main features
to that found by TW21 (see their figure 10), including the previously
noted correlations among the iron-peak element residuals, positive
correlations among O, Si, and S, and positive correlations among Ca, Na, and
Al (extending here to include K and Cr).  
Since conditioning on Mg and Fe
has much in common with fitting the 2-process model, we would expect
residual correlations to be similar, but details of our analysis are
entirely different and independent, so we consider this agreement an
encouraging indication that the correlations are qualitatively 
robust to these details.  Many of these correlations are quite strong,
e.g. 0.15 or larger, and would be stronger still if we used the percentile-based
intrinsic variance estimate.

Figure~\ref{fig:correl}b shows the correlations for the simulated data set.
This shows that measurement aberration can induce substantial spurious
correlations even with six-element fitting.  The primary effects are
a positive correlation between Mg and Fe and anti-correlations between
these elements and the other fit elements (O, Si, Ca, Ni), and to a lesser
degree among those elements themselves.  In principle our subtraction
of the simulated covariance matrix from the data covariance matrix should
have removed these artificial correlations from our estimate of the
intrinsic correlation matrix.  However, given the uncertainties in this
procedure (primarily our imperfect knowledge of the observational error
distributions), the inferred correlations involving the fit elements
should be treated with some caution.  If we used only Mg and Fe to
infer $\Acc$ and $\AIa$, then the off-diagonal correlations for the
simulated data set would be comparable in typical magnitude but nearly all 
positive.

In sum, the measured covariance of residual abundances is larger than
expected from observational errors alone, demonstrating the existence
of intrinsic physical correlations in the residual abundance patterns.
The most prominent of these are correlations among Ca, Na, Al, K, Cr, and Ce
and correlations among Ni, V, Mn, and Co.

\subsection{Correlations with age and kinematics}
\label{sec:residuals_correlations}

\begin{figure*}
\centerline{
\includegraphics[width=6.5truein]{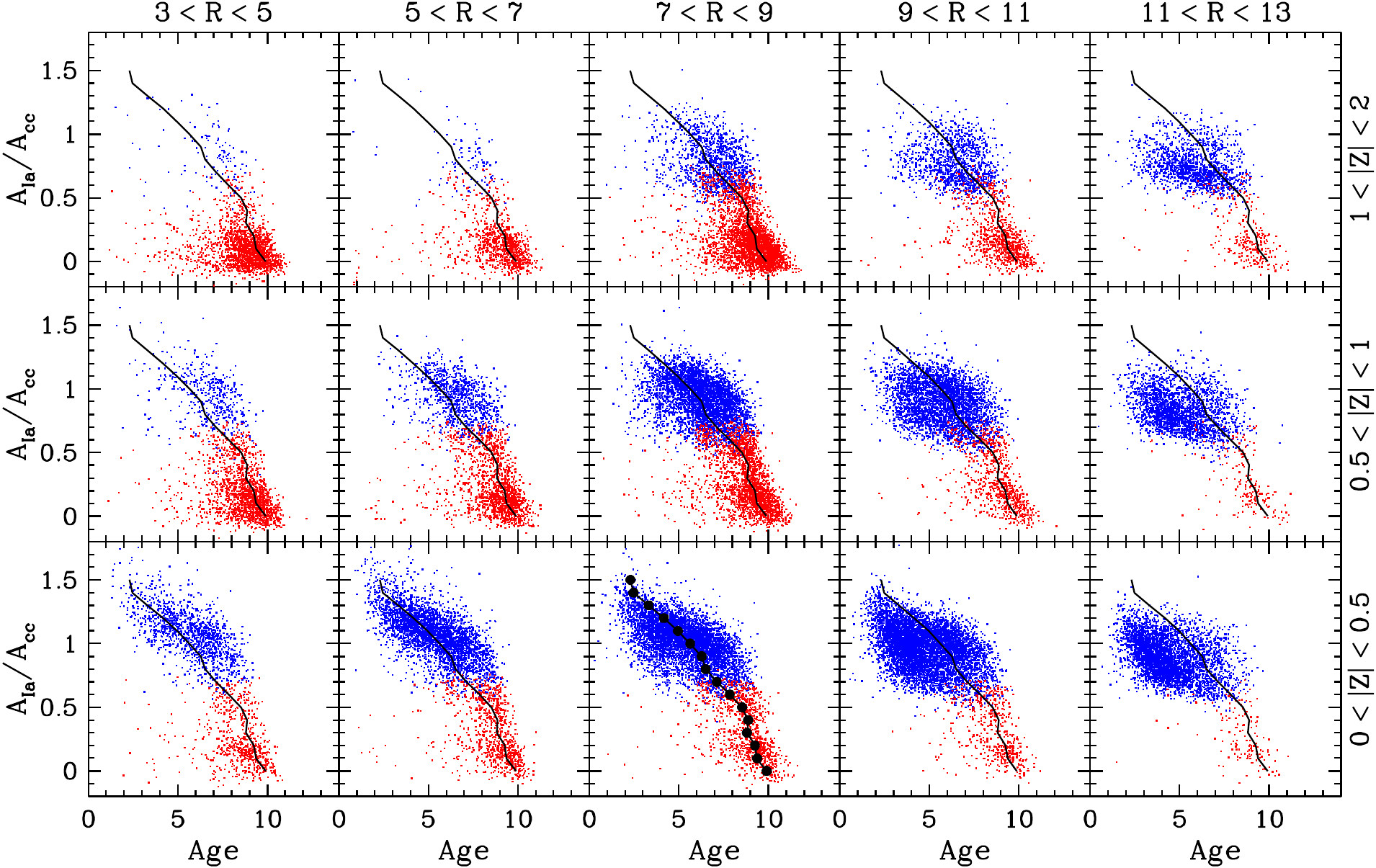}
}
\caption{
Correlation of the 2-process amplitude ratio $\AIa/\Acc$ with stellar
age estimated from the APOGEE spectra by AstroNN, in zones of $R$ (columns)
and $|Z|$ (rows) as labeled.  Red and blue points show stars in the
low-Ia and high-Ia population, respectively.  In the $R=7-9\kpc$, $|Z|<0.5\kpc$
panel, black circles show the median age in bins of width 0.1 in $\AIa/\Acc$.
Black lines repeat this median sequence and are the same in all panels.
To improve coverage of the inner Galaxy, this plot uses the SN100 sample.
}
\label{fig:aratio_age_map}
\end{figure*}

Figure~\ref{fig:aratio_age_map} plots the amplitude ratio $\AIa/\Acc$
inferred from our 2-process fits against the stellar age inferred from 
the APOGEE spectra by AstroNN \citep{Leung2019a,Mackereth2019}, a Bayesian
neural network model trained on a subset of APOGEE stars with asteroseismic
ages.  We use the DR17 AstroNN Value Added Catalog, which will be made
available with the DR17 public release.
Specifically we use the {\tt age\_lowess\_correct} ages, which correct the
raw neural network ages for biases at low and high age using a lowess
smoothing regression (see \citealt{Mackereth2017}).
We adopt the same Galactic zones shown previously in 
Figure~\ref{fig:aratio_acc_map} and again use the SN100 sample to
improve coverage of the inner Galaxy.  In the solar neighborhood
($R=7-9\kpc$, $|Z|<0.5\kpc$) we compute the median age in narrow
bins of $\AIa/\Acc$, and we repeat this median sequence in other 
panels for visual reference.  We use this order of binning 
because $\AIa/\Acc$ 
is measured much more precisely than age, so the median $\AIa/\Acc$ in
bins of age cannot be determined as reliably.  

Not surprisingly, the low-Ia stars are systematically older than the
high-Ia stars.  There is, furthermore, a continuous trend of age with
$\AIa/\Acc$ within both the low-Ia and high-Ia populations, and 
although these two populations are
separated in $\afe$ the age trend is continuous across them.
The trend is similar in different Galactic zones,
though in the high-Ia
population at $|Z|<1\kpc$ the stars are systematically older in the
inner Galaxy and younger in the outer Galaxy at fixed $\AIa/\Acc$,
by roughly 1-2 Gyr.  The correlation of age with $\afe$ 
within the high-Ia population is visible in previous studies 
\citep{Martig2016,Miglio2021}, though it is perhaps more obvious
in this representation.

The primary spectroscopic diagnostic
of age in APOGEE spectra comes from features that trace the C/N ratio
\citep{Masseron2015,Martig2016},
because the surface C and N abundances are changed by dredge-up on
the giant branch in a way that depends on stellar mass, and for red
giants the age (slightly longer than the main sequence lifetime) is
tightly correlated with mass.
Although AstroNN works directly on spectra, it responds primarily to
C and N features and returns large age uncertainties when these features
are masked.  It is therefore unlikely that it is ``learning'' a
correlation between age and other abundance ratios from its asteroseismic
training set and then applying that to other stars.  However, the birth
[C/N] ratio likely depends on stellar metallicity and [$\alpha$/Fe]
\citep{Vincenzo2021b}, and these birth abundance trends could induce
systematic age errors that correlate with abundances.

\begin{figure*}
\centerline{
\includegraphics[width=6.5truein]{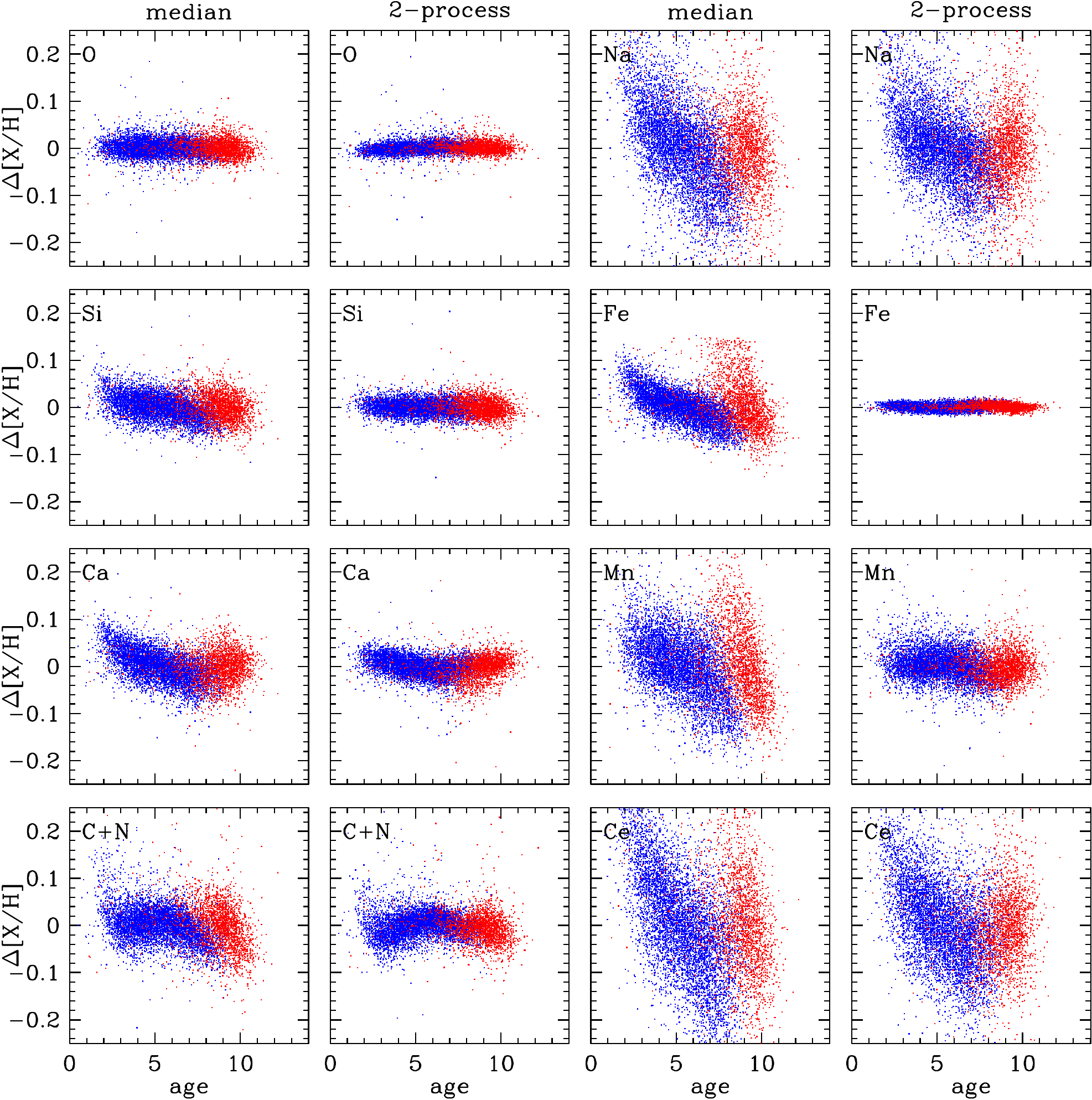}
}
\caption{
Correlation of the abundance residual $\Delta\xh$ (in dex) with stellar age
for eight selected elements as labeled.  In the first and third columns,
residuals are computed with respect to the median sequences, while in the
second and fourth columns residuals are computed with respect to the
2-process model predictions.  Red and blue points show stars in the
low-Ia and high-Ia population, respectively.  To reduce crowding, we
plot only 25\% of the stars.
}
\label{fig:delta_vs_age}
\end{figure*}

Figure~\ref{fig:delta_vs_age} plots residual abundances vs. AstroNN age.
We return to using the high SNR-threshold sample to reduce observational
contributions to the residual scatter, and we have selected a subset of
elements that illustrate a range of behaviors.  In the first and third
columns, $\Delta\xh$ is computed relative to the median sequence of the
low-Ia or high-Ia population.  We see a clear correlation between
$\Delta\feh$ and age in the high-Ia population and a weaker correlation
in the low-Ia population.  This correlation indicates that even though
the scatter about the median sequence at fixed $\mgh$ is small
(about 0.04 dex in $\femg$, see \citealt{Vincenzo2021}), it is correlated
with age in the expected sense, with younger stars showing greater Fe
enrichment.  Mn, which we infer to have the largest SNIa contribution of
all APOGEE elements (Figure~\ref{fig:dataq_peakodd}), shows a similarly
strong correlation, and Na and Ce show similar correlations in the high-Ia
population despite the larger scatter from observational errors.
O residuals show no correlation with age, but Si and Ca residuals do,
consistent with our inference of a significant SNIa contribution to
these two $\alpha$-elements (Figure~\ref{fig:dataq_alpha}).  Residual
correlations for C+N are weak, though there is a population
of older high-Ia stars that have below-median C+N.

In the second and fourth columns, $\Delta\xh$ is computed relative to
the predictions of the 2-process model.  The residual scatter is smaller
for the well measured elements, as seen previously in 
Figures~\ref{fig:delta_dist} and~\ref{fig:sigma}.
The age trends seen previously for Fe, Mn, and Si are removed, and
the trend for Ca is reduced though not entirely eliminated.
We regard this flattening of age trends as evidence for the physical
validity of the 2-process model, which is constructed with no knowledge
of the stellar ages.  However, residuals from the 2-process model could
correlate with age if they are caused by other enrichment processes 
that have a different time dependence than SNIa.  We see a slight correlation
of C+N residual with age in the high-Ia population,
though there is some risk of a systematic effect because
of the central role of these elements in the age determinations.

More strikingly, we see a clear trend of Ce residuals and, to a lesser
extent, Na residuals with age in the high-Ia population.
Stars with $\Delta\xxmg{Ce} > 0.1$ have typical AstroNN ages
of 2-3 Gyr, while stars with $\Delta\xxmg{Ce} \approx 0$ have
typical ages of 4-6 Gyr.  Within the high-Ia population, the trend
continues to negative $\Delta\xxmg{Ce}$ and older ages, though
there is no clear trend within the low-Ia population.
These results are qualitatively consistent with the findings
of Sales-Silva et al.\ (2021) that APOGEE's [Ce/Fe] and [Ce/$\alpha$] ratios
for open clusters increase with decreasing cluster age over at
least the past $\sim 4\Gyr$.  Previous studies
(e.g., \citealt{daSilva2012,Nissen2015,Feltzing2017}) have also
shown that abundances of s-process elements are well correlated
with age in thin-disk stars.
We have already noted (\S\ref{sec:median}) that the separation of
median sequences for [Na/Mg] is larger than expected based on
the yield models used by \cite{Andrews2017} and \cite{Rybizki2017}.
The qualitative similarity of age trends for Na and Ce residuals suggests 
that a common source, presumably AGB stars, makes important contributions
to both elements.

In future work we will directly examine trends with asteroseismic ages,
instead of the spectroscopic estimates trained on them.
However, the current sample is not ideally suited for this purpose because
our $\logg$ and $\Teff$ cuts eliminate many of the APOGEE stars for
which asteroseismic parameters from {\it Kepler} are available
\citep{Pinsonneault2018}.

\begin{figure*}
\centerline{
\includegraphics[width=6.5truein]{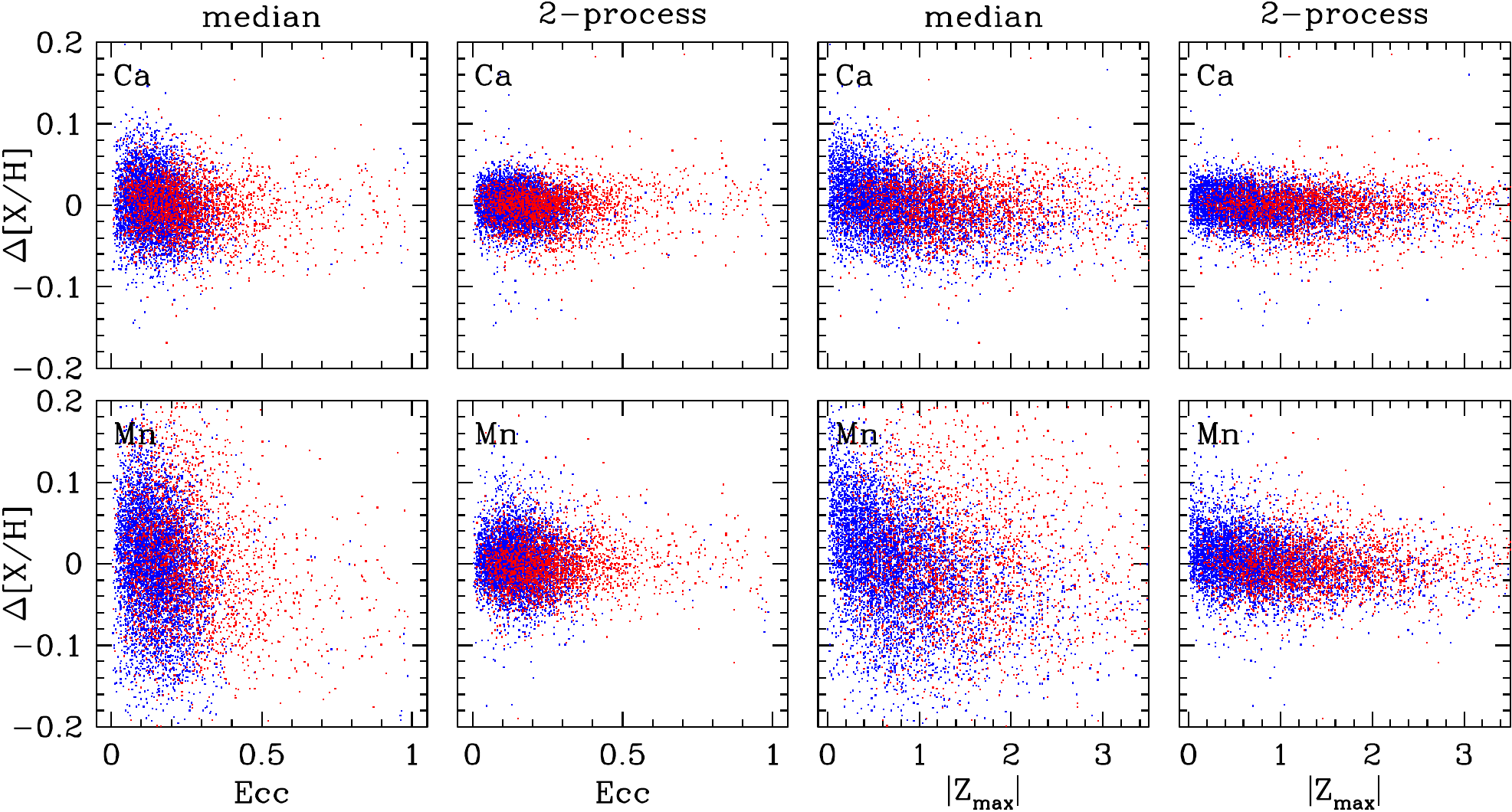}
}
\caption{
Correlation of the $\Delta\xxh{Ca}$ and $\Delta\xxh{Mn}$ residual abundances
with orbital eccentricity (left) and maximum midplane distance $|Z_{\rm max}|$
(right).  As in Fig.~\ref{fig:delta_vs_age}, residuals in the 
the first and third columns are computed with respect to the median sequences, 
while those in the second and fourth columns are computed with respect to the
2-process model predictions.  Red and blue points show stars in the
low-Ia and high-Ia population, respectively.  To reduce crowding, only
25\% of the stars are plotted.
}
\label{fig:delta_vs_kinematics}
\end{figure*}

Figure~\ref{fig:delta_vs_kinematics} plots residual abundances for Ca and Mn
against orbital parameters derived by AstroNN from APOGEE and {\it Gaia}
data, computed using the fast method of \cite{Mackereth2018} implemented
in \texttt{galpy} \citep{Bovy2015}, assuming the \texttt{MWPotential2014}
gravitational potential from \cite{Bovy2015}.
The two left columns show residuals vs.\ eccentricity.  Not surprisingly,
low-Ia (``thick disk'') 
stars are more likely to have high orbital eccentricity than high-Ia
stars.  Within each population, there is a slight tendency for the highest
eccentricity stars to have lower [Ca/H] and [Mn/H] relative to the median
sequence, but this trend is weak, and it vanishes when examining 
residuals from the 2-process fits instead of from the median sequence.
The right two columns show residuals vs.\ $|Z_{\rm max}|$, the maximum
distance a star's orbit reaches from the midplane.  Low-Ia ``thick disk''
stars are more likely to have high $|Z_{\rm max}|$, though a number
of stars in the high-Ia population have inferred $|Z_{\rm max}| > 2\kpc$.
Trends (or lack thereof) are similar to those seen for eccentricity but
somewhat more pronounced.  In particular, high-Ia stars with 
$|Z_{\rm max}| < 0.5\kpc$ tend to have higher [Ca/H] and [Mn/H] relative 
to the median sequences, an effect that is subtle (0.02-0.05 dex) but
discernible with a large sample.  The coldest ``thin disk'' stars tend
to have higher fractions of elements with SNIa contributions, as expected
from the trend of $\AIa/\Acc$ with age (Fig.~\ref{fig:aratio_age_map})
and the age-velocity relation.

As with eccentricity, shifting from median residuals to 2-process residuals
removes these weak correlations.  We have examined other elements and 
find no convincing correlations of individual 2-process residuals with 
kinematics or with Galactic position.  
As discussed in \S\ref{sec:beyond} below, grouping correlated elements
sharpens sensitivity and reveals weak correlations that are difficult
to discern in individual element plots like Figures~\ref{fig:delta_vs_age}
and~\ref{fig:delta_vs_kinematics}.  In \S\ref{sec:populations} we give
examples of stellar populations whose median residual abundances clearly
depart from those of the main disk sample.


In sum, deviations from median sequences show weak but expected correlations
with age and kinematics, with the stars that have higher SNIa/CCSN ratios 
within each population also having younger ages and colder kinematics.
Changing to 2-process residuals removes most of these correlations, but
within the high-Ia population the Ce and Na residuals show significant age
correlations, with younger stars exhibiting higher abundances of both elements
relative to stars with similar $\Acc$ and $\AIa$.

\section{High-$\chi^2$ Stars}
\label{sec:outliers}

The 2-process model fits the APOGEE abundances of most disk stars to an
accuracy that is comparable to the reported observational uncertainties.
However, 
the estimated {\it intrinsic} scatter about the 2-process predictions exceeds
0.01 dex for most elements (Figure~\ref{fig:sigma}), and the off-diagonal
covariance of abundance residuals demonstrates the physical reality of
intrinsic deviations even among stars that appear individually well 
described by the model (Figures~\ref{fig:tefftrend} and~\ref{fig:covar}).
In this section we examine a selection of stars whose measured abundances
are poorly described by the 2-process model, i.e., with high values of 
$\chi^2$.  We refer to these stars as outliers, but we note that with
sufficiently precise measurements it is likely that most stars would
show statistically robust deviations from the 2-process fit.
Some of these outlier stars may simply be extreme examples of the
same correlated deviations present in the main stellar population, 
offering clues to the physical drivers of these deviations.  In other
cases, unusual abundances may arise from rare physical processes that
do not affect most stars.  Yet other high-$\chi^2$ cases arise from
measurement errors that are much larger than the reported observational
uncertainty, for reasons that may be simple (e.g., a poorly deblended line)
or subtle (e.g., inaccurate interpolations in a grid of synthetic spectra
at an unusual location in abundance space).

\begin{figure*}
\centerline{\includegraphics[width=5.5truein]{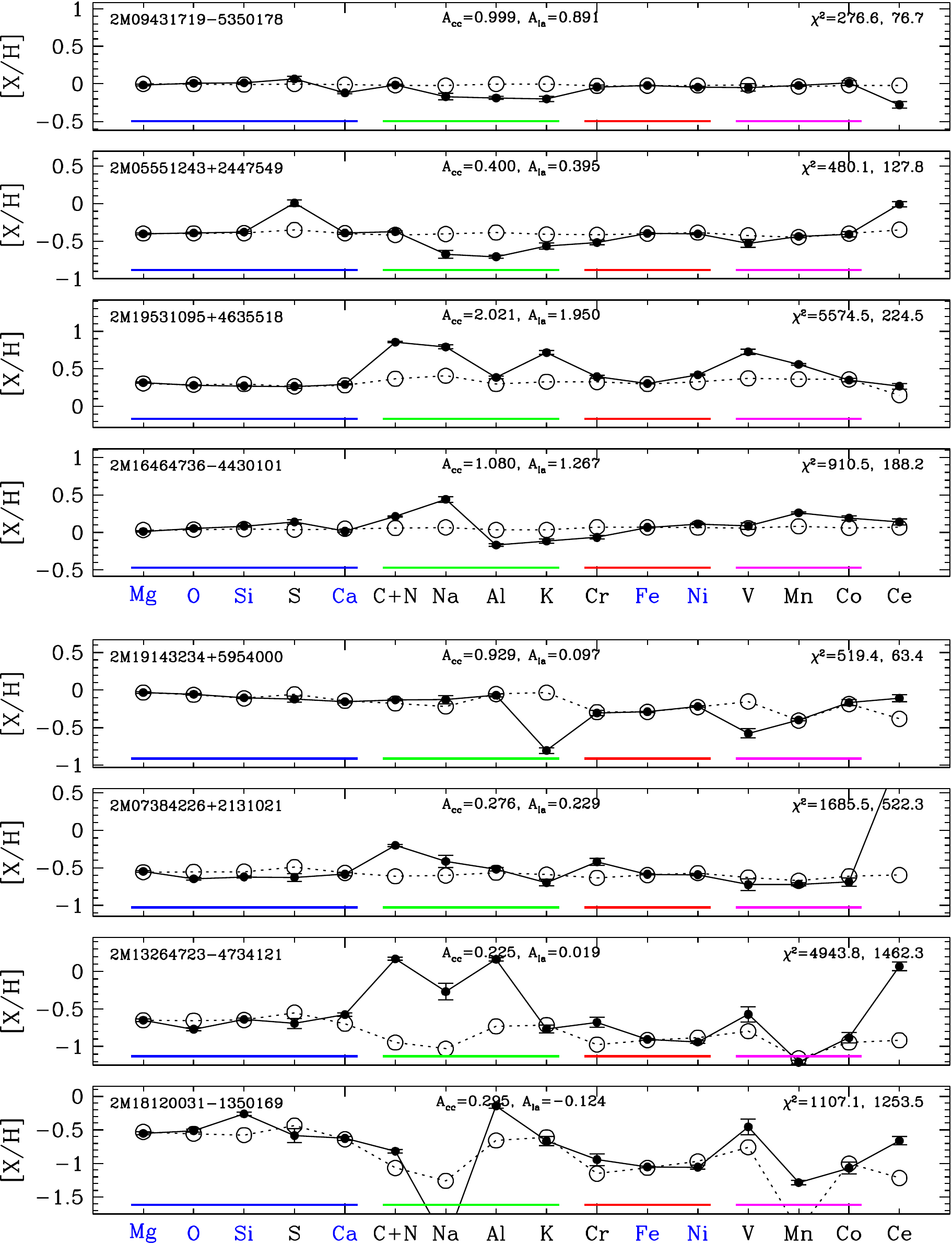}}
\caption{Examples of stars that are poorly fit by the 2-process model,
in a format similar to Fig.~\ref{fig:elem_stars1}.
In each panel,
filled circles with error bars show the APOGEE abundance measurements
and open circles show the abundances predicted by the best 2-process 
model fit.  Each panel lists the star's best-fit $\Acc$ and $\AIa$,
the $\chi^2$ of this fit, and the (usually lower)
modified $\chi^2$ described in the text.  
}
\label{fig:elem_highchi2}
\end{figure*}

High $\chi^2$ values can arise from single deviant measurements, which may
have a variety of mundane observational causes.  
To preferentially select genuine
physical outliers, we have used a modified $\chi^2$ criterion in which
(a) we use the total scatter (filled circles in Figure~\ref{fig:sigma})
rather than the observational uncertainty, and (b) for each star, we omit
the element that makes the single largest contribution to $\chi^2$.
This criterion thus requires at least two anomalous abundances, and it
downweights the impact of elements that more frequently have observational
errors much larger than the reported uncertainties.
Figure~\ref{fig:elem_highchi2} shows a selection of eight stars drawn from the
top 2\% of this modified $\chi^2$ distribution.
We list both the original $\chi^2$ and the
modified $\chi^2$ for each star.  For reference, the 98\%, 99\%, and 99.5\%
highest values of the modified $\chi^2$ are 59.9, 97.0, and 154.7, respectively.
We selected these eight stars after examining $\sim 40$ examples in 
the top 2\%, illustrating a few of the common themes that we find within
this high-$\chi^2$ population.

{\it 2M09431719-5350178} has nearly solar values of $\mgh$ and $\femg$, but 
it has low values (relative to the 2-process predictions) of Ca, Na, Al, K,
and Ce, by $\sim 0.25$ dex for Ce and 0.1-0.2 dex for the other elements.
This star individually exemplifies the pattern shown by the block of 
observed correlations in Figure~\ref{fig:covar}d, and we find similar 
behavior in some other high-$\chi^2$ stars.  These examples and the
residual correlations themselves
hint at a common physical source that contributes to these 
elements and is deficient in some stars.  
However, we do not have a clear physical interpretation of this pattern.
We have not noticed comparably
clean examples in which all of these elements are high, though 
Figure~\ref{fig:elem_d3d4} (discussed in \S\ref{sec:beyond}) shows two examples
with high average deviations among these elements.

{\it 2M05551243+2447549} is a lower metallicity, high-Ia star
($\mgh=-0.40$, $\femg=0.01$) that shows a similar deficiency of Na, Al, and K
but a Ce abundance that is {\it enhanced} by 0.34 dex, demonstrating that
large deviations among these elements do not necessarily move in lockstep.
This star also shows a large (0.36-dex) enhancement of S relative to the 
predicted, near-solar [S/Mg].
This star has broader lines than the synthetic spectral fit, suggesting
high rotation, and it has a radial velocity spread of $\sim 40\kms$ over
the 100 days that it was observed, implying a binary companion.
While these properties could be connected to unusual abundances, it is
also possible that high rotation is causing systematic errors in the ASPCAP
abundance measurements, and in the spectroscopic $\logg$, which is low
(by about 0.4 dex) relative to most stars of similar metallicity and $\Teff$.

{\it 2M19531095+4635518} is a metal-rich, high-Ia star 
($\mgh=0.32$, $\femg=-0.01$) that is an extreme outlier in both its
standard $\chi^2=5575$ and its modified $\chi^2=225$.  
This star has an extremely high C+N residual, with ASPCAP values of 
[C/Fe]=0.32 and [N/Fe]=1.04.  The [O/Fe] from ASPCAP is 0.045,
implying a C/O number ratio of 1.01 that puts this star just over
the boundary into the carbon star regime, where the stellar spectral
features become very different from those of typical, ${\rm C/O} < 1$ stars.
The large deviations in Na, K, V, and Mn may be physical, but they
could also be affected by the very strong blends that occur in the
carbon star regime, such that any inaccuracies in the ASPCAP synthesis
could lead to poor fits or incorrect abundance values.

{\it 2M16464736-4430101} displays a pattern that we have found in multiple
examples of high-$\chi^2$ stars of solar or super-solar metallicity.
Consistently in these stars, the Na abundance is far above the 2-process 
prediction (by 0.37 dex in this case), the C+N abundance is moderately 
elevated ($\sim 0.1-0.2$ dex), the Al, K, and Cr abundances are moderately
depressed, and the V, Mn, and Co abundances are slightly enhanced.
Although this could be a distinctive class of chemically peculiar
stars, we suspect that this pattern arises from an artifact of ASPCAP
abundance determinations, in part because we find anomalous structures
in [X/Fe]-[Fe/H] diagrams for Na, Al, and Mn that do not appear physical.
We do not understand the origin of this artifact, though the elevated
C+N hints that it could be related to inaccurate interpolation across the
model grid near the carbon-star regime, where spectral syntheses change
rapidly with stellar parameters.

{\it 2M19143234+5954000} has near-solar $\mgh=-0.03$ and an $\femg=-0.26$
that places it below the median of the low-Ia population at this metallicity
(Figure~\ref{fig:dataq_peakeven}).  Its high $\chi^2$ value is driven
by extremely low K and low V, and to a lesser extent by elevated Ce.
Low K values and to some degree low V values are fairly common among
high-$\chi^2$ stars, and the residual distributions for both elements
(especially K) are asymmetric towards negative values 
(Figure~\ref{fig:delta_dist}).  K and V both have weak, sometimes blended
features in APOGEE spectra, so the abundances are more subject to 
statistical and systematic errors.  Furthermore, follow-up of the 
low K population shows that many of them (including this star) have
a heliocentric velocity of $-70\kms$ that happens to place two of 
APOGEE's K lines on top of stronger telluric features.  It therefore
seems likely that the low K abundances are a consequence of imperfect
telluric subtraction.

{\it 2M07384226+2131021} is a low metallicity,\footnote{We refer to stars with
$\mgh<-0.5$ as low metallicity because they lie at the metal poor end of
our sample and of disk populations in general, though of course halo 
populations reach to much lower metallicity.} high-Ia star ($\mgh=-0.55$,
$\femg=-0.04$) with highly elevated C+N (0.41 dex) and extremely elevated 
Ce (1.59 dex), as well as moderate enhancements ($\sim 0.2$ dex) in Na
and Cr.  This star is a member of the old open cluster NGC 2420, and
it was identified by \cite{Smith1987} as an extreme example of a ``barium
star'' based on its strong excess abundances of s-process elements.
The extreme Ce enhancement is in line with these previous findings.
ASPCAP's individual C and N abundances are [C/Fe]=0.27 and [N/Fe]=0.73.  
We find numerous examples of stars with large enhancements of both C+N
and Ce, as discussed further below.  These enhancements may arise from
binary mass transfer from an AGB companion, or from internal AGB enrichment
in star clusters, or both.  These s-process enhanced stars are
frequently referred to as barium stars at high metallicity and CH 
or CEMP-s stars at low metallicity (e.g., \citealt{McClure1985,Lucatello2005}).

{\it 2M13264723-4734121} is a low metallicity, low-Ia star
($\mgh=-0.65$, $\femg=-0.26$) with 0.7-1.1 dex enhancements in
C+N, Na, Al, and Ce.  The ASPCAP values of [C/Fe] and [N/Fe] are
-0.17 and 1.78, respectively, so the elevated C+N is driven entirely
by the extreme N enhancement.  This star is a member of $\omega\,$Cen, a globular
cluster that is often hypothesized to be the stripped core of a
dwarf galaxy because of its large internal $\feh$ spread
(e.g., \citealt{Smith2000}).  The well established and distinctive
pattern of enhanced N, Na, Al, and s-process elements is thought to
be a signature of self-enrichment by the cluster's evolved AGB stars
\citep{Smith2000,Johnson2010,Meszaros2020,Meszaros2021}.  
A significant number of the most extreme $\chi^2$ stars in our sample
are members of $\omega\,$Cen, and we discuss the abundance pattern of
these stars further in \S\ref{sec:populations}.

{\it 2M18120031-1350169} is a low metallicity star ($\mgh=-0.55$) with
unusual abundances for many elements.  Its $\femg=-0.50$ lies well below
the low-Ia plateau at $-0.3$, so the 2-process fit assigns it a negative
value of $\AIa$.  However, even if we set $\AIa=0$ its abundances would
depart strongly from the 2-process prediction, especially the high Al,
high Ce, low Na, and unusual (0.31-dex) enhancement of Si.
The C+N of of this star is only moderately (0.25 dex) above the 2-process
prediction, but this enhancement is dominated by an unusual N abundance,
with ASPCAP values of [C/Fe]=0.01 and [N/Fe]=0.72.
\cite{Schiavon2017} and \cite{Fernandez-Trincado2020} have previously
highlighted 2M18120031-1350169 as a N-rich star that is a likely escapee
from a globular cluster, part of an extensive population of such stars
identified in APOGEE 
\citep{Schiavon2017,Fernandez-Trincado2016,Fernandez-Trincado2017,
Fernandez-Trincado2019,Fernandez-Trincado2020b,Fernandez-Trincado2020}.
The pattern of high N, Al, and Ce resembles that found for $\omega\,$Cen
and could reflect a similar self-enrichment process.  However, this star
does not show Na enhancement, and the Si enhancement seen here does not
appear in our $\omega\,$Cen stars (see Figure~\ref{fig:populations} below),
though enhanced Si is found in a population of field stars in the 
inner halo (dubbed ``Jurassic''; 
\citealt{Fernandez-Trincado2019b,Fernandez-Trincado2020c}), which may
arise from tidally disrupted globular clusters.
Intriguingly, \cite{Masseron2020a} identified 2M18120031-1350169 as
one of 15 APOGEE stars with extreme P enhancement, finding [P/Fe]=1.65
using a custom analysis of the APOGEE spectrum (rather than the ASPCAP
abundance).  The unusually high [X/Fe] values for Mg, O, Si, and Al 
are also found in the other members of this P-rich population
(see Figure~9 of \citealt{Masseron2020a}), and the high [Ce/Fe] accords
with the enhanced s-process abundances found in follow-up optical
spectroscopy by \cite{Masseron2020b}.
From the overall abundance patterns, \cite{Masseron2020a,Masseron2020b}
argue that the chemical peculiarities of these stars do not originate
in globular clusters or binary mass transfer, and they are a challenge
to explain with existing stellar nucleosynthesis models.
The source of 2M18120031-1350169's unusual enrichment is unclear,
but it is encouraging that a simple $\chi^2$ analysis readily turns
up some of the most interesting stars found in independent studies.

Instead of selecting stars based on $\chi^2$ values, one can look for
specific abundance anomaly patterns.  For example, motivated by examples
like 2M07384226+2131021, we have searched for stars that have unusual
enhancements of both C+N and Ce.  With high thresholds of 0.2 dex in C+N
and 0.8 dex in Ce, we find 24 such stars in our disk sample, six of
which are members of $\omega\,$Cen.  Further investigation of several of
these cases shows evidence of velocity variations among the multiple
APOGEE visits, supporting the idea that some of these anomalous patterns
arise in binary systems.  If we lower the thresholds to 0.15 dex
and 0.5 dex, the number of high-(C+N)/high-Ce stars rises to 87, 
and to 127 if we select from the larger SN100 sample.  We have not yet
carried out a systematic census to assess the frequency of likely binaries
or of cluster members other than $\omega\,$Cen.

Stars with anomalous abundances of single elements may also be interesting,
but these require careful individual vetting.  For example, we found two
stars with unusually high Ca abundances that arose because a particular
combination of radial velocity and APOGEE fiber placed one of the Ca lines
on previously unrecognized bad pixels in one of the APOGEE spectrograph
detectors.  Some other cases of anomalous abundances appear to arise from
high rotation broadening weak features in a way that affects multiple
elements.  Others arise in stars that appear to be double-lined spectroscopic
binaries.  Rare outliers can diagnose unusual problems in data reduction
and abundance measurements as well as physically unusual systems.
As the above examples show, it is not always easy to tell one from the other.
None of the eight stars in Figure~\ref{fig:elem_highchi2} has obvious
problems in its APOGEE spectrum, but they could nonetheless be affected
by abundance measurement systematics.

\begin{figure*}
\centerline{\includegraphics[width=5.5truein]{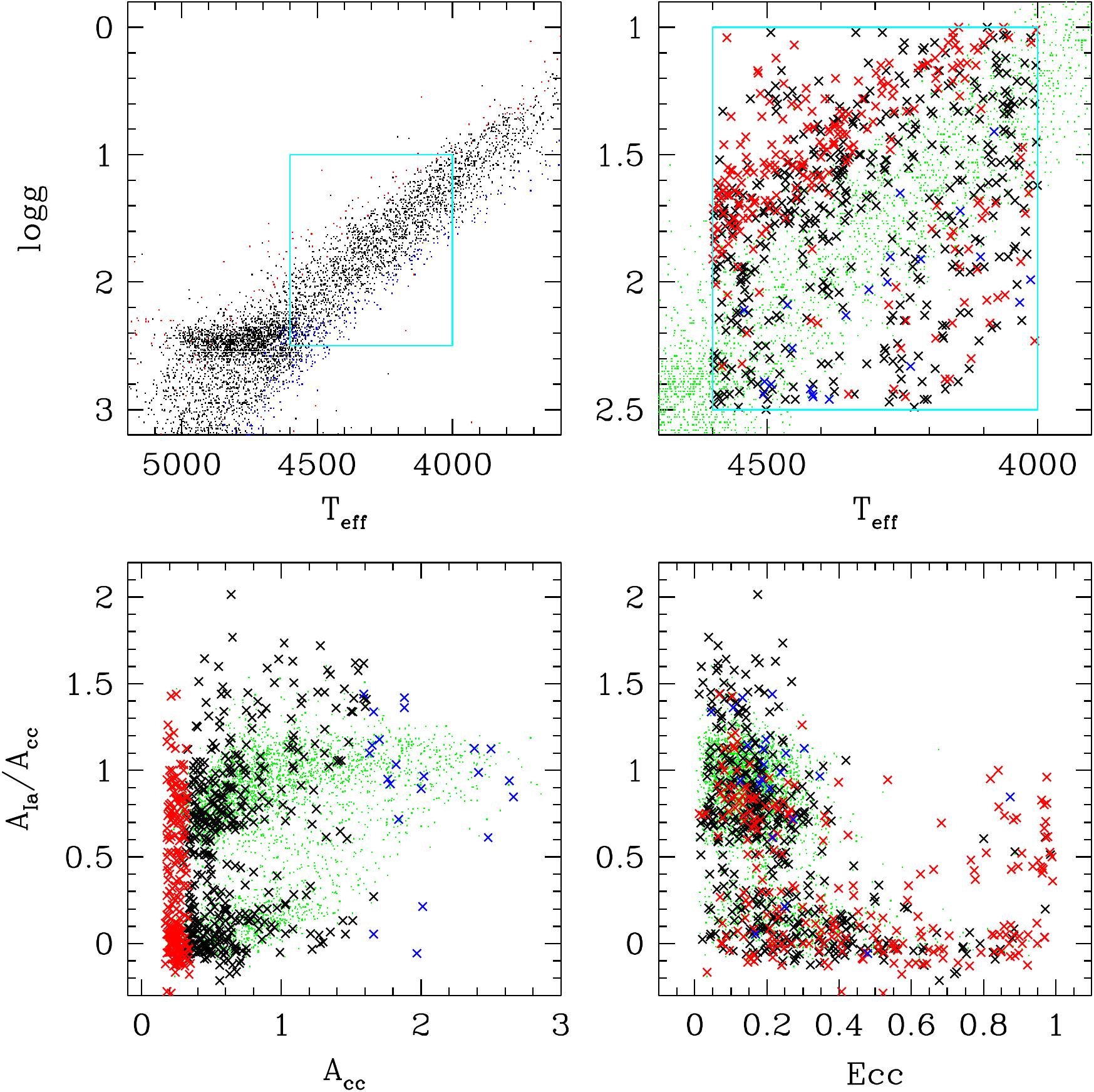}}
\caption{Properties of the 689 stars with modified $\chi^2$ values in 
the top 2\% of the cumulative distribution.  In all panels, red points
represent stars with $\mgh < -0.5$, blue points represent stars with
$\mgh > 0.2$, and black points represent stars with intermediate $\mgh$.
For context, the upper left panel shows the logg-$\Teff$ distribution for 
a sample of 5000 stars that satisfy the selection criteria for our main disk 
sample but span a wider range of $\logg$ and $\Teff$.  The cyan box
indicates our sample selection.  The upper right panel shows 
$\logg$ vs.\ $\Teff$ for the high-$\chi^2$ stars, with the background
sample represented by green dots.  The lower panels plot the $\AIa/\Acc$
ratio against $\Acc$ (left) and orbital eccentricity (right), with
green dots showing a random 10\% of the full disk sample.
}
\label{fig:chisqmap}
\end{figure*}

Figure~\ref{fig:chisqmap} presents a more global view of the high-$\chi^2$
stars, selected as the sample members with modified $\chi^2$ values in
the highest 2\%, showing their distributions in $\logg$ vs.\ $\Teff$,
$\AIa/\Acc$ vs.\ $\Acc$, and $\AIa/\Acc$ vs. orbital eccentricity 
(taken from the DR17 AstroNN catalog).  The high-$\chi^2$ stars span 
the sample's entire range in these parameters, but they do not follow
the same distribution as the background stars (green dots, a random 
subset of our full sample).  The high-$\chi^2$ stars are preferentially
low metallicity, which is physically plausible because it is easier
to perturb abundances (e.g., with mass transfer) if they are low to
begin with, but which could also be a sign of measurement errors when
features are weak.  In the $\logg$-$\Teff$ diagram, most of the high-$\chi^2$
stars have low $\logg$ for their $\Teff$, which is an expected consequence
of their preferentially low metallicity.  However, the low metallicity
stars that have high $\logg$ are likely to be cases where anomalous
abundance patterns are affecting the spectroscopic $\logg$ estimates or
where unusual properties of the spectrum (e.g., broad lines from high
rotation) are producing erroneous values of $\logg$ and perhaps 
of the abundances as well.
We find high-$\chi^2$ stars throughout the thin and thick disk populations,
and they are clearly overrepresented among the high eccentricity, 
high-Ia population that likely corresponds to accreted halo stars.
We return to this point in subsequent sections.

The literature on chemically peculiar stars is voluminous and rich.
The examples in Figure~\ref{fig:elem_highchi2} illustrate the possibilities
for pursuing such studies with 2-process residual abundances.
For our high-SNR disk star sample, the top 2\% of the $\chi^2$ distribution
already corresponds to nearly 700 stars, so exploiting this approach will
be a substantial research effort in its own right.
While there are many ways to find chemically anomalous stars, 2-process
residuals have the virtue of automatically relating a star's abundances to
values that are typical for its metallicity and [$\alpha$/Fe].  
This normalization makes it easier to identify stars that have moderate
deviations across multiple elements but no single extreme values, such
as the first example in Figure~\ref{fig:elem_highchi2}.  
Using machine learning techniques to pick out stars
whose abundance patterns have low conditional probability given their
values of $\mgh$ and $\mgfe$ is another potentially powerful approach
to this problem, well suited to take advantage of large homogeneous data
sets like APOGEE (TW21).

\section{Residual abundances of selected stellar populations}
\label{sec:populations}

One goal of 2-process modeling is to assist the identification of 
chemically distinctive stellar populations, generically referred to
as chemical tagging.  Describing a star's $N$ elemental abundance
measurements with two parameters and $N-2$ residuals does not create new
information, but it may improve the effectiveness of tagging algorithms by
extracting two dimensions that vary widely in the disk, bulge, and halo
populations and that shift many abundances in a strongly correlated, non-linear
way.  The 2-process+residual decomposition also prevents one from 
multi-counting abundance deviations that all reflect the same underlying
changes in the bulk levels of a star's CCSN and SNIa enrichment.
We plan to pursue chemical tagging with residual abundances in future work.
Here we illustrate prospects with the simpler but related exercise of 
examining residual abundances of selected stellar populations.

\begin{figure*}
\centerline{\includegraphics[width=5.5truein]{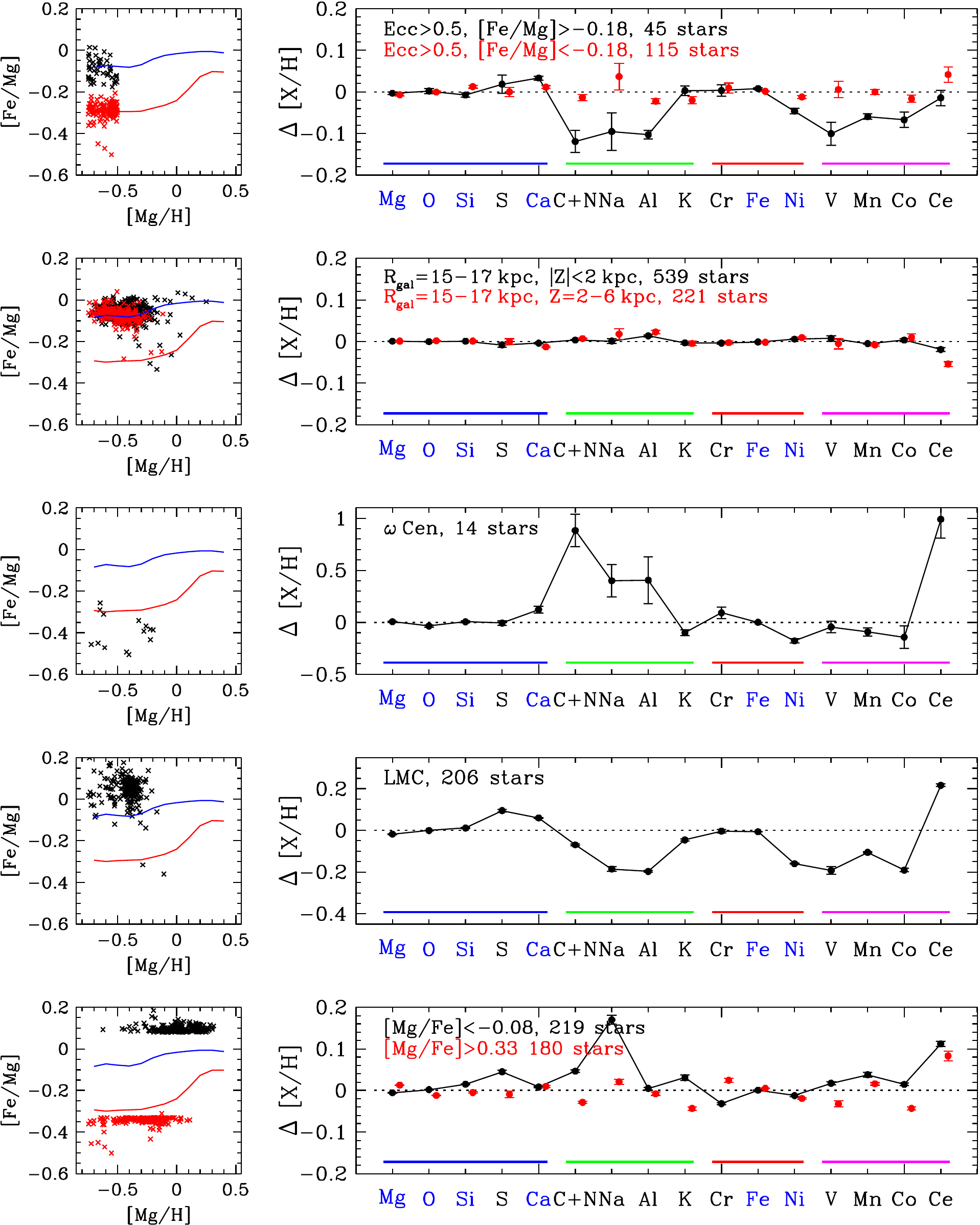}}
\caption{Median deviations from the best-fit 2-process model for stars in 
selected populations.  Left panels show the location of the population
in [Fe/Mg] vs.\ [Mg/H], with median low-Ia (red) and high-Ia (blue)
sequences for the full sample shown for reference.  Points with error
bars in the right panels show the median deviation and the $1\sigma$
uncertainty in the median computed from 1000 bootstrap resamplings
of the population.  ({\it Top}) Disk stars with $\mgh<-0.5$ and
eccentricity greater than 0.5, with $\femg>-0.18$ (black) or 
$\femg<-0.18$ (red).  ({\it Second row}) Disk stars in
the outer Galaxy, with $15 \leq R \leq 17\kpc$ and $|Z|\leq 2\kpc$ (black)
or $15 \leq R \leq 17\kpc$ and $Z=2-6\kpc$ (red).
({\it Third row}) Stars that are probable members of the $\omega\,$Cen
cluster based on angular position and radial velocity.  
({\it Fourth row}) Stars that are probable members of the 
LMC.  
({\it Fifth row}) Stars that have unusually low (black) or high (red)
values of [Mg/Fe].
For all populations, we adopt an
SNR cut of 100 and impose our usual $\logg$ and $\Teff$ cuts, except
that we retain the standard SNR cut for the fifth row.
Note that different panels have different vertical ranges.
}
\label{fig:populations}
\end{figure*}

Figure~\ref{fig:chisqmap}, and Figure~\ref{fig:d3d4_map} below, suggest
that the high eccentricity population may have distinctive abundances,
particularly those high eccentricity stars that lie significantly above
the $\femg$ plateau.  This high-Ia (low-$\alpha$), high eccentricity
population was identified by \cite{Nissen2010} as the likely remnant
of a disrupted dwarf galaxy.  Evidence for a dynamically distinct population
became much stronger with {\it Gaia} data, and the accreted population
is now identified as the remnant of the relatively massive 
``Gaia Sausage/Enceladus'' (GSE) dwarf galaxy that merged early
in the Milky Way's history \citep{Belokurov2018,Helmi2018}.  In the top right
panel of Figure~\ref{fig:populations}, black points show the median
residual abundances of the 45 stars in our SN100 disk sample that have
orbital eccentricity $e > 0.5$, $\mgh < -0.5$, and $\femg > -0.18$.
We estimate uncertainties in these medians as the dispersion of medians
of 1000 bootstrap resamplings, i.e., for each resampling we choose 45
stars from the sample with replacement and compute the median, then take the
standard deviation of these medians as the representative error bar.
The median abundances of C+N, Na, Al, Ni, V, Mn, and Co are all depressed
by 0.05-0.12 dex relative to other disk stars with matched values of
$\AIa/\Acc$.  The median Ca abundance is elevated by a small but
statistically significant 0.03 dex.  By contrast, the
median residual abundances for the high-eccentricity low-Ia stars
($\femg<-0.18$) are statistically compatible with zero (red points).
GSE stars are chemically distinct from other high-eccentricity
stars in their $\afe$ ratios, and they are distinct from disk
stars with similar metallicity {\it and} $\afe$ in their abundances of
multiple odd-$Z$ elements.
The distinctive abundances of these stars may contribute to the drop
of median [X/Mg] ratios in the lowest $\mgh$ bin of the high-Ia population,
seen in Figures~\ref{fig:dataq_oddz}-\ref{fig:dataq_peakodd} for C+N,
Na, Al, Ni, Mn, and Co.

The second row shows stars with $R = 15-17\kpc$ and $|Z|<2\kpc$
satisfying our usual $\logg$, $\Teff$, and $\mgh$ cuts (\S\ref{sec:data})
and $\SNR\geq 100$.  The 2-process model is ``trained'' using stars with
$R=3-13\kpc$, so none of the stars at $R=15-17\kpc$ contributed
to calibrating the process vectors $\qxcc(z)$ and $\qxIa(z)$.
Nonetheless, the median residual abundances of all elements in this 
population are within 0.02 dex (and mostly within 0.01 dex) of zero.
Despite their presence in the outer reaches
of the stellar disk, these stars have APOGEE abundances very close to 
those of low metallicity, high-Ia stars in the rest of the disk.
This similarity could indicate that these stars were born at smaller $R$
and migrated outward, or it could simply indicate that their enrichment
history was similar despite their distinctive location.
The outer disk is warped, with substantially more stars in the
anti-center direction at these radii 
residing at large positive $Z$ than at large negative $Z$.  
Red points show stars in the same radial range
with $Z=2-6\kpc$.  This population also has median abundances within
0.02 dex of the main disk population, except for Ce which is depressed
by 0.05 dex.  At larger $R$ ($\approx 18-30\kpc$), \cite{Hayes2018}
have found that stars in the ``Triangulum-Andromeda'' overdensity
\citep{Majewski2004,Rocha-Pinto2004,Sheffield2014} also have APOGEE
abundance ratios similar to those of normal Milky Way disk stars.

One of the most dramatic abundance outliers in Figure~\ref{fig:elem_highchi2}
is a member of $\omega\,$Cen,
and we first noticed $\omega\,$Cen as a distinctive population in our analysis
because many of the extreme high-$\chi^2$ stars at low metallicity had
similar sky coordinates.  In the third row of Figure~\ref{fig:populations}
we have selected all stars in the SN100 disk sample that have angular
coordinates within $1^\circ$ of the cluster center at
${\rm RA}=201.7^\circ$, $\delta = -47.48^\circ$ and heliocentric velocity
$v > 200\kms$.  The 14 stars selected have a mean $v=234\kms$ with
a dispersion of $10\kms$, while other sample stars that satisfy the
angular selection have heliocentric velocities of $-92\kms$ to $+108\kms$.
The $\omega\,$Cen stars have $\mgh$ values ranging from our sample cutoff
of $-0.75$ up to $-0.2$.  Like the star shown in Figure~\ref{fig:elem_highchi2},
their median residual abundances of C+N and Ce are extremely elevated
(by 0.9-1 dex), and their median Na and Al residuals are +0.4 dex.
Ca, K, Ni, and Co all show median deviations at the 0.1-0.2 dex level.
Many of these stars have $\femg$ below the plateau value of $-0.3$, so 
they are assigned (unphysical) negative values of $\AIa$.  The negative
median deviations of most iron-peak elements may be a consequence of
the 2-process predictions extrapolating poorly to this regime.
Three of the $\omega\,$Cen stars have $\femg$ near the median sequence of
the low-Ia disk population.  Like the other $\omega\,$Cen members, these 
three 
stars all have extremely elevated (0.7-1.1 dex) C+N, Na, Al, and Ce,
positive Ca residuals (0.1-0.2 dex) and negative Ni residuals (0.05-0.25 dex).
\cite{Meszaros2021} present an APOGEE analysis of a much larger sample
(982 stars) of $\omega\,$Cen members, identifying multiple sub-populations
in the (Fe,Al,Mg) distribution and examining abundance ratio trends
in detail (see also \citealt{Johnson2010}).

In the fourth row we show residual abundances for stars
identified as probable members of the LMC by 
Hasselquist et al.\ (2021), drawn from several different
APOGEE programs targeting LMC stars 
(\citealt{Nidever2020}; Santana et al.\ 2021).
For this sample we drop our geometrical cuts, but we do apply the same
cuts in $\logg$, $\Teff$, $\mgh$, and SNR to ensure a fair comparison
to stars in our disk sample.  We caution that of the 10655 LMC candidates
in our original sample only 207 pass our $\logg$ and $\mgh$ cuts.
Most stars are lower $\logg$ because they must be luminous in order for
APOGEE to obtain high SNR spectra at the distances of the LMC.
It is possible that stars passing our cut are on the tail of the $\logg$
error distribution and have systematic abundance errors as a result.

Taking the measurements at face value, we note first that the 2-process
model trained on Milky Way disk stars predicts the median LMC 
abundances of many APOGEE elements to 0.1 dex or better, which is an impressive
degree of similarity given the radically different star formation environments
and enrichment histories.  
However, several elements show median depressions of 0.15-0.2 dex 
(Na, Al, Ni, V, Co), and C+N and Mn show median depressions of 0.07 and 0.11
dex, respectively.  The largest deviation is a 0.22-dex enhancement of Ce,
and S and Ca show median enhancements of 0.10 and 0.06 dex.
Similar deviations are found by Hasselquist et al.\ (2021) comparing
the median [X/Mg] ratios of the LMC to values for the high-Ia Milky Way 
disk in the overlapping metallicity range. 
The $\afe-\feh$ tracks of the LMC imply a low star formation efficiency at
early times, and an upward turn in $\afe$ at high $\feh$ suggests a 
substantial increase of star formation $\sim 2-4\Gyr$ in the past
(\citealt{Nidever2020}; Hasselquist et al.\ 2021), in qualitative 
agreement with photometric studies \citep{Harris2009,Weisz2013,Nidever2021}.
The different enrichment history of the LMC has left its imprint on
the relative abundances of Ce, Ni, and multiple odd-$Z$ elements in
addition to the $\afe$ ratios.  

Comparison of disk and LMC abundances
can be improved by selecting a disk sample with the same $\logg$
distribution as the LMC sample, as for the disk-bulge comparison 
by \cite{Griffith2021a}.  This approach can also be applied to APOGEE
observations of the Sgr dwarf and tidal stream \citep{Hayes2020},
and with lower metallicity samples it can be used to compare the 
Milky Way disk and halo to other dwarf satellites observed by APOGEE
(Hasselquist et al.\ 2021) and to compare the satellites among themselves.
Interpretation of these results would be aided by chemical evolution 
models that predict relative enrichment patterns for AGB elements and elements
with metallicity-dependent yields in different regimes of star formation
efficiency and star formation history.

All of the above populations are selected based on geometric and kinematic
criteria.  The final row of Figure~\ref{fig:populations} shows residual
abundances for stars in our main disk sample selected to have unusually
high or low values of [Mg/Fe], roughly $200/34,410 \approx 0.6\%$ of
the sample in each case (see Figure~\ref{fig:mgfe} for reference).
Red points show stars that lie at least 0.03 dex
above our adopted plateau value of $\mgfepl=0.30$.
The other elements that enter the 2-process fit (O, Si, Ca, Ni) have
median deviations below 0.02 dex, so this population does not seem to
arise from unusual values of Mg or Fe in isolation.
The median {\it residual} [Mg/Fe] is 0.008, significantly smaller
than the $> 0.03$-dex offset from $\mgfepl$.
The 2-process model assigns negative (unphysical) values of $\AIa$
to these stars, and the $\sim 0.03$-dex median residuals for many of the
elements with large $\qxIa$ (0.08 dex for Ce)
may be a consequence of extrapolating the model to this extreme regime.
K shows an intriguing $-0.04$-dex median residual even though it
has $\qxIa \approx 0$.
A plausible scenario is that the enrichment of these rare high-[Mg/Fe]
disk stars is dominated by CCSN that have moderately lower Fe and Ni
yields relative to $\alpha$ elements, perhaps just from stochastic
sampling of the IMF.
As previously noted, many $\omega\,$Cen stars have high [Mg/Fe], but
the high-[Mg/Fe] population as a whole does not show the extreme
residual abundances of the $\omega\,$Cen population.
The AstroNN parameters for these stars indicate preferentially old
ages and a wide range of eccentricities, as one would expect from
their high [$\alpha$/Fe] ratios, but they exhibit no obvious clumping
in $R$ and $Z$.

The stars with $\mgfe < -0.08$ (black points) do show distinctive
abundances, most notably for Na (0.17 dex) and Ce (0.11 dex).
The small residuals for O, Si, Ca, and Ni again implies that this
population is not produced by poor Mg or Fe measurements or by
isolated variations of these two elements.  These stars have
a mean AstroNN age of only $2.7\Gyr$, as expected based on
Figure~\ref{fig:aratio_age_map} and their high values of $\AIa/\Acc$.
The elevated Na and Ce residuals of this population are thus
another facet of the correlation of these residuals with age,
seen previously in Figure~\ref{fig:delta_vs_age}.
However, S and C+N residuals do not show strong correlations
with age but nonetheless exhibit $\sim 0.04$-dex enhancements
in this low-[Mg/Fe] population.
If we consider the far more numerous ($\sim 8000$) stars with
$-0.05 \leq \mgfe < 0.0$, the median residuals of Na and Ce
are only 0.014 dex and 0.020 dex, respectively, and median
residuals of all other elements are smaller than 0.01 dex.
Thus, the very low [Mg/Fe] stars do appear to be a distinct population,
in both age and abundance patterns.  
These stars are preferentially low eccentricity and close to the
Galactic plane, as expected for a young population, but they also
do not exhibit obvious clumping in $R$ and $Z$.

\section{Beyond two processes}
\label{sec:beyond}

The covariance of residuals demonstrated in Figure~\ref{fig:covar}
(and by TW21) implies that we should do more than simply look at
residuals element-by-element.
Theoretically, we would like to describe stellar abundances in 
terms of all of the astrophysical processes that contribute significantly
to their origin.  Empirically, we can describe star-by-star variations
in terms of components that vary multiple elements in concert.
The latter approach is similar in spirit to applying principal component
analysis to stellar abundances \citep{Andrews2012,Ting2012,Andrews2017}, 
but focusing on residuals allows us to first remove the CCSN and SNIa
processes that we know make dominant contributions to most APOGEE elements.
Both approaches are connected to the underlying question of the 
dimensionality of the stellar distribution in chemical abundance space:
if we have measurements of $M$ abundances for every star, how well can
the full distribution (not just the mean trends)
of those abundances in $M$-dimensional space be 
approximated by a 1-dimensional curve, a 2-dimensional surface, 
a 3-dimensional hypersurface, etc.? 
(For related discussion see \S 5.1 of TW21.)
In this section we first discuss the generalization of the 2-process
model to additional processes (\S\ref{sec:beyond_model}) and the
relation between process fluctuations and residual correlations 
(\S\ref{sec:beyond_fluctuations}). We then turn to an empirical approach
of fitting correlated residual components (\S\ref{sec:beyond_components})
and look for correlations of those components with age and kinematics
(\S\ref{sec:beyond_correlations}).

\subsection{An N-process model of abundances}
\label{sec:beyond_model}

As a mathematical exercise, it is trivial to generalize the 2-process
model of \S\ref{sec:2process} to $\mu=1,...,N$ processes.  The abundance
of element $\Xj$ in a star is given by
\begin{equation}
x_j \equiv  {\left(\Xj/{\rm H}\right)\phantom{_\odot} 
              \over \left(\Xj/{\rm H}\right)_\odot }
    = \sum_{\mu=1}^N A_\mu \qmuj~.
\label{eqn:xj}
\end{equation}
We use greek subscripts to denote processes and latin subscripts to denote
elements, and for compactness we have omitted the $z$-dependence of $\qmuj$
and have not introduced a separate index to denote the star.  
As with the 2-process model, the process vectors
$\qmuj$ are taken to be universal at a given metallicity
across all stars in the population, while
the amplitudes $A_\mu$ vary from star-to-star and are defined to be 
$A_\mu=1$ for a star with solar abundances.  The generalizations of equations 
(\ref{eqn:xhratios},\ref{eqn:xmgratios},\ref{eqn:fxcc},\ref{eqn:fxccsun}) are:
\begin{eqnarray}
\Xhj &=& \log_{10} \left({\sum A_\mu \qmuj}\right)~, \label{eqn:np_xh} \\
\Xmgj &=& \log_{10} \left({{\sum A_\mu \qmuj} \over \Acc} \right)~, 
          \label{eqn:np_xmg} \\
f^{X_j}_{\rm cc} &=& {\Acc q_{{\rm cc},j} \over \sum A_\mu \qmuj}~,
          \label{eqn:np_fxcc} \\
f^{X_j}_{{\rm cc},\odot} &=& {q_{{\rm cc},j}(z=1) \over \sum \qmuj(z=1)} 
          \label{eqn:np_fxccfun}~,
\end{eqnarray}
where all sums are over $\mu=1,...N$.  

In our discussion below we will take $\mu=1$ to represent CCSN and $\mu=2$
to represent SNIa.  The obvious choice for $\mu=3$ is AGB enrichment,
while larger $\mu$ could represent rarer processes that are important for
some elements, such as neutron star mergers, magnetar winds, etc.
However, we caution that partitioning enrichment channels into a moderate
number of discrete processes is an approximate exercise, and a characterization
that is adequate for one stellar abundance sample may become inadequate for
a sample with higher measurement precision or a different range of stellar
populations.  For example, at one level of precision it may be fine to 
treat CCSN enrichment as a single IMF-averaged process, while at higher 
precision or for a metal-poor stellar population one may need to consider 
stochastic variations in IMF sampling.  The mass dependence of AGB yields
is different for different elements, and because the lifetimes of stars
depend strongly on mass it may not be adequate to describe AGB enrichment
in terms of a single IMF-averaged process.  For any source (CCSN, SNIa,
AGB, etc.), the relative production of elements that have very different
metallicity dependence will change to some degree with the enrichment
history of the stellar population.

Despite these caveats, an N-process description offers a powerful way to
isolate two largely distinct aspects of Galactic chemical evolution (GCE)
models: nucleosynthetic yields and enrichment history.
While the enrichment history --- which is itself affected by accretion,
star formation, and gas flows --- can strongly affect metallicity
distribution functions, it has much more restricted impact on element
ratios.  In the N-process language, the enrichment history
determines the joint distribution of process amplitudes
$p(\{A_\mu\})$ and its trends with age and kinematics,
but the nucleosynthetic yields determine the process vectors $\qmuj$
with little dependence on enrichment history.
For 2-process modeling with APOGEE data, we have the advantage
that some elements (O, Mg) are expected to arise almost entirely
from CCSN and that Fe and Ni provide well measured diagnostics of
SNIa enrichment.  To characterize a third process, we would like one or
more well measured elements that have minimal contributions from SNIa
to serve as markers of this process.  There are no ideal candidates
in current APOGEE data, though Ce is a possibility,
and if C and N could be individually
corrected to birth abundance values then they might provide a further foothold
for quantifying AGB enrichment.  With GALAH data or joint APOGEE-GALAH
data sets \citep{Nandakumar2021}, neutron capture elements such as Ba, Y,
and Eu may provide valuable diagnostics of additional processes.

\subsection{Process fluctuations and residual correlations}
\label{sec:beyond_fluctuations}

The N-process model provides a conceptual language for thinking about
residual abundance correlations like those shown in Figs.~\ref{fig:covar}
and~\ref{fig:correl}.  First, we adjust equation~(\ref{eqn:xj}) to
allow ``intrinsic noise'' in individual abundances that is not described
by the N-process model,
\begin{equation}
x_j = \sum_{\mu=1}^N A_\mu \qmuj + \eta_j x_j~,
\label{eqn:xj2}
\end{equation}
where $\langle \eta_j^2 \rangle$ would be the fractional variance of
the residual abundance of element $\Xj$ if we knew each star's $A_\mu$
exactly, and we assume $\langle \eta_j \rangle = 0$ and 
$\langle \eta_j\eta_k \rangle = 0$ for $j \neq k$.
If the model includes all processes that are important for the production
of element $\Xj$ then we expect $\langle \eta_j^2 \rangle \ll 1$.
The 2-process fit is applied to elements that we expect to be dominated
by $\mu=1$, 2 (CCSN and SNIa).  At given values of $A_1$, $A_2$, the stellar
population has mean values of the process amplitudes $\bar{A}_\mu$ for
$\mu > 2$.  The mean abundances in the population are
\begin{equation}
\bar{x}_j(A_1,A_2) = A_1 q_{1,j} + A_2 q_{2,j} + \sum_{\mu>2} 
  \bar{A}_\mu \qmuj~.
\label{eqn:xjbar}
\end{equation}

To predict the correlations of observed abundance residuals, we must allow 
for the fact that we do not know each star's true values of $A_1$ and $A_2$
but instead have estimates of these quantities, which we denote by
$\AOhat$ and $\AThat$.  Our abundance 
measurements are also affected by observational noise
\begin{equation}
\xjhat = x_j + \epsilon_j x_j~,
\label{eqn:xjhat}
\end{equation}
where $\langle \epsilon_j^2 \rangle$ is the fractional variance of the
measurement errors.  The residual abundances for a given star are
\begin{equation}
\Delta x_j = \xjhat - \bar{x}_j(\AOhat,\AThat)~.
\label{eqn:dxj}
\end{equation}
To approximate these residuals we introduce
\begin{equation}
\delta A_1 \equiv \AOhat-A_1, \qquad \delta A_2 \equiv \AThat-A_2
\label{eqn:deltaA}
\end{equation}
and
\begin{equation}
\Delta A_\mu \equiv A_\mu - \bar{A}_\mu(A_1,A_2), \qquad \mu > 2~,
\label{eqn:DeltaA}
\end{equation}
using $\delta,\Delta$ to represent observational and intrinsic 
differences, respectively.
We make the first-order Taylor expansion
\begin{equation}
\bar{A}_\mu(\AOhat,\AThat) \approx \bar{A}_\mu(A_1,A_2) + 
  {\partial\bar{A}_\mu \over \partial A_1}\delta A_1 +
  {\partial\bar{A}_\mu \over \partial A_2}\delta A_2 ~.
\label{eqn:Abarmu}
\end{equation}
Writing
\begin{equation}
\bar{x}_j(\AOhat,\AThat) = \AOhat q_{1,j} + \AThat q_{2,j} + 
  \sum_{\mu>2} \bar{A}_\mu(\AOhat,\AThat)\qmuj
\label{eqn:xjbar2}
\end{equation}
and
\begin{equation}
\xjhat = A_1 q_{1,j} + A_2 q_{2,j} + \sum_{\mu>2} A_\mu \qmuj +
  \eta_j x_j + \epsilon_j x_j
\label{eqn:xjhat2}
\end{equation}
and applying equations~(\ref{eqn:deltaA}-\ref{eqn:xjhat2}) to
equation~(\ref{eqn:dxj}) yields,
after some manipulation,
\begin{eqnarray}
\Delta x_j &=& (\eta_j+\epsilon_j)x_j + \sum_{\mu>2} \Delta A_\mu \qmuj 
  \nonumber \\
&& +\, \delta A_1 q_{1,j} + \delta A_2 q_{2,j}  \nonumber \\
&& -\,\delta A_1 
    \left(\sum_{\mu>2} {\partial\bar{A}_\mu\over \partial A_1}\qmuj\right)
   -\delta A_2 
    \left(\sum_{\mu>2} {\partial\bar{A}_\mu\over \partial A_2}\qmuj\right) ~.
\label{eqn:dxj2}
\end{eqnarray}
The first term represents the sum of ``intrinsic noise'' and observational 
noise.  The second term is the most physically interesting, showing the
impact of random fluctuations in additional processes beyond CCSN and SNIa.
The last four terms represent the ``measurement aberration'' discussed
by TW21 and in \S\ref{sec:residuals_covariance} above.

To obtain an expression for covariance that is tractable enough to be 
conceptually useful, we ignore the last two terms of 
equation~(\ref{eqn:dxj2}), and we assume that 
$\langle \delta A_1 \delta A_2\rangle = 0$, that
$\langle \delta A_\mu \Delta A_\nu\rangle = 0$, and that
$\langle \Delta A_\mu \Delta A_\nu\rangle = 0$ for $\mu \neq \nu$.
It is not clear that any of these approximations is accurate in a realistic
case, but the resulting expression does illuminate several of the 
effects that influence the covariance of residuals:
\begin{eqnarray}
\langle \Delta x_j \Delta x_k \rangle &\approx&
  \langle \eta_j^2 + \epsilon_j^2 \rangle x_j^2 \delta^{\rm Kron}_{jk} 
  \nonumber \\
  && + \sum_{\mu>2} \langle (\Delta A_\mu)^2 \rangle \qmuj q_{\mu,k} 
  \nonumber \\
  && +
  \langle (\delta A_1)^2\rangle q_{1,j}^2 +
  \langle (\delta A_2)^2\rangle q_{2,j}^2~.
\label{eqn:dxjdxk}
\end{eqnarray}
If measurement aberration is small enough to be neglected, then 
off-diagonal covariances all arise from the second term.
These off-diagonal covariances can be small either because the variation
in process amplitudes at fixed $(A_1,A_2)$ is small, so that
$\langle (\Delta A_\mu)^2 \rangle \ll 1$,
or because the $\mu > 2$ processes make small contributions to the abundances
of elements $x_j$ or $x_k$, so that $\qmuj q_{\mu,k} \ll 1$.
Covariances alone offer no way to distinguish these two cases.
However, if the second term dominates over the other three, then the
off-diagonal {\it correlation} will be large for elements that come largely
from a single $\mu > 2$ process, even if the covariance is small because
$\langle (\Delta A_\mu)^2 \rangle \ll 1$.

As a concrete example of this point, consider a pair of elements whose
production is dominated by AGB stars.  Because AGB enrichment is delayed
in time like SNIa, we expect $\Aagb$ to increase with both $A_1$ and $A_2$,
and at solar abundances we expect $\Aagb \approx 1$.  Even if the variance
of $\Aagb$ is small, it is responsible for most of the variation in the
two elements, so the correlation coefficient
$\langle \Delta x_j \Delta x_k \rangle / 
  \sqrt{\langle (\Delta x_j)^2 \rangle \langle (\Delta x_k)^2 \rangle}$
will be near unity even though $\langle \Delta x_j \Delta x_k \rangle$
itself is small.  In fact the correlation coefficient can be large even if the
elements themselves have large contributions from CCSN and SNIa, because
the {\it variation} at fixed $(A_1,A_2)$ still comes from other processes.
However, in this case it is more challenging to distinguish the true
correlations caused by additional processes from the artificial correlations
induced by measurement aberration (non-zero $\delta A_1$ and $\delta A_2$).

In light of this discussion, the large correlation coefficients seen in
Figure~\ref{fig:correl} or in Figure 8 of TW21 are not surprising.
Even when element abundances are predicted to high accuracy by a 2-parameter
model, or by conditioning on two elements, the intrinsic correlations of
the residual abundances will be high if they are dominated by a small
number of additional processes.

An interesting feature of equation~(\ref{eqn:dxjdxk}) is that it generates
only positive correlations if the $\qmuj$ are positive.  Anti-correlations
can arise if the process amplitudes themselves are anti-correlated,
$\langle \Delta A_\mu \Delta A_\nu \rangle < 0$, a possibility that
(for simplicity) we did not
allow in deriving equation~(\ref{eqn:dxjdxk}).  They could also arise from
processes that deplete some elements (negative $\qmuj$) but produce others,
which could happen in unusual circumstances.
Measurement aberration may easily lead to 
$\langle \delta A_1 \delta A_2 \rangle < 0$, since one is fitting parameters
to abundances that have contributions from both processes.
The largest anti-correlations in Figure~\ref{fig:correl} involve elements
that contribute to the $(A_1,A_2)$ fit, and similar features appear
in the simulated data, which suggests that these anti-correlations
are dominated by measurement aberration.
If intrinsic anti-correlations can be well established empirically then they
could be quite physically informative, since they are not easy to produce.

\subsection{Fitting additional components}
\label{sec:beyond_components}

While we would ideally like to infer values of $\qmuj$ for additional 
processes from the 2-process residuals, then fit to obtain values of
$A_\mu$ for individual stars as we did for $\Acc$ and $\AIa$, 
it is not clear that there is any practical way to do this without
theoretical priors on what elements to assign to what processes.
For the current APOGEE data, the challenge is exacerbated by the
fact that the residuals from the 2-process predictions are usually
not much larger than the estimated observational noise, and the observational
error distribution is itself uncertain.  Correlation of residuals
can be measured at high significance in a large sample, but the residual
abundances of individual stars are mostly measured at low or moderate
significance.  In future work, we will use chemical evolution simulations
that incorporate multiple enrichment channels and stochastic variations
to guide strategies for isolating additional processes from observed
abundance distributions.

\begin{figure*}
\centerline{\includegraphics[width=5.5truein]{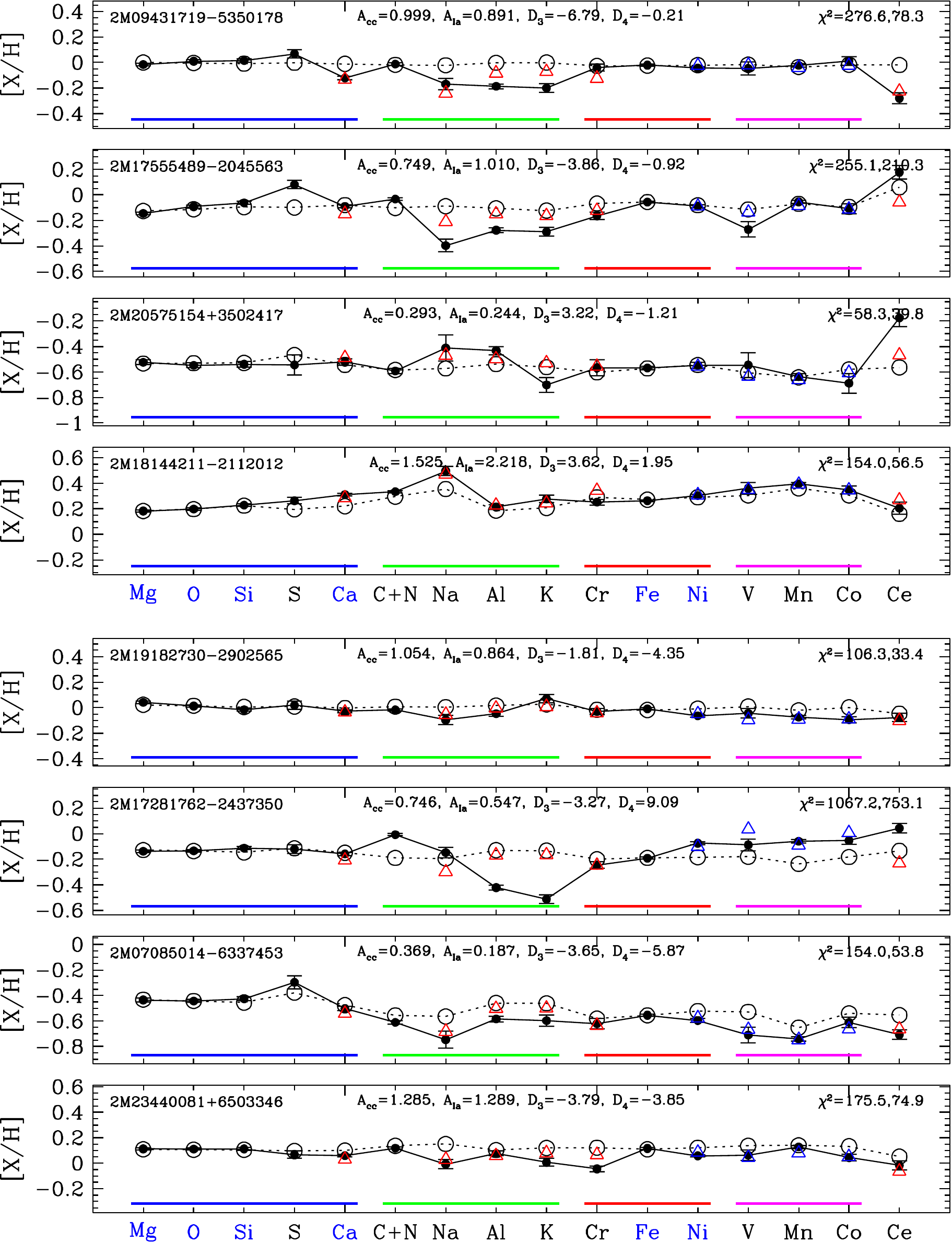}}
\caption{Examples of 4-process fits to element abundance ratios,
in a format similar to Fig.~\ref{fig:elem_stars1}.  In each panel,
filled circles with error bars show the APOGEE abundance measurements
and open circles show the abundances predicted by the best 2-process 
model fit.  Red triangles show the abundances predicted after fitting
a component of amplitude $D_3$ to the residual abundances of Ca, Na, Al,
Cr, and Ce.  Blue triangles show the predicted abundances after fitting
a component of amplitude $D_4$ to the residual abundances of Ni, V, Mn, 
and Co.  The change in an element's predicted abundance in dex
is the product of $D_\mu$ with the corresponding value of $r_{\mu,j}$
in Table~\ref{tbl:d3} or~\ref{tbl:d4}.
Each panel lists the $\chi^2$ values for the 2-process and 4-process
fits.  The first two stars have unusually low values of $D_3$; the next
two have unusually high values of $D_3$; the next two have unusually
low and high values of $D_4$; and the final two have unusually low
values of both $D_3$ and $D_4$.  
}
\label{fig:elem_d3d4}
\end{figure*}

For a data-driven approach, the most obvious tack is to apply principal
component analysis (PCA) to our estimate of the intrinsic covariance matrix
of residual abundances in Figure~\ref{fig:covar}d.  The new components
would be the eigenvectors of this matrix that have the largest eigenvalues
and thus explain the largest fraction of the variance.
However, there is no reason to expect the physical enrichment processes to
produce orthogonal components in abundance space, so even if the intrinsic
covariance matrix were perfectly known the eigenvectors would represent
mixtures of the physical processes.  We have also found that the results
of PCA are sensitive to minor details of how we treat the data and
measure the covariance, making physical interpretation difficult.
For now we adopt a simpler approach that is loosely motivated by the
discussion in \S\ref{sec:beyond_fluctuations}.

We pick a subset of elements that show significant correlations and that
we suspect on physical grounds should be treated as a group.
For each group, we have a covariance matrix of residual abundances
$C_{jk}=\langle \Delta_j\Delta_k\rangle$.  Suppose that the residuals within
this element group arise from a {\it single} process $\mu$ plus 
uncorrelated ``noise'' that may include both observational noise
and intrinsic element-by-element scatter:
\begin{equation}
\Delta_{j,*} = D_{\mu,*} r_{\mu,j} + \epsilon_{j,*}~.
\label{eqn:djstar}
\end{equation}
In contrast to our notation in \S\ref{sec:beyond_fluctuations},
$\Delta_{j,*}$, $\epsilon_{j,*}$, and $r_{\mu,j}$ are all in dex,
and we use $D_\mu$ in place of $A_\mu$ because it represents a deviation
from the mean amplitude $\bar{A}_\mu(\Acc,\AIa)$ rather than an
amplitude that is defined to be unity at solar abundances.

Under these assumptions, the predicted covariance matrix of these elements is
\begin{equation}
C_{jk,{\rm pred}} = \langle D_\mu^2 \rangle r_{\mu,j} r_{\mu,k} +
                    s_j^2 \delta^{\rm Kron}_{jk}~,
\label{eqn:cjkpred}
\end{equation}
where $s_j^2 = \langle \epsilon_j^2 \rangle$ represents the variance in
the residual abundance $\Delta_j$ that is not explained by the correlated
deviations.  As previously noted, from covariances alone we cannot
distinguish between large $\{r_{\mu,j}\}$ with small $\langle D_\mu^2 \rangle$
and small $\{r_{\mu,j}\}$ with large $\langle D_\mu^2 \rangle$.  We arbitrarily
take $\langle D_\mu^2 \rangle = 1$ and infer the corresponding values
of $r_{\mu,j}$ by minimizing the cost function
\begin{equation}
{\rm cost} = \sum_{j=1}^{N_\mu-1} \sum_{k=j+1}^{N_\mu} 
  \left(C_{jk,{\rm pred}}-C_{jk,{\rm obs}}\right)^2~,
\label{eqn:cost}
\end{equation}
i.e., by minimizing the squared deviation between the predicted and
observed {\it off-diagonal} values of the covariance matrix for the $N_\mu$
elements in the group.  The values of $s_j$ then follow from matching 
the predicted and observed diagonal components (equation~\ref{eqn:cjkpred}).
We require $N_\mu \geq 4$ to have sufficient off-diagonal constraints
$N_\mu(N_\mu-1)/2$ to determine the $N_\mu$ values of $r_{\mu,j}$.
Since we have assumed $\langle D_\mu^2 \rangle = 1$, we see from
equation~(\ref{eqn:djstar}) that if $r_{\mu,j} > s_j$ then the typical
residual abundances of $X_j$ can be explained predominantly by the
correlated fluctuations with other elements in the group, while if
$s_j > r_{\mu,j}$ then independent fluctuations dominate over this
correlated contribution.  The relative values of $r_{\mu,j}$ indicate
the relative deviations of elements $X_j$ associated with process
fluctuations $D_\mu$.

Based on Figure~\ref{fig:correl}a we have selected two element groups,
one ($\mu=3$) comprised of Ca, Na, Al, K, Cr, and Ce, and the second
($\mu=4$) comprised of Ni, V, Mn, and Co.  There is some arbitrariness 
in this choice.  For example, Na and K show significant
correlations with the iron-peak group in addition to Ca, Al, and Ce,
and Al is (weakly) anti-correlated with Ce and (weakly)
positively correlated with Ni, V, and Co.  One must be cautious about
naively applying nucleosynthesis intuition to {\it residual} abundances
because we have already removed the main effects of CCSN and SNIa through
the 2-process fit.  For example, even though K comes mainly from CCSN
and Mn comes mainly from SNIa, the {\it deviations} from typical K and Mn
abundances at a given $\Acc$ and $\AIa$ could be physically linked.

Conceptually, we could imagine that Ca, Na, Al, K, Cr, and Ce all
have contributions from AGB stars, and that positive and negative values
of $D_3$ represent stars that have more or less than the average 
amount of AGB enrichment relative to stars with the same $\Acc$
and $\AIa$ (2nd term on the r.h.s.\ of equation~\ref{eqn:dxj2}).
The $\mu=4$ component could be driven by a subset of SNIa (or even CCSN)
that have higher yields of Ni, V, Mn, and Co, and positive and negative
values of $D_4$ would represent stars enriched by more or fewer than
the average number of such unusual supernovae.
However, this physical interpretation is by no means unique.
These caveats notwithstanding, our approach offers a plausible way to 
combine physical expectations with data-driven lessons to search for
correlated element deviations on a star-by-star basis.

\begin{deluxetable}{lrrrr}[]
\tablecaption{Coefficients of component $\mu=3$\label{tbl:d3}}
\tablehead{
\colhead{Elem} & \colhead{$r_3$} & \colhead{$s_3$} & \colhead{$\sigma_{68}$} &
  \colhead{$r_3/r_{3,{\rm Ce}}$}
}
\startdata
Ca  & $0.0171$ & $0.0125$ & $0.0193$ & $0.563$ \\
Na  & $0.0292$ & $0.0873$ & $0.0699$ & $0.961$ \\
Al  & $0.0065$ & $0.0346$ & $0.0302$ & $0.214$ \\
K   & $0.0109$ & $0.0647$ & $0.0542$ & $0.359$ \\
Cr  & $0.0102$ & $0.0434$ & $0.0346$ & $0.336$ \\
Ce  & $0.0304$ & $0.0850$ & $0.0849$ & $1.000$ \\
\enddata
\tablecomments{
Coefficients for the elements comprising component 3.
For a star with amplitude deviation $D_3$ the model prediction
of $[{\rm X}/{\rm H}]$ changes by $D_3 r_{3}$ dex
(equation~\ref{eqn:djstar}).  The variance not explained
by the correlated contribution is $s_{3}^2$ (equation~\ref{eqn:cjkpred}).
When fitting $D_3$ values for individual stars, elements are
weighted by the inverse-square of $\sigma_{68}$ (equation~\ref{eqn:dmuhat}).
}
\end{deluxetable}

\begin{deluxetable}{lrrrr}[]
\tablecaption{Coefficients of component $\mu=4$\label{tbl:d4}}
\tablehead{
\colhead{Elem} & \colhead{$r_4$} & \colhead{$s_4$} & \colhead{$\sigma_{68}$} &
  \colhead{$r_4/r_{4,{\rm V}}$}
}
\startdata
Ni   & $0.0087$ & $0.0151$ & $0.0158$ & $0.364$ \\
V    & $0.0239$ & $0.0787$ & $0.0650$ & $1.000$ \\
Mn   & $0.0143$ & $0.0256$ & $0.0271$ & $0.598$ \\
Co   & $0.0197$ & $0.0385$ & $0.0363$ & $0.825$ 
\enddata
\tablecomments{
Coefficients for the elements comprising component 4.
}
\end{deluxetable}

Tables~\ref{tbl:d3} and~\ref{tbl:d4} report our inferred values of $r_{\mu,j}$
and $s_j$ for these two components.  The sum of $r_{\mu,j}^2$
and $s_{j}^2$ is equal to the variance of the element's residual
deviations from the 2-process fit (equation~\ref{eqn:cjkpred}).
For a star with a given
value of $D_\mu$, the change in the predicted $[{\rm X}_j/{\rm H}]$
from adding component $\mu$ is $D_\mu r_{\mu,j}$ 
(equation~\ref{eqn:djstar}).  
We estimate the value of $D_\mu$
for each star from the weighted average
\begin{equation}
\hat{D}_{\mu,*} = 
  {\sum_{j=1}^{N_\mu} \Delta_{j,*} r_{\mu,j}/\sigma_{68,j}^2 \over
   \sum_{j=1}^{N_\mu} r_{\mu,j}^2/\sigma_{68,j}^2 }~,
\label{eqn:dmuhat}
\end{equation}
which minimizes 
$\sum (D_{\mu,*}r_{\mu,j}-\Delta_{j,*})^2/\sigma_{68,j}^2$.
We weight by the inverse of the variance estimated from
the 16-84\% percentile range of the observed residual abundance distribution,
which is less sensitive to outliers than the variance itself;
we list the values of $\sigma_{68}$ 
in the fourth column of Tables~\ref{tbl:d3} and~\ref{tbl:d4}.
This choice weights the elements more uniformly than if we used
the observational error estimates.  We omit elements that have
flagged data values for a given star.  
The final column of the tables gives the relative change of the 
elements associated with each component.  
A given value of $D_3$ changes the predicted Ce and Na abundances by
about twice as much as the predicted Ca and K abundances and by
$\sim 3$-5 times as much as the predicted Cr and Al abundances.
The range of $r_4$ values is somewhat smaller, with V the element
most sensitive to $D_4$ and Ni the least sensitive.
With the exception of Ca,
the values of $s_{j}$ exceed those of $r_{\mu,j}$, indicating
that the correlated deviations associated with these two processes
explain only a small portion of the observed residual abundance
variance for these elements.
This finding is consistent with the results of TW21, who concluded
that at least five ``components'' (implemented there as individual 
conditioning elements) beyond Mg and Fe are needed to reduce residual
fluctuations in APOGEE abundances to a level consistent with
observational uncertainties alone.

Figure~\ref{fig:elem_d3d4} shows examples of fits to eight stars
that have unusually large values (in the outer 2\% tails) of $|D_3|$
or $|D_4|$ or both.  In each of these cases, the large $D_3$ or $D_4$
reduces coherent residuals across most or all of the elements in
the component, typically 0.05 dex or larger.  However, there are 
also examples (not shown) where a single highly discrepant abundance
drives a large component amplitude.  Not surprisingly, for these stars
selected to have large $|D_3|$ or $|D_4|$ the 4-process fit achieves
a large $\chi^2$ reduction relative to the 2-process fit, but the
median reduction across the whole sample is only 4.8.
The first star shown in
Figure~\ref{fig:elem_d3d4} is also the first star shown in the
selection of high-$\chi^2$ stars in Figure~\ref{fig:elem_highchi2}.
The addition of $D_3$ and (unimportant in this case) $D_4$ reduces
$\chi^2$ from 277 to 78, though it still does not produce agreement
within the reported observational uncertainties for all of the 
deviant elements.  This pattern, a substantial $\chi^2$ reduction but
significant remaining deviations after the 4-process fit, holds for
most of the $D_3$ examples,
though the $D_4$ component typically explains the deviations (usually smaller)
in Ni, V, Mn, and Co fairly well.
The final two stars in Figure~\ref{fig:elem_d3d4} have unusually large
negative values of both $D_3$ and $D_4$.

\subsection{Correlations with age and kinematics}
\label{sec:beyond_correlations}

\begin{figure}
\centerline{\includegraphics[width=3.2truein]{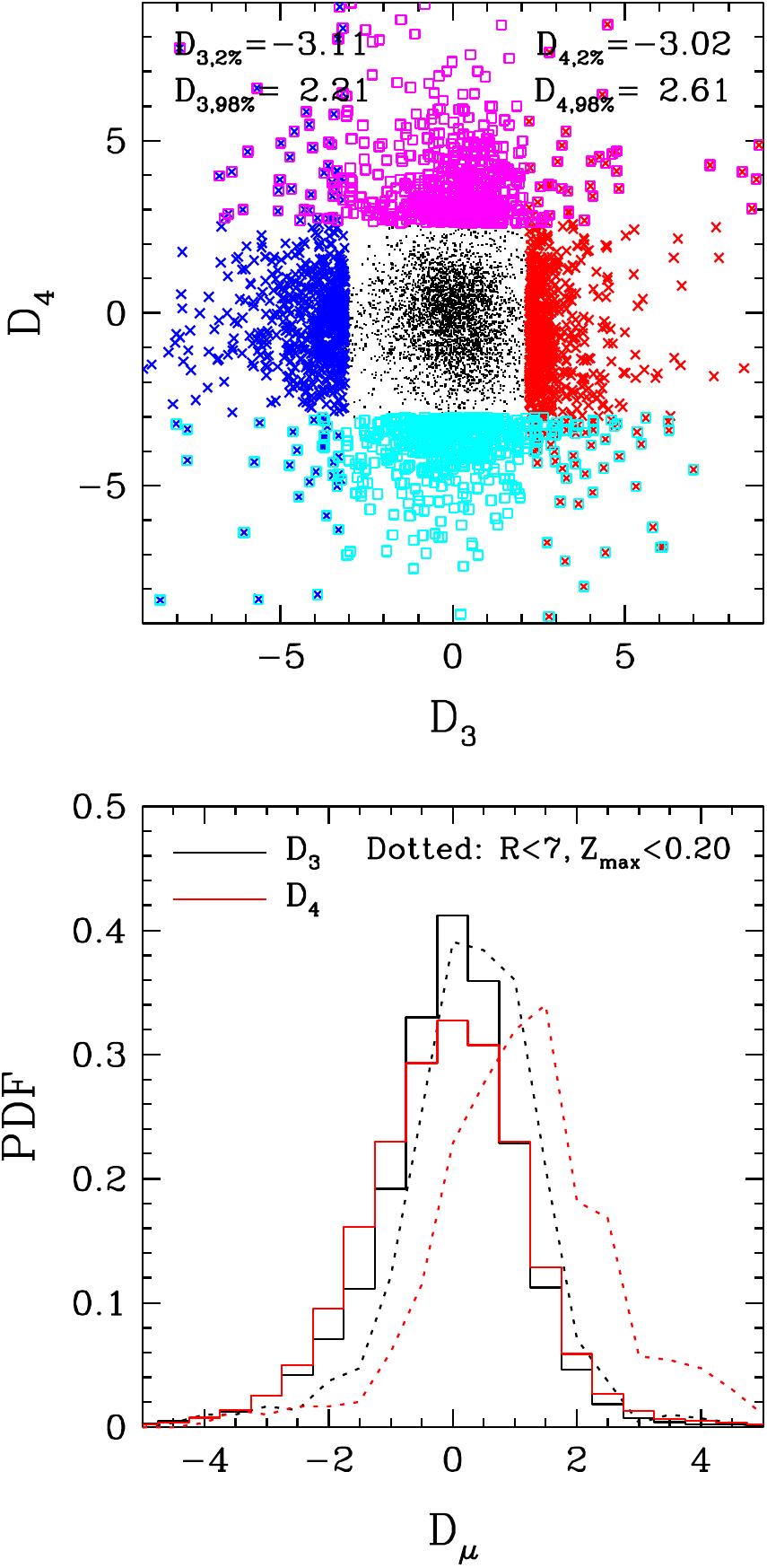}}
\caption{{\it Top:} Distribution of stars in the plane of component
amplitudes $D_3$, $D_4$.  Black dots show a 25\% random sampling 
of stars in the 2\%-98\% range of each distribution, while colored
points show stars outside the listed 2\% and 98\% boundaries.
A value of $D_3=1$ corresponds to a deviation of $D_3 r_3 = 0.0304$ dex
for the most sensitive component element (Ce) and 0.0065 dex 
for the least sensitive (Al).
A value of $D_4=1$ corresponds to a deviation of $D_4 r_4 = 0.0239$ dex
for the most sensitive component element (V) and 0.0087 dex 
for the least sensitive (Ni).
{\it Bottom:} Distributions of $D_3$ (black) and $D_4$ (red) for
the full sample (solid histograms) and for the subset of stars
with $R<7\kpc$ and $Z_{\rm max}<0.2\kpc$ (dashed curves).
Other geometric cuts produce
distributions similar to the solid histograms, but the inner thin
disk stars tend to have slightly higher values of $D_3$ and $D_4$.
}
\label{fig:d3d4_dist}
\end{figure}

In the top panel of Figure~\ref{fig:d3d4_dist}, colored points show
$(D_3,D_4)$ for stars in the outer 2\% tails of the $D_3$ or $D_4$
distributions, and black dots show a random sampling of stars in the
inner 96\% of both distributions.  The values of $D_3$ and $D_4$ are
essentially uncorrelated, with a Pearson correlation coefficient of $-0.04$,
changing to $-0.07$ if we restrict to the inner 96\%.
The lower panel plots the distribution of component amplitudes, which
are slightly skew-negative for both $D_3$ and $D_4$.  For the
most part, we find no obvious trends of $D_3$ or $D_4$ with Galactic
position.  However, stars near the midplane in the
inner Galaxy ($R=3-7\kpc$, $Z_{\rm max}<0.2\kpc$) 
tend to have slightly higher values of 
$D_3$ and $D_4$, as shown by the shifted distributions in this panel.
In other words, stars of the inner thin disk tend to have slightly
elevated values of the ten elements that contribute to these components,
relative to other stars with the same values of $\Acc$ and $\AIa$.
The mean values of $D_3$ and $D_4$ for this population are higher
by 0.44 and 1.31, corresponding to mean differences $D_3 r_3$ 
and $D_4 r_4$ of only 0.013 dex and 0.031 dex for the two most sensitive
elements (Ce and V, respectively), and smaller shifts for other elements.
This subtle change
of chemistry is detectable because we have many stars to average over and
have controlled through 2-process fitting for the much larger differences
between the inner thin disk and the full sample
in $\Acc$ and $\AIa/\Acc$ (mean offsets of 0.56 and 0.42, respectively,
corresponding to shifts of $\sim 0.2$ dex in $\mgh$ and $\sim 0.1$ dex
in $\afe$).
If we define the inner thin disk based on the current midplane distance
rather than dynamically estimated maximum distance,
i.e., by $|Z|<0.2\kpc$ instead of $Z_{\rm max}<0.2\kpc$,
then the shift of the $D_4$ distribution is similar but the
shift of the $D_3$ distribution is weaker, with a mean offset of only 0.25
instead of 0.44.

\begin{figure*}
\centerline{\includegraphics[width=5.0truein]{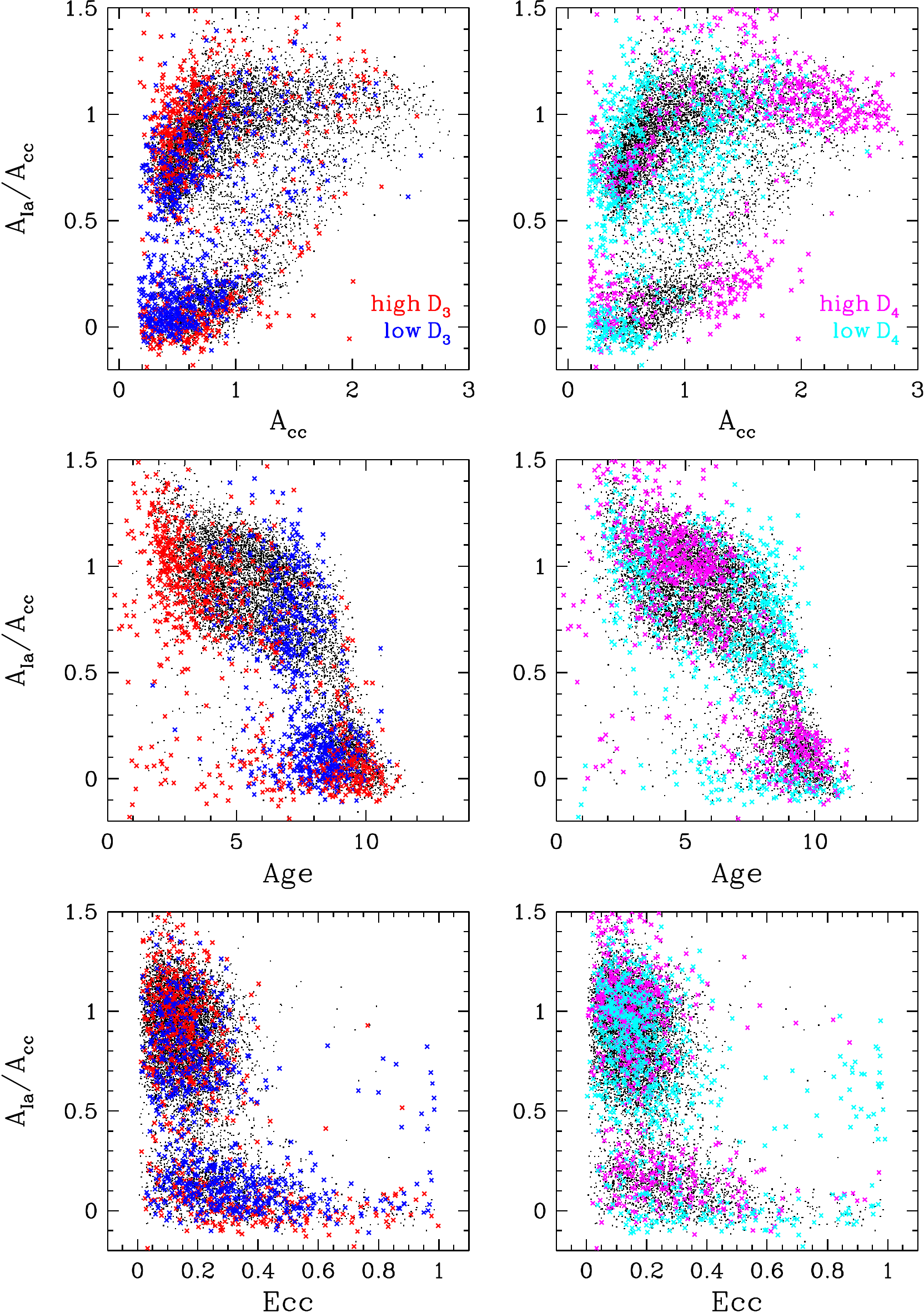}}
\caption{Distributions of stars in the 2\%-tails of the $D_3$ distribution
(left column) or $D_4$ distribution (right column), with the same
color coding as in Fig.~\ref{fig:d3d4_dist}.  Black dots show a 25\%
random sampling of the full distribution.  Top panels show the
2-process plane $\AIa/\Acc$ vs.\ $\AIa$.  Middle and bottom rows
plot $\AIa/\Acc$ vs.\ AstroNN values of stellar age and orbital
eccentricity, respectively.  Stars with extreme values of $D_3$ 
and $D_4$ are found throughout these distributions.  Low $D_3$ stars
tend to have younger ages than the full population.
The population of high eccentricity stars with $\AIa/\Acc > 0.2$ 
tends to have low values of $D_3$ and $D_4$; these ``accreted halo'' 
stars also have
values of $\mgh$ (and thus $\Acc$) near the minimum of our sample.
}
\label{fig:d3d4_map}
\end{figure*}

Figure~\ref{fig:d3d4_map} plots the $D_3$ outliers (left) and the
$D_4$ outliers (right) in the planes of $\AIa/\Acc$ vs. $\Acc$,
$\AIa/\Acc$ vs.\ age, and $\AIa/\Acc$ vs.\ eccentricity, with a
random subset of the full sample plotted for comparison.
The outliers arise throughout the $(\Acc,\AIa/\Acc)$ distribution
with no obvious clustering, though there is some overrepresentation
of low-$D_4$ outliers among metal-rich stars that are in between
the low-Ia and high-Ia populations ($\Acc \approx 0.8-1.5$,
$\AIa/\Acc \approx 0.5-0.9$).  The outliers are present at all ages,
though there is a clear tendency for high-$D_3$ stars to have younger
ages (by $\sim 2-3\Gyr$) in the high-Ia population, and a significant
number of high-$D_3$ stars have low $\AIa/\Acc$ and young estimated ages.
There is also a concentration of low-$D_4$ stars at ages of 6-8 Gyr.

Outliers are also widely distributed in the plane of $\AIa/\Acc$
vs.\ eccentricity.  However, there is a clear excess of stars with
low $D_3$ {\it and} low $D_4$ that have high eccentricity and 
elevated values of $\AIa/\Acc$ relative to other high eccentricity
disk stars.  This population also has low values of $\mgh$ (and
thus of $\Acc$), near the lower boundary of our sample.  The 
second-to-last star in Figure~\ref{fig:elem_d3d4} is a
member of this high-eccentricity population, with $e=0.985$,
though it was chosen for this plot based on
its $D_3$ and $D_4$ values alone.  We have already seen this population
stand out in the sample of high-$\chi^2$ stars (Figure~\ref{fig:chisqmap}),
and the top row of Figure~\ref{fig:populations} shows that the 
``accreted halo'' stars (a.k.a.\ GSE stars) have negative residuals
of all four elements in the $D_4$ component and of two of the elements
(Na and Al) in the $D_3$ component.
The extreme $D_3$ and $D_4$ values are another signature of the 
distinctive abundance patterns of this population.

We view this analysis as a first step in exploiting the information
encoded by correlated patterns of residual abundances.
The component formalism introduced here offers a data-motivated
way to compute average deviations of correlated elements with
appropriate relative weights, obtaining measurements that are 
higher SNR than the deviations of individual elements.
The geometric, age, and kinematic patterns in 
Figures~\ref{fig:d3d4_dist} and~\ref{fig:d3d4_map} demonstrate
that the $D_3$ and $D_4$ component amplitudes are capturing genuine
physical distinctions among stellar populations.  However, extreme
values of these components arise in stars
throughout the disk with a wide range of ages, kinematics, and
CCSN and SNIa enrichment levels.

\section{Conclusions}
\label{sec:conclusions}

We have developed a novel approach to statistical analysis of multi-element
abundance distributions of large stellar samples and applied it to the final
(DR17) data release of APOGEE-2 (from SDSS-IV), which includes a homogeneous
re-analysis of spectra from APOGEE in SDSS-III.  Our primary sample
consists of 34,410 stars with $3\kpc \leq R \leq 13\kpc$, $|Z| \leq 2\kpc$,
$-0.75 \leq \mgh \leq 0.45$, $1 \leq \logg \leq 2.5$, and
$4000\K \leq \Teff \leq 4600\K$, with the last two cuts adopted to limit
the impact of differential systematics on abundance measurements.
We consider the $\alpha$-elements Mg, O, Si, S, and Ca, the light odd-$Z$
elements Na, Al, and K, the even-$Z$ iron-peak elements Cr, Fe, and Ni,
the odd-$Z$ iron-peak elements V, Mn, and Co, the s-process element Ce,
and the element combination C+N, employed because C+N is conserved during
dredge-up processes that change the individual C and N surface abundances 
in the convection zones of red giants.  Following W19 and \cite{Griffith2019},
we fit the median $\xmg-\mgh$ trends of low-Ia and high-Ia populations with a
2-process model that approximates stellar abundance patterns as the sum
of a CCSN contribution that tracks Mg enrichment and an SNIa contribution
that tracks the SNIa Fe enrichment.  For elements with substantial 
contributions from processes other than CCSN and SNIa, the 2-process model
approximately separates a ``prompt'' and ``delayed'' enrichment component.
With the global model parameters ($\qxcc$ and $\qxIa$ for each element X
in 0.1-dex bins of $\mgh$) determined from the median sequences, we proceed
to fit each sample star's measured abundances with two free parameters
($\Acc$ and $\AIa$) that scale the amplitude of the two processes
(equation~\ref{eqn:2processq}; Figure~\ref{fig:2pro_explain}).
We characterize each star by its values of $\Acc$ and $\AIa$ {\it and}
the residuals $\Delta\xh$ from this 2-process fit.

\subsection{Median sequences and their implications}

For the 14 elements in common with W19's analysis (based on DR14), we
find similar results for median sequences and thus draw similar conclusions
about the relative CCSN and SNIa contributions.  Among the $\alpha$-elements,
Si and Ca are inferred to have significant SNIa contributions, though not
as large as those of iron-peak elements.  Among the light odd-$Z$ elements,
Al and K appear to be dominated by CCSN, but the low-Ia and high-Ia 
populations have substantially different [Na/Mg] ratios, implying
a large delayed contribution to Na that could be associated with SNIa
or AGB sources.  Among the iron-peak elements, Mn is inferred to have
the largest SNIa contribution.  The most significant differences from DR14 are
that the increasing metallicity trend of [Al/Mg] becomes flat in DR17
and that the steeply rising trends of [V/Mg] with metallicity become
shallower.  While W19 fit the median trends with power-law 
metallicity dependence
for the CCSN and SNIa processes, here we adopt a generalized metallicity
dependence in bins of [Mg/H] such that the 2-process model reproduces 
the observed [X/Mg] sequences exactly.  Several elements --- Na, V, Mn, Co,
and to a lesser extent Ni --- show evidence of rapidy rising SNIa yields
for $\mgh > 0$, though this conclusion is sensitive to the accuracy of
APOGEE's abundances in the super-solar metallicity regime.

For [(C+N)/Mg] and [Ce/Mg], both new to this study, we find substantial
gaps between the median sequences of low-Ia and high-Ia stars, implying
a substantial contribution from delayed sources.  For these elements,
the delayed source is probably AGB enrichment rather than SNIa.
The metallicity dependence of the high-Ia [Ce/Mg] sequence is 
non-monotonic, peaking at $\mgh \approx -0.2$, similar to the behavior
seen in GALAH DR2 for the neutron-capture elements Y, Ba, and La
\citep{Griffith2019}.  The rising trend at low [Mg/H] can be understood
from the increase of seed nuclei for neutron capture, which shifts to
a falling trend when the ratio of seed nuclei to free neutrons becomes
too large to allow the s-process to reach heavy nuclei \citep{Gallino1998}.
The low-Ia/high-Ia median trends for [(C+N)/Mg] and [Ce/Mg] are a powerful
empirical test for supernova and AGB yield predictions.  The $\qxcc$ and
$\qxIa$ values that we derive for other elements allow tests of supernova
yield models (e.g., \citealt{Griffith2021b}) that are insensitive to
uncertainties in other aspects of disk chemical evolution.

\subsection{Residual abundance scatter and correlations}

Turning to residual abundances, we find that the distribution of 
$\Delta\xh$ residuals from the 2-process predictions is narrower than
the distribution of residuals from the observed median sequences for
all of the elements that APOGEE measures well (i.e., with mean
observational uncertainties below 0.03 dex; see Figure~\ref{fig:delta_dist}).
This reduction implies that much of the observed scatter in [Mg/Fe] at
fixed [Mg/H] {\it within} the low-Ia and high-Ia populations is intrinsic
\citep{Bertran2016,Vincenzo2021}, reflecting real variations in SNIa/CCSN
enrichment ratios, and that accounting for these variations correctly
predicts variations in other elements.  Similarly, we find that using
residuals from the 2-process predictions rather than residuals from
median sequences largely removes trends with stellar age and orbital
parameters (Figures~\ref{fig:delta_vs_age} and~\ref{fig:delta_vs_kinematics}).
However, Ce and Na residuals both show clear correlations with age in 
the high-Ia population, with the youngest stars showing higher abundances
of both elements relative to other stars with similar $\Acc$ and $\AIa$.

After subtracting the observational uncertainties reported by ASPCAP from
the observed $\Delta\xh$ scatter, we infer rms intrinsic scatter in the
2-process residuals ranging from $\sim 0.005$ dex to $\sim 0.04$ dex
for most elements, with values up to $\sim 0.08$ dex for Na, K, V, and Ce
(Figure~\ref{fig:sigma}).  Our estimates of the characteristic intrinsic
scatter and of the relative scatter among different elements agree quite
well with the estimates of TW21 for scatter in abundances conditioned
on [Fe/H] and [Mg/Fe], and with those of \cite{Ness2019} for scatter
conditioned on [Fe/H] and age.

More informative than the element-by-element scatter is the covariance of
residual abundances between elements (equation~\ref{eqn:cij}).  We find
significant off-diagonal covariances among many elements, with many values
clearly exceeding the expected covariance from observational errors alone
(Figure~\ref{fig:covar}).  Our estimates of 2-process residual correlations
(Figure~\ref{fig:correl}) agree qualitatively with those found by TW21
for conditional abundance residuals despite many differences in methodology,
a reassuring indication of the robustness of the results.
Correcting the observed covariances for observational contributions remains
uncertain because the observational error distributions are not fully
understood.  The clearest findings are two ``blocks'' of correlated
residuals, one involving Ca, Na, Al, K, Cr, and Ce and the other comprised
of Ni, V, Mn, and Co.  For most correlated element pairs, the bi-variate
distribution of residuals shows a consistent slope between the core of
the distribution and the tails (see Figure~\ref{fig:tefftrend} for examples).
This structure suggests that the residuals are mostly driven by a 
continuous spectrum of variations, e.g., by the relative contribution of
processes beyond CCSN and SNIa, or by stochastic sampling of the CCSN
and SNIa populations combined with imperfect mixing in the ISM.
The one striking exception to this rule is the (C+N)-Ce correlation
(Figure~\ref{fig:CN_Ce}), where the core of the distribution shows a
clear anti-correlation but a population of rare outliers exhibits strong
positive deviations of both (C+N) and Ce.  These highly enhanced stars
could be a consequence of mass transfer from AGB companions or of 
second-generation AGB enrichment in star clusters.

\subsection{High-$\chi^2$ stars and selected populations}

By automatically normalizing a star's abundances to those of other stars
with similar [Mg/H] and [Mg/Fe], 2-process fitting makes it easy to identify
outlier stars with unusual measured abundance patterns.  This approach
is especially valuable for cases with moderate deviations (e.g., 0.05-0.1 dex)
across multiple elements, which might be difficult to pick out in an
eyeball scan of [X/Fe]-[Fe/H] diagrams.  Unfortunately, easy identification
does not mean easy interpretation, and a key challenge is distinguishing
physical outliers from cases where measurement errors are much larger than
the reported observational uncertainties.  Among the physical outliers,
some may be extreme examples of the same variations that produce residual
correlations in the bulk of the population, while others may arise from
rare physical processes that affect only a small fraction of stars.

Figure~\ref{fig:elem_highchi2} presents a selection of eight stars from the
$\sim 700$ that comprise the top 2\% of the residual $\chi^2$ distribution.
These examples include two stars with depressed Na, Al, and K abundances
and low or high Ce, a carbon star that also has high measured abundances
of Na, Al, and V, a ``barium'' star first identified by \cite{Smith1987}
that is one of the extreme (C+N)-Ce outliers, a member of the $\omega\,$Cen
globular cluster with 0.5-1 dex enhancements in C+N, Na, Al, and Ce, and
a N-rich star with elevated Al, Ce, and Si, which has been independently 
identified both as a possible globular cluster escapee
\citep{Schiavon2017,Fernandez-Trincado2017,Fernandez-Trincado2019,
Fernandez-Trincado2020,Fernandez-Trincado2020b} and as a member of
a small population of chemically peculiar stars with extreme P
enhancement \citep{Masseron2020a,Masseron2020b}.
Another star shows strong deficiencies of K and V, an effect that we see
in multiple stars but that may be a consequence of radial velocity placing
stellar features over strong telluric lines that are difficult to subtract
precisely.  Another shows a distinctive pattern of enhanced Na, elevated
C+N and Mn, and depressed Al, K, and Cr.  We also see this pattern in
multiple outlier stars, but we remain unsure whether it represents an
unusual physical abundance pattern or a subtle observational systematic.

Residual abundances may prove to be a powerful tool for chemical tagging
studies, i.e., for identifying groups of stars that share distinctive
abundance patterns suggesting a common birth environment.  In this paper
we have illustrated these prospects with the much simpler exercise of
computing the median residual abundances of a few select stellar
populations (Figure~\ref{fig:populations}).  Stars with high eccentricity,
low metallicity ($\mgh \la -0.5$), and relatively low $\afe$
($\mgfe \la 0.2$) have been previously identified as ``accreted halo''
stars \citep{Nissen2010}, probably 
formed in the ``Gaia-Sausage/Enceladus'' dwarf
\citep{Belokurov2018,Helmi2018}.  Relative to 2-process model predictions,
these stars have C+N, Na, and Al abundances that are low by about 0.1 dex
and Ni, V, Mn, and Co abundances that are low by $\sim 0.05$-0.1 dex.
However, high eccentricity stars in the same $\mgh$ range with $\mgfe>0.18$
have median abundance residuals consistent with zero.  Stars observed
by APOGEE in the LMC that overlap our disk star metallicity, $\logg$, 
and $\Teff$ range show a similar abundance pattern to the GSE stars,
and a 0.2-dex enhancement of Ce.  The 14 $\omega\,$Cen members that fall in our
disk sample show extreme ($\sim 1$-dex) enhancements of C+N and Ce and large
($\sim 0.4$-dex) enhancements of Na and Al.  Stars in the outer disk
($R=15-17\kpc$), either near the midplane ($|Z| \leq 2\kpc$) or well
above it ($Z=2-6\kpc$) have abundances entirely consistent with those of
our $R=3-13\kpc$ sample, with the slight exception of a 0.05-dex median
depression of Ce in the high-$Z$ population.  Each of these results is
a target for chemical evolution models of these populations, and many other
populations can be studied in similar fashion.

\subsection{Beyond 2-process}

We do not expect the 2-process model to provide a complete description of
stellar abundances, and the intrinsic scatter of $\Delta\xh$, the
element-to-element correlations among residuals, the outlier stars,
and the distinctive patterns of selected populations all demonstrate
empirically that it does not.  In \S\ref{sec:beyond} we have taken some
first steps towards a more general ``N-process'' model.  On the theoretical
side, we have proposed a natural generalization of the 2-process formalism
that can encompass an arbitrary number of additional processes, and we have
shown, approximately, how variations in the relative amplitudes of those
processes would translate into correlated residuals from the 2-process
fits (equations~\ref{eqn:xjbar2}-\ref{eqn:dxjdxk}).
On the observational side, we have used the observed covariance matrix
of residual abundances (Figure~\ref{fig:covar}) to define two new
``components'' with weighted contributions of Ca, Na, Al, K, Cr, Ce
(component 3) and Ni, V, Mn, Co (component 4).  We then fit amplitudes
$D_3$ and $D_4$ defining the deviations of these components to all
stars, with $\langle D_3 \rangle \approx \langle D_4 \rangle \approx 0$
by construction.  We find stars with high and low values of $D_3$ and
$D_4$ throughout the disk and widely spread in $\Acc$, $\AIa/\Acc$, age,
and kinematics (Figure~\ref{fig:d3d4_map}).  However, the GSE population
has low $D_3$ and $D_4$, the high-$D_3$ stars have preferentially young
ages in both the low-Ia and high-Ia populations, and the coldest subset
of the inner thin disk ($R=3-7\kpc$, $Z_{\rm max}<0.2\kpc$) has slightly
elevated mean values of $D_3$ and $D_4$.

\subsection{Prospects and challenges}

The combination of 2-process fitting and residual abundance analysis
is a potentially powerful new tool for interpreting multi-element abundance
measurements in large spectroscopic surveys such as APOGEE, GALAH, and 
SDSS-V.\footnote{This approach may also prove valuable for lower resolution
surveys such as LAMOST and DESI, but its natural application is to data
sets that achieve precision of 0.01-0.05 dex or better for multiple elements
that probe a variety of nucleosynthetic pathways.}
This method has much in common with the conditional PDF method of TW21,
in which one matches stars in [Fe/H], [Mg/Fe], and other abundances or
parameters as desired.  Each method may have practical advantages for
some applications.  The greatest challenge to exploiting these approaches
is fully characterizing the observational contributions to abundance
residuals, to their correlations and systematic trends, and 
to abundance outliers.
One way forward is to make more comprehensive use of repeat observations
(see \S 5.4 of \citealt{Jonsson2020}) to map out the distribution and
correlations of ``statistical'' errors, which arise from photon noise
but also from effects such as telluric line contamination and varying
line spread functions that are difficult to predict from models and simulations.
A second is to exhaustively follow up a large sample of 2-process outliers
and run to ground any observational systematics that give rise to them.
A third is to compare results from different abundance analysis pipelines
to determine which residual abundance correlations and outlier populations
are robust and which are sensitive to analysis choices.  Samples of stars
with observations and abundance measurements from two separate surveys,
such as APOGEE and GALAH, allow a complete end-to-end comparison for
elements in common, as well as extending the number of elements that
trace different astrophysical sources.  For some investigations, 
high-resolution, high-SNR observations of smaller samples that are matched
in stellar parameters, such as ``solar twin'' studies 
\citep{Ramirez2009,Nissen2015,Bedell2018},
may be a valuable complement to the larger samples from massive surveys.

Residual abundance analysis imposes stiff demands on the accuracy of
stellar abundance pipelines.  Even after restricting our sample to
$1 \leq \logg \leq 2.5$ and $4000\K \leq \Teff \leq 4600\K$, we find trends of
residuals with $\Teff$ that we must remove before measuring element-to-element
correlations (Figure~\ref{fig:tefftrend}).  To compare distinct populations
such as bulge and disk or disk and satellites, one must create comparison
samples that are matched in $\logg$ (e.g., \citealt{Griffith2021a};
Hasselquist et al.\ 2021) and/or condition on $\logg$ and $\Teff$ as
variables in addition to abundances (TW21).  Such comparisons would
become more straightforward if the $\logg/\Teff$ systematics in APOGEE
abundances were removed either by empirical calibration or, preferably,
by identifying and correcting the effects that give rise to them.

There are numerous natural follow-ons to this initial effort in residual
abundance cartography, some that can be done with the existing sample,
some requiring similar analysis of different APOGEE subsets, and some
involving new or different observational data.  Systematic examination
of the high-$\chi^2$ population should turn up a variety of physically
unusual stars, perhaps including previously unknown categories.
Comparison of samples with and without binarity signatures in their
radial velocity variations could reveal more subtle impacts than the
C+N/Ce outliers already identified.  Residual abundances offer new ground
for clustering searches in the high-dimensional space of chemistry and
kinematics, especially useful for uncovering populations that could span
a range of $\mgh$ and $\mgfe$.  With the 2-process model ``trained'' on
samples with matched $\logg$ and $\mgh$ ranges, one can compare
residual abundance patterns among the disk, bulge, halo, dwarf satellites,
and star clusters, building on the results of \cite{Griffith2021a} and
Hasselquist et al.\ (2021) and the examples in Figure~\ref{fig:populations}.
A third generation of the APOKASC catalog 
(\citealt{Pinsonneault2014,Pinsonneault2018}; Pinsonneault et al.\ in prep.)
will soon provide asteroseismic masses, ages, and evolutionary states
for $\sim 15,000$ APOGEE stars in DR17.  This sample can be
used to look for more subtle trends of residual abundances with age, to
look for trends with evolutionary state or internal rotation that could
be signatures of non-standard mixing processes, and to disentangle C+N
into separate C and N components (see \citealt{Vincenzo2021b}).
Combinations of APOGEE and GALAH data will provide cross-checks on common
elements and a wider range of elements tracing a greater variety of
nucleosynthetic origins.  In combination with {\it Gaia} space velocities,
residual abundances should be well suited to the program of
Orbital Torus Imaging \citep{Price-Whelan2021}, which exploits the
fact that stellar abundance patterns in steady state may depend on
orbital actions but should be invariant with respect to their 
conjugate angles.
The Milky Way Mapper program of SDSS-V will
obtain APOGEE spectra for an order of magnitude more stars than DR17,
enabling much more comprehensive mapping of disk, bulge, and halo abundance
patterns and much more powerful constraints on clustering in
chemo-dynamical space.

Theoretically, this approach would benefit from a new generation of 
Galactic chemical evolution models that predict joint distributions of
multiple elements from multiple astrophysical sources.  Models that
combine stellar radial migration with
radially dependent gas accretion, star formation, and outflow
histories have achieved impressive (but not
complete) success in reproducing many aspects of the observed joint
distributions of metallicity, $\afe$, age, $R$, and $|Z|$ 
(e.g., 
\citealt{Schoenrich2009,Minchev2013,Minchev2014,Minchev2017,Johnson2021}).
A natural next step is to extend these models to additional elements, using
yields that are theoretically motivated but also empirically constrained
to reproduce observed median trends.  Radial mixing of populations with 
different enrichment histories will then produce fluctuations
in abundances at fixed $\Acc$ and $\AIa$ (or $\mgh$ and $\mgfe$).  
These ``mixture'' models will provide useful guidance for
extending the 2-process formalism, sharpening the ideas outlined
in \S\ref{sec:beyond}.

We suspect that stellar migration alone will prove insufficient
to explain the observed level of residual fluctuations and their correlations.
Radial gas flows and galactic fountains may also be important ingredients
in chemical evolution (e.g., \citealt{Bilitewski2012,Pezzulli2016}), 
but we again suspect that
they will alter mean trends without adding scatter in residual abundances.
Instead we expect that explaining the observed residual covariances will
require models that incorporate localized star formation and gradual
ISM mixing, and it may also require stochastic sampling of the supernova
and AGB populations.  Recent galactic evolution models offer steps in
this direction 
\citep{Armillotta2018,Krumholz2018,Kamdar2019}.
Our results provide a quantitative
testing ground for such models.

Over a decade of observations and increasingly sophisticated data analysis,
APOGEE has obtained an unprecedented trove of high-precision, high-dimensional
stellar abundance data, probing all components of the Milky Way and several
of its closest neighbors.  The combination of 2-process modeling and residual
abundance analysis is one way to exploit the rich complexity of this data set,
taking advantage of its high dimensionality and helping to disentangle the
intertwined impacts of nucleosynthetic yields and Galactic enrichment history.
Systematic application of these tools to APOGEE and its brethren, and
comparison to a range of theoretical models, will teach us much about the 
physics of nucleosynthesis in stars and supernovae, about the processes that
distribute elements through the ISM and into new stellar generations,
and about the particular events that have shaped our galactic home.

\begin{acknowledgments}

DHW gratefully acknowledges the hospitality of the IAS and financial support
of the W.M.\ Keck and Hendricks Foundations during much of this work.
DHW and JAJ are also supported by NSF grant AST-1909841.
JAH acknowledges the support of NSF grant AST-1909897.
YST gratefully acknowledges support of NASA Hubble Fellowship grant
HST-HF2-51425.001 awarded by the Space Telescope Science Institute.
DAGH acknowledges support from the State Research Agency (AEI) of the Spanish
Ministry of Science, Innovation and Universities (MCIU) and the European
Regional Development Fund (FEDER) under grant AYA2017-88254-P.

Funding for the Sloan Digital Sky Survey IV has been provided by the 
Alfred P.\ Sloan Foundation, the U.S.\ Department of Energy Office of
Science, and the Participating Institutions.

SDSS-IV acknowledges support and resources from the Center for High
Performance Computing  at the University of Utah. The SDSS
website is www.sdss.org.

SDSS-IV is managed by the
Astrophysical Research Consortium
for the Participating Institutions
of the SDSS Collaboration including
the Brazilian Participation Group,
the Carnegie Institution for Science,
Carnegie Mellon University, Center for
Astrophysics | Harvard \&
Smithsonian, the Chilean Participation
Group, the French Participation Group,
Instituto de Astrof\'isica de
Canarias, The Johns Hopkins
University, Kavli Institute for the
Physics and Mathematics of the
Universe (IPMU) / University of
Tokyo, the Korean Participation Group,
Lawrence Berkeley National Laboratory,
Leibniz Institut f\"ur Astrophysik
Potsdam (AIP),  Max-Planck-Institut
f\"ur Astronomie (MPIA Heidelberg),
Max-Planck-Institut f\"ur
Astrophysik (MPA Garching),
Max-Planck-Institut f\"ur
Extraterrestrische Physik (MPE),
National Astronomical Observatories of
China, New Mexico State University,
New York University, University of
Notre Dame, Observat\'ario
Nacional / MCTI, The Ohio State
University, Pennsylvania State
University, Shanghai
Astronomical Observatory, United
Kingdom Participation Group,
Universidad Nacional Aut\'onoma
de M\'exico, University of Arizona,
University of Colorado Boulder,
University of Oxford, University of
Portsmouth, University of Utah,
University of Virginia, University
of Washington, University of
Wisconsin, Vanderbilt University,
and Yale University.

\end{acknowledgments}

\bibliographystyle{apj} 
\bibliography{residual_cartography}

\appendix
\section{Tables of $\qxcc$ and $\qxIa$}

Tables~\ref{tbl:qcc} and~\ref{tbl:qIa} report our inferred values of
$\qxcc$ and $\qxIa$, respectively, for all 16 abundances and all 12
bins of $\mgh$.  The $\mgh=0$ and $\mgh=-0.5$ vectors (column 9 and
column 4 of these tables) are plotted in Figure~\ref{fig:2pro_explain}.
Table~\ref{tbl:aratio} gives the ratio of $\AIa/\Acc$ along the
low-Ia and high-Ia sequences, inferred from equation~(\ref{eqn:Aratio})
using the measured median $\femg$ values plotted in Figure~\ref{fig:mgfe}.
These ratios and the values of $\qxcc$ and $\qxIa$ can be used in
equation~(\ref{eqn:xmgratios}) to exactly reproduce the median
$\xmg$ vs.\ $\mgh$ sequences shown by the red and blue points in 
the left panels of Figures~\ref{fig:dataq_alpha}-\ref{fig:dataq_peakodd}.

\begin{deluxetable}{lrrrrrrrrrrrr}[h]
\tablecaption{Values of $\qxcc$}
\label{tbl:qcc}
\tablehead{
\colhead{Elem} & 
\colhead{$\mgh=-0.7$} &
\colhead{-0.6} &
\colhead{-0.5} &
\colhead{-0.4} &
\colhead{-0.3} &
\colhead{-0.2} &
\colhead{-0.1} &
\colhead{0.0} &
\colhead{0.1} &
\colhead{0.2} &
\colhead{0.3} &
\colhead{0.4} 
}
\startdata
Mg & 1.000 &  1.000 &  1.000 &  1.000 &  1.000 &  1.000 &  1.000 &  1.000 &  1.000 &  1.000 &  1.000 &  1.000 \\
O  & 0.994 &  0.973 &  0.954 &  0.943 &  0.934 &  0.929 &  0.926 &  0.923 &  0.906 &  0.870 &  0.844 &  0.800 \\
Si & 1.034 &  0.975 &  0.899 &  0.872 &  0.852 &  0.833 &  0.824 &  0.814 &  0.784 &  0.747 &  0.703 &  0.726 \\
S  & 1.228 &  1.268 &  1.173 &  1.120 &  1.087 &  1.032 &  0.974 &  0.923 &  0.853 &  0.780 &  0.711 &  0.636 \\
Ca & 0.901 &  0.868 &  0.821 &  0.792 &  0.784 &  0.767 &  0.754 &  0.744 &  0.717 &  0.689 &  0.686 &  0.683 \\
C+N & 0.466 &  0.470 &  0.512 &  0.549 &  0.580 &  0.615 &  0.655 &  0.700 &  0.718 &  0.671 &  0.643 &  0.531 \\
Na & 0.346 &  0.410 &  0.446 &  0.483 &  0.523 &  0.552 &  0.594 &  0.620 &  0.582 &  0.419 &  0.275 &  0.279 \\
Al & 0.847 &  0.825 &  0.829 &  0.854 &  0.887 &  0.917 &  0.941 &  0.955 &  0.968 &  0.946 &  0.906 &  0.966 \\
K  & 0.871 &  0.848 &  0.886 &  0.913 &  0.923 &  0.949 &  0.980 &  1.006 &  1.023 &  1.007 &  1.017 &  0.934 \\
Cr & 0.434 &  0.441 &  0.442 &  0.462 &  0.472 &  0.485 &  0.493 &  0.496 &  0.459 &  0.510 &  0.561 &  0.630 \\
Fe & 0.501 &  0.501 &  0.501 &  0.501 &  0.501 &  0.501 &  0.501 &  0.501 &  0.501 &  0.501 &  0.501 &  0.501 \\
Ni & 0.537 &  0.558 &  0.580 &  0.585 &  0.596 &  0.600 &  0.601 &  0.597 &  0.546 &  0.503 &  0.482 &  0.454 \\
V  & 0.717 &  0.679 &  0.678 &  0.678 &  0.679 &  0.703 &  0.729 &  0.735 &  0.692 &  0.588 &  0.573 &  0.739 \\
Mn & 0.265 &  0.272 &  0.292 &  0.310 &  0.332 &  0.341 &  0.354 &  0.360 &  0.320 &  0.207 &  0.165 &  0.272 \\
Co & 0.445 &  0.500 &  0.546 &  0.566 &  0.611 &  0.627 &  0.665 &  0.672 &  0.626 &  0.535 &  0.519 &  0.580 \\
Ce & 0.531 &  0.478 &  0.410 &  0.395 &  0.373 &  0.352 &  0.351 &  0.387 &  0.404 &  0.453 &  0.498 &  1.097 \\
\enddata
\end{deluxetable}

\begin{deluxetable}{lrrrrrrrrrrrr}[]
\tablecaption{Values of $\qxIa$}
\label{tbl:qIa}
\tablehead{
\colhead{Elem} & 
\colhead{$\mgh=-0.7$} &
\colhead{-0.6} &
\colhead{-0.5} &
\colhead{-0.4} &
\colhead{-0.3} &
\colhead{-0.2} &
\colhead{-0.1} &
\colhead{0.0} &
\colhead{0.1} &
\colhead{0.2} &
\colhead{0.3} &
\colhead{0.4} 
}
\startdata
Mg & 0.000 &  0.000 &  0.000 &  0.000 &  0.000 &  0.000 &  0.000 &  0.000 &  0.000 &  0.000 &  0.000 &  0.000 \\
O  & 0.105 &  0.064 &  0.056 &  0.064 &  0.069 &  0.076 &  0.075 &  0.077 &  0.089 &  0.112 &  0.124 &  0.161 \\
Si & 0.018 &  0.061 &  0.131 &  0.141 &  0.158 &  0.180 &  0.181 &  0.186 &  0.208 &  0.245 &  0.289 &  0.264 \\
S  & 0.413 & -0.099 & -0.016 & -0.004 & -0.007 &  0.034 &  0.059 &  0.077 &  0.121 &  0.170 &  0.211 &  0.265 \\
Ca & 0.206 &  0.143 &  0.183 &  0.218 &  0.243 &  0.261 &  0.262 &  0.256 &  0.262 &  0.274 &  0.269 &  0.269 \\
C+N & 0.366 &  0.490 &  0.460 &  0.411 &  0.367 &  0.325 &  0.299 &  0.300 &  0.342 &  0.441 &  0.518 &  0.667 \\
Na & 0.260 &  0.573 &  0.575 &  0.505 &  0.456 &  0.409 &  0.361 &  0.380 &  0.508 &  0.757 &  1.001 &  1.155 \\
Al & -0.353 &  0.178 &  0.202 &  0.179 &  0.148 &  0.105 &  0.073 &  0.045 &  0.020 &  0.041 &  0.077 &  0.018 \\
K  & 0.066 &  0.085 &  0.068 &  0.062 &  0.065 &  0.043 &  0.016 & -0.006 &  0.000 &  0.039 &  0.036 &  0.144 \\
Cr & 0.386 &  0.460 &  0.504 &  0.507 &  0.505 &  0.491 &  0.482 &  0.504 &  0.568 &  0.530 &  0.498 &  0.463 \\
Fe & 0.499 &  0.499 &  0.499 &  0.499 &  0.499 &  0.499 &  0.499 &  0.499 &  0.499 &  0.499 &  0.499 &  0.499 \\
Ni & 0.368 &  0.504 &  0.472 &  0.447 &  0.404 &  0.376 &  0.377 &  0.403 &  0.480 &  0.539 &  0.580 &  0.629 \\
V  & 0.232 &  0.205 &  0.207 &  0.269 &  0.280 &  0.241 &  0.220 &  0.265 &  0.367 &  0.520 &  0.604 &  0.533 \\
Mn & 0.373 &  0.582 &  0.596 &  0.591 &  0.573 &  0.581 &  0.588 &  0.640 &  0.750 &  0.906 &  0.996 &  0.987 \\
Co & 0.256 &  0.438 &  0.435 &  0.439 &  0.371 &  0.340 &  0.300 &  0.328 &  0.424 &  0.556 &  0.624 &  0.649 \\
Ce & 0.369 &  0.504 &  0.640 &  0.738 &  0.883 &  0.950 &  0.815 &  0.613 &  0.487 &  0.317 &  0.205 & -0.407 \\
\enddata
\end{deluxetable}

\begin{deluxetable}{lrrrrrrrrrrrr}[]
\tablecaption{Ratio of $\AIa/\Acc$ along the low-Ia and high-Ia sequences}
\label{tbl:aratio}
\tablehead{
\colhead{Sequence} & 
\colhead{$\mgh=-0.7$} &
\colhead{-0.6} &
\colhead{-0.5} &
\colhead{-0.4} &
\colhead{-0.3} &
\colhead{-0.2} &
\colhead{-0.1} &
\colhead{0.0} &
\colhead{0.1} &
\colhead{0.2} &
\colhead{0.3} &
\colhead{0.4} 
}
\startdata
Low-Ia & 
0.055 &
0.036 &
0.051 &
0.053 &
0.058 &
0.089 &
0.128 &
0.189 &
0.350 &
0.548 &
0.636 &
0.632 \\
High-Ia & 
0.710 &
0.753 &
0.734 &
0.719 &
0.766 &
0.875 &
0.960 &
1.000 &
1.028 &
1.042 &
1.042 &
1.018 \\
\enddata
\end{deluxetable}

\end{CJK*}
\end{document}